\documentclass[12pt,a4paper]{amsart}

\usepackage{empheq}
\usepackage{framed}
\usepackage{amsmath}
\usepackage[latin9]{inputenc}
\usepackage{amstext}
\usepackage{amssymb}
\usepackage{caption}
\usepackage{longtable}
\usepackage{lscape}
\usepackage{tabularx}
\usepackage{varioref}
\usepackage{amsthm}
\usepackage{graphicx}
\usepackage{txfonts}
\usepackage{pxfonts}
\usepackage{marginnote}
\usepackage{epic}
\usepackage{eepic}
\usepackage{float}
\usepackage{rotating}
\usepackage{epsfig}
\usepackage{indentfirst}
\usepackage{array}
\usepackage{varioref}
\usepackage{tikz}
\usepackage{marginnote}
\usepackage{pgf}
\usepackage{bbm}
\usepackage{appendix}
\usepackage{amsbsy}
\usepackage{latexsym}
\usepackage{amsfonts}
\usepackage{pstricks}
\usepackage{color}
\usepackage[numbers,square,comma, compress]{natbib}
\usepackage{eurosym}
\usepackage[a4paper]{geometry}
\usepackage{tikz}
\usepackage{pdfpages}
\usepackage{wasysym}
\usepackage{caption}
\usepackage{subcaption}

\makeatletter

\pdfpageheight\paperheight
\pdfpagewidth\paperwidth

\numberwithin{equation}{section}
\numberwithin{figure}{section}

\numberwithin{equation}{section}
\numberwithin{figure}{section}

\let\emptyset\varnothing

\def\ll{\left\lgroup}
\def\rr{\right\rgroup}

\def\leq{\leqslant}
\def\geq{\geqslant}

\def\ba{\pmb{a}}
\def\bx{\pmb{x}}
\def\by{\pmb{y}}

\def\bV{\pmb{V}}
\def\bW{\pmb{W}}
\def\bY{\pmb{Y}}

\def\bi{\pmb{\iota }}
\def\bj{\pmb{\jmath}} 

\def\i{\iota}

\def\x{x^{(1)}}

\def\ll{ \left\lgroup}
\def\rr{\right\rgroup}

\newcommand{\cA}{\mathcal{A}}
\newcommand{\cB}{\mathcal{B}}
\newcommand{\cC}{\mathcal{C}}

\newcommand{\cM}{\mathcal{M}}
\newcommand{\cN}{\mathcal{N}}
\newcommand{\cO}{\mathcal{O}}
\newcommand{\cP}{\mathcal{P}}

\newcommand{\cR}{\mathcal{R}}
\newcommand{\cS}{\mathcal{S}}

\newcommand{\cW}{\mathcal{W}}

\newcommand{\RR}{\mathbbm{R}}

\newcommand{\eps}{\epsilon}

\newcommand{\pq}{\phantom{q}}
\newcommand{\pt}{\phantom{t}}

\newcommand{\bD}{\pmb{\Delta}}

\newcommand{\1}{{\bf 1.}}
\newcommand{\2}{{\bf 2.}}
\newcommand{\3}{{\bf 3.}}
\newcommand{\4}{{\bf 4.}}
\newcommand{\5}{{\bf 5.}}
\newcommand{\6}{{\bf 6.}}
\newcommand{\7}{{\bf 7.}}
\newcommand{\8}{{\bf 8.}}

\newcommand{\wsq}{\square}
\newcommand{\bsq}{\blacksquare}

\newcommand{\bea}{\begin{eqnarray}\displaystyle}
\newcommand{\eea}{\end{eqnarray}}

\restylefloat{figure}

\begin{document} 
	
\title[]{A Macdonald refined topological vertex}
\dedicatory{To the memory of Professor Petr P Kulish}
	
\author[]{Omar Foda    \!$^{{\scriptstyle {\, 1      }}}$ and 
		  Jian-feng Wu \!$^{{\scriptstyle {\, 2,\, 3}}}$}
		  
	\address{
\!\!\!\!\!\!\!\!\!
$^{{\scriptstyle 1}}$
Mathematics and Statistics, University of Melbourne,
Royal Parade, Parkville, VIC 3010, Australia
\newline
$^{{\scriptstyle 2}}$
Mathematics and Statistics, Henan University,
Minglun Street, Kaifeng city, Henan, China
\newline
$^{{\scriptstyle 3}}$
Institute of Theoretical Physics and Mathematics,
3rd Shangdi Street, Beijing, China}

\email{omar.foda@unimelb.edu.au, muchen.wu@hotmail.com}

	
\begin{abstract}
We consider the refined topological vertex of Iqbal {\it et al.} \cite{iqbal.kozcaz.vafa}, 
as a function of two parameters $\ll x, y \rr$, and deform it by introducing the Macdonald 
parameters $\ll q, t \rr$, as in the work of Vuleti\'c on plane partitions \cite{vuletic.02}, 
to obtain {\it \lq a Macdonald refined topological vertex\rq}. 
In the limit $q \rightarrow t$, we recover the refined topological vertex of Iqbal {\it et al.}
and in the limit $x \rightarrow y$, we obtain a $q t$-deformation of the original topological 
vertex of Aganagic {\it et al.} \cite{aganagic.klemm.marino.vafa}. 
Copies of the vertex can be glued to obtain $q t$-deformed 5D instanton partition functions 
that have well-defined 4D limits and, for generic values of $\ll q, \, t\rr$, contain 
infinite-towers of poles for every pole present in the limit $q \rightarrow t$.   
\end{abstract}
	
\maketitle
	
\section{Introduction}
\label{1}
{\it 
\noindent We start with a motivation for the study of topological vertices, then collect basic 
definitions from combinatorics, and basic facts related to the known topological vertices, their 
connection to partition functions of weighted plane partitions, and outline the purpose of the 
present work.
}
\smallskip  

\subsection{Motivation}
The topological vertex is a combinatorial object that has come to play an increasingly 
important and surprisingly ubiquitous role in mathematical physics since its introduction, 
almost 25 years ago
\footnote{\,
For reviews of topological strings that include an introduction to the topological vertex, 
see \cite{marino.01, marino.02, marino.book, hori.book}, and for a thorough discussion of 
the refined topological vertex, see \cite{kozcaz.thesis}. 
}. 
In theoretical high energy physics, it is the building block of topological string partition 
functions on local toric Calabi-Yau threefolds.  
In enumerative combinatorics, it is the generating function of topological invariants.  
In 2D quantum integrable models, it is the most elementary component in the structure of 
the correlation functions. 
To motivate the present work, we briefly explain the latter point. 
The most important problem in 2D critical phenomena is the computation of the correlation 
functions. This problem was essentially solved in a series of breakthroughs. 

\subsubsection{From 2D correlation functions to conformal blocks} In 1984, Belavin, 
Polyakov and Zamo\-lod\-chi\-kov \cite{belavin.polyakov.zamolodchikov} showed that 
correlation 
functions in conformally-invariant 2D critical phenomena split into holomorphic and 
anti-holomorphic conformal blocks, which are fully-determined by the constraints 
imposed by the infinite-dimensional 2D conformal symmetry. 
However, aside from simple cases such as 4-point blocks that can be computed in terms 
of hypergeometric functions, computing conformal blocks is far from an easy problem, 
if one is interested in explicit expressions.

\subsubsection{From conformal blocks to matrix elements} In 2009, Alday, Gaiotto and 
Tachikawa \cite{alday.gaiotto.tachikawa} conjectured that 2D conformal blocks in 
conformal field theories with Virasoro symmetry are equal to 4D instanton partition 
functions
\footnote{\, 
For reviews of the Alday-Gaiotto-Tachikawa conjecture and developments based on it, 
see \cite{teschner.review}.
}, 
which have been computed in power-series form using localisation \cite{nekrasov}
\footnote{\,  
For reviews of localisation methods, see \cite{pestun.zabzine.review}.
}.
This conjecture has since been proven \cite{alba.fateev.litvinov.tarnopolskiy}, 
and extended to conformal field theories with higher-rank infinite-dimensional algebras 
\cite{wyllard, mironov.morozov}. From the point of view of computing conformal blocks, 
the simplification that Alday {\it et al.} bring is that, by introducing an auxiliary 
scalar free field (whose contributions factorise at the end of a calculation) such that 
the conformal algebra is slightly extended, any conformal block splits into a product of 
matrix elements of a primary fields between arbitrary states. These matrix elements are 
basic Nekrasov partition functions that are completely known in convergent power series form
\footnote{\, 
These power series may not converge as fast as, let's say, Zamolodchikov's recursion 
relations for the 4-point conformal blocks \cite{zamolodchikov.01, zamolodchikov.02}, 
but they are known as simple power sums labeled by Young diagrams. Taking products of 
these matrix elements leads to an explicit power series expression for any conformal 
block, while Zamolodchikov's recursion relations are (currently) known only for the 
4-point blocks.
}.
But, since the matrix elements of Alday {\it et al.} are instanton partition functions, 
they split even further, which takes us to topological strings. 

\subsubsection{From matrix elements to topological vertices} In 2003, Aganagic, Klemm, 
Marino and Vafa \cite{aganagic.klemm.marino.vafa} showed that instanton partition functions 
split into topological vertices
\footnote{\, 
Soon thereafter, Okounkov, Reshetikhin and Vafa \cite{okounkov.reshetikhin.vafa} showed 
that a topological vertex is the generating function of plane partitions that satisfy specific 
boundary conditions, as will be discussed in the sequel. In other words, a topological vertex 
splits into weighted 3D cells, and the difficult problem of computing 2D correlation functions 
is reduced to the combinatorial problem of counting 3D cells with specific boundary conditions!
}. 
In 2005, Awata and Kanno \cite{awata.kanno.01, awata.kanno.02} introduced a refined version 
of the topological vertex, and in 2007, Iqbal, Kozcaz and Vafa \cite{iqbal.kozcaz.vafa} 
introduced another, equivalent version
\footnote{\,
This is the version that we refer to as {\it \lq\lq the refined topological vertex\rq \rq}
in this work.}. 
As we discuss in more detail below, the original topological vertex of Aganagic {\it et al.} 
can be used to construct conformal blocks in Gaussian conformal field theory, while the refined 
topological vertex can be used to construct conformal blocks in more general conformal field 
theories.

\subsubsection{A further-refined topological vertex} In the present work, we derive a topological 
vertex with further refinement. The geometric meaning of these refinements (in terms of 
topological strings on some Calabi-Yau threefold) remains to be understood, and their physical 
applications (what is computed when one glues many copies of these vertices) remains to be studied. 
Since the new refined topological vertex reduces to the known refined vertex in a suitable limit, 
it is clear that it leads to deformations of the results obtained using the refined vertex. However, 
the precise physical meaning of these deformations is outside the scope of this work.

\subsection{Notation}
\subsubsection{Boldface variables}
$\bi$ and also $\bj$ is the set of positive, non-zero integers $\ll 1,   2,   \cdots \rr$,
$\bx = \ll x_1, x_2, \cdots \rr$ and 
$\by = \ll y_1, y_2, \cdots \rr$ are sets of possibly infinitely-many variables, 
$\ba_{-} = \ll a_{-1}, a_{-2}, \cdots \rr$ and   
$\ba_{+} = \ll a_{ 1}, a_{ 2}, \cdots \rr$ are 
the free-boson creation and annihilation modes,
$\bY = \ll Y_1, Y_2 \rr$,        
$\bV = \ll V_1, V_2 \rr$ and 
$\bW = \ll W_1, W_2 \rr$ 
are pairs of Young diagrams, and 
$\pmb{\emptyset} = \ll \emptyset, \emptyset \rr$ is a set of two empty Young diagrams. 

\subsubsection{Primed variables}
Given the variables
$\ll x,        y,        q,        t,        \cdots \rr$, we use 
$\ll x^{\, \prime}, y^{\, \prime}, q^{\, \prime}, t^{\, \prime}, \cdots \rr$, 
for the inverse variables, $x^{\, \prime} = x^{-1}, \cdots$
Given a set of variables 
$\, \bx             = \ll x_1,        x_2,        \cdots \rr$, we use  
$\, \bx^{\, \prime} = \ll x_1^{\, \prime}, x_2^{\, \prime}, \cdots \rr$ 
for the set of inverse variables. 
The Young diagram $Y^{\, \prime}$ is the transpose of $Y$.  
 
\subsubsection{Parameters}
Our parameters $\ll x, y\rr$ are the parameters $\ll q, t \rr$ in \cite{iqbal.kozcaz.vafa}
\footnote{\,
More precisely, the conventions that we use to construct a topological vertex 
are such that our $x$ is $t$, and our $y$ is $q$ in \cite{iqbal.kozcaz.vafa}
}, 
and our $\ll q, t\rr$ are the same Macdonald parameters $\ll q, t\rr$ in 
\cite{vuletic.02}.  

\subsection{Combinatorics}

\subsubsection{Young diagrams}
A Young diagram $Y = \ll y_1, \cdots \rr$, $y_i \geq y_{i+1} \geq 0$, 
is a 2D graphical representation of a partition of an integer 
$|Y| = \sum_{i=1}^N y_i$, see Figure {\bf \ref{A.Young.diagram}}.
It consists of rows $\ll y_1, y_2, \cdots \rr$, 
the $i$-row consists of $y_i$ cells with coordinates $\ll i, j \rr$, that is 
$\ll \wsq_{i, 1}, \wsq_{i, 2}, \cdots, \wsq_{i, y_i} \rr$. 
A generic cells $\wsq_{i j} \in Y$ has coordinates 
$1 \leq i \leq y_1$, 
$1 \leq j \leq   N$, where $N$ is the number of non-zero rows in $Y$.
The Young diagram 
$Y^{\, \prime} = \ll y^{\, \prime}_1, y^{\, \prime}_2, \cdots, y^{\, \prime}_{\, Y_1} \rr$ 
is the transpose of $Y$, where $y_1$ is the length of the top row in $Y$. 

\subsubsection{The arms and legs of a cell}
Consider a cell $\wsq_{i j}$ with coordinates $\ll i, j \rr$ in $\RR^2$, not necessarily 
inside a Young diagram $Y$. We define the lengths of
the           arm $A^{}_\wsq$ and 
the           leg $L^{}_\wsq$
of $\wsq_{i j}$, with respect to the Young diagram $Y$, to be, 

\begin{equation}
A_\wsq  =  y_i        -\, j, 
\quad
L_\wsq  =  y^{\, \prime}_j -\, i,
\label{arm.leg}
\end{equation}

\noindent where $y^{\, \prime}_j$ is the length of the $j$-row in $Y^{\, \prime}$, 
which is the $j$-row in $Y$. We also use the notation, 

\begin{equation}
A^+_\wsq  =  A_\wsq + 1, 
\quad
L^+_\wsq  =  L_\wsq + 1
\label{extended.arm.leg}
\end{equation}

\subsubsection{Remark} $A_{\wsq,Y}$ and $L_{\wsq,Y}$ are both negative for 
$\wsq \notin Y$.

\subsubsection{Sequences}
\label{sequences}
Given a Young diagram $Y$ that consists of an infinite sequence of rows
$Y       = \ll  y_1,     y_2,    \cdots \rr$, such that only finitely-many 
row-lengths are non-zero, together with an infinite sequence of integers
$\bi  = \ll 1, 2, \cdots \rr$, and 
two variables $u$ and $v$,
we define the exponentiated sequences $u^{\, \bi}$ and $v^{\, \pm Y}$,
and the product sequences 
$u^{\, \bi    }\, v^{\, \pm Y}$ and  
$u^{\, \bi - 1}\, v^{\, \pm Y}$, {\it etc.}

\begin{multline}
u^{\, \bi} = \ll u, u^2, \cdots \rr,
\quad
u^{\, \bi - 1} = \ll 1, u,   \cdots \rr,
\quad  
v^{\, \pm Y}    = \ll v^{\pm y_1}, v^{\pm y_2}, \cdots \rr,
\\ 
u^{\, \bi}\, v^{\, \pm Y}  = \ll u  \, v^{\, \pm y_1}, u^2\, v^{\, \pm y_2} \cdots \rr, 
\quad
u^{\, \bi - 1}\, v^{\, \pm Y}  = \ll v^{\, \pm y_1}, u\, v^{\, \pm y_2} \cdots \rr, 
\quad
\textit{etc.}
\end{multline}

%FIGURE.1.1
\begin{figure}
\begin{tikzpicture}[scale=.8]
\draw [thick] (0, 0) rectangle (1,1);
\draw [thick] (1, 0) rectangle (1,1);
\draw [thick] (2, 0) rectangle (1,1);
\draw [thick] (3, 0) rectangle (1,1);
\draw [thick] (4, 0) rectangle (1,1);
\draw [thick] (5, 0) rectangle (1,1);
\draw [thick] (0,-1) rectangle (1,1);
\draw [thick] (1,-1) rectangle (1,1);
\draw [thick] (2,-1) rectangle (1,1);
\draw [thick] (3,-1) rectangle (1,1);
\draw [thick] (4,-1) rectangle (1,1);
\draw [thick] (0,-2) rectangle (1,1);
\draw [thick] (1,-2) rectangle (1,1);
\draw [thick] (2,-2) rectangle (1,1);

\node at (0.5,0.5) {\tiny $1, 1$};
\node at (1.5,0.5) {\tiny $1, 2$};
\node at (2.5,0.5) {\tiny $1, 3$};
\node at (3.5,0.5) {\tiny $1, 4$};
\node at (4.5,0.5) {\tiny $1, 5$};

\node at (0.5,-0.5) {\tiny $2, 1$};
\node at (1.5,-0.5) {\tiny $2, 2$};
\node at (2.5,-0.5) {\tiny $2, 3$};
\node at (3.5,-0.5) {\tiny $2, 4$};
\node at (4.5,-0.5) {$\checkmark$};

\node at (0.5,-1.5) {\tiny $3, 1$};
\node at (1.5,-1.5) {\tiny $3, 2$};

\end{tikzpicture}
\caption{
{\it
The Young diagram $Y$ that corresponds to the partition 
$11 = 5 + 4 + 2$.
The rows are numbered from top to bottom.
The columns are numbered from left to right.
The cell $\wsq_{2, 2} \in Y$ has
$A_{\wsq} = 2$, $A^+_{\wsq} = 3$, $L_{\wsq} = 1$, and $L^+_{\wsq} = 2$.
The cell $\wsq_{2, 5} \notin Y$, indicated with $\checkmark$, has
$A_{\wsq} = -1$, $A^+_{\wsq} = 0$, 
$L_{\wsq} = -1$, and $L^+_{\wsq} = 0$.
The transpose $Y^{\, \prime}$ is the Young diagram that corresponds 
to the partition 
$11 = 3 + 3 + 2 + 2 + 1$
}
}
\label{A.Young.diagram}
\end{figure}

\subsubsection{Plane partitions}
Consider a Young diagram $Y$ and assign each cell $\wsq_{i,\, j}$ 
a non-negative integer $h_{i,\, j}$, such that,  

\begin{equation}
h_{i,\, j} \geq h_{{i + 1},\,j}, 
\quad
h_{i,\, j} \geq h_{i,\,{j+1}}
\end{equation}

\begin{figure}
\centering
\begin{subfigure}{.5\textwidth}
  \centering
  \includegraphics[width=.6\linewidth]{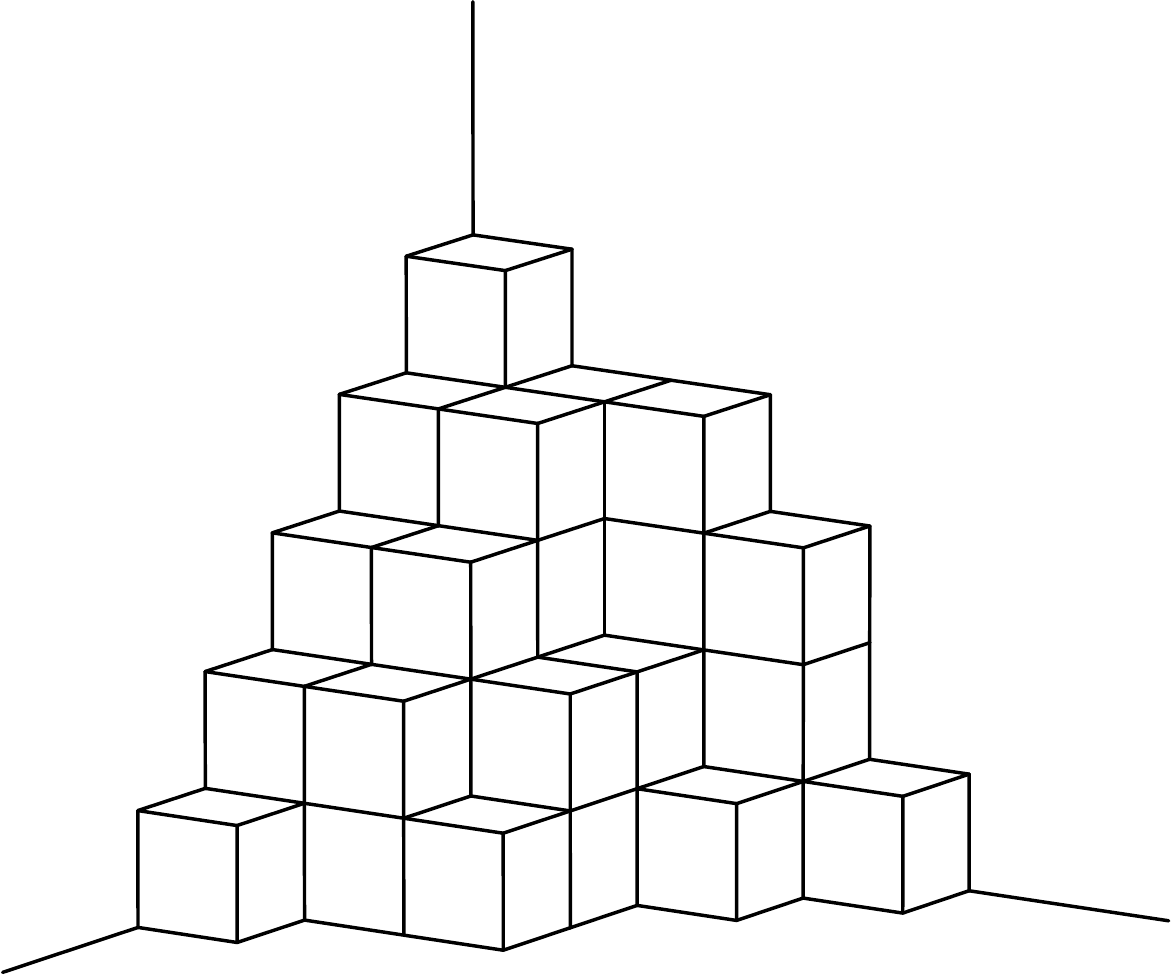}
  \caption{ \it A plane partition}
  \label{figure.plane.partition}
\end{subfigure}%
\begin{subfigure}{.5\textwidth}
  \centering
  \includegraphics[width=.6\linewidth]{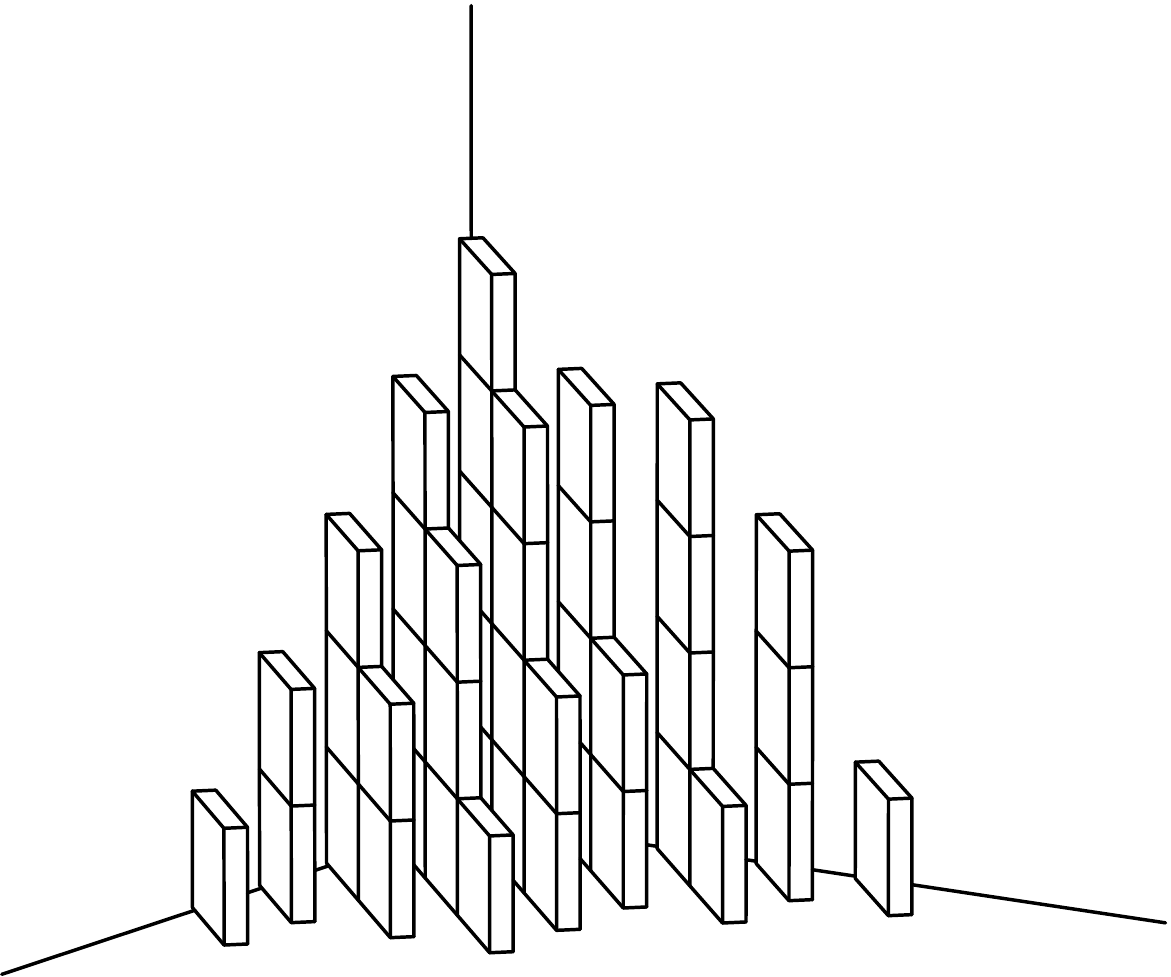}
  \caption{ \it The plane partition on left, diagonally sliced}
  \label{plane.partition.slices}
\end{subfigure}
\caption{}
\label{fig:test}
\end{figure}

\noindent If we consider $h_{i,\, j}$ as a height variable that counts 
the number of boxes placed on top of the cell $\wsq_{i,\, j}$, we obtain 
a plane partition $\pi$, see Figure \ref{figure.plane.partition}

%FIGURE.1.2

\subsubsection{Plane partitions, Young diagrams and vertex operators}
A plane partition $\pi$ can be sliced diagonally into a set of interlacing Young diagrams 
\cite{okounkov.reshetikhin}, see Figure \ref{plane.partition.slices} In \cite{okounkov.reshetikhin}, Okounkov and Reshetikhin 
introduced a method to construct plane partitions {\it via} the action of vertex operators
\footnote{\, 
Expressions for (more general versions of) these operators are given in Section \ref{2}.
},
$\Gamma_{+} \ll x^{\, \prime}_i \rr$, $i \in \ll \cdots, -2, -1 \rr$, and
$\Gamma_{-} \ll x_i             \rr$, $i \in \ll 0, 1, 2, \cdots \rr$ 
\cite{okounkov.reshetikhin}.  These vertex operators act on a Young diagram to generate a sum over all 
possible interlacing Young diagrams. Setting the variables $x_i$ to powers of a single 
parameter $x$, the weight of a box in a plane partition, one generates plane partitions, 
where each partition is weighted by the number of its boxes
\footnote{\
In \cite{okounkov.reshetikhin}, the box-weight parameter is $q$. In the present work, we use $x$, 
and reserve $q$ for one of the two Macdonald deformation parameters.
}. 

\subsection{Topological vertices}

\subsubsection{The original topological vertex}

In \cite{aganagic.klemm.marino.vafa}, Aganagic, Klemm, Marino and Vafa introduced the topological vertex as 
a building block of topological string partition functions
\footnote{\
For a pedagogical introduction to topological strings and the topological vertex, see 
\cite{marino.01, marino.02, marino.book, hori.book, kozcaz.thesis}.
}, 
which are equal to instanton partition functions in 5D $\cN \! = \! 2$ supersymmetric 
Yang-Mills theories with Nekrasov parameters $\eps_1$ and $\eps_2$, such 
that $\eps_1 + \eps_2 = 0$ 
\cite{nekrasov, 
iqbal.kashani.poor.01, 
iqbal.kashani.poor.02, 
eguchi.kanno, 
hollowood.iqbal.vafa}.

\subsubsection{The topological vertex and plane partitions}	
In \cite{okounkov.reshetikhin.vafa}, Okounkov, Reshetikhin and Vafa showed that the topological 
vertex can be constructed in terms of the vertex operators, and that the topological vertex is 
a partition function that counts the number of plane partitions bounded by three Young 
diagrams $Y_1$, $Y_2$ and $Y_3$. It depends on the parameter $x$, the weight of a box 
in a plane partition.
	
\subsubsection{Perpendicular {\it vs.} diagonal boundaries}
\label{slicing}
We consider plane partitions that live in the positive octant of $\RR^3$ with coordinates 
$x, y, z \geq 0$. These plane partitions are bounded by three Young diagrams, $Y_1, Y_2$ 
and $Y_3$ that intersects the $x$-, $y$- and $z$-axis respectively. There are different 
possible choices for the exact ways that these Young diagrams intersect the respective 
axes. For example, the Young diagram $Y_1$ on the right boundary can be either perpendicular 
to the $x$-axis, or parallel to the main diagonal in the $xy$-plane.
In the present work, we choose 
$Y_1$ and $Y_2$ to intersect the $x$-axis and the $y$-axis respectively, while parallel 
to the main diagonal in the $xy$-plane, and choose $Y_3$ to intersect the $z$-axis 
while parallel to the $xy$-plane. Any other choice, particularly the choice in which 
$Y_1, Y_2$ and $Y_3$ intersect the $x$-, $y$- and $z$-axis respectively, diagonally,
leads expressions for the topological vertices with extra {\it \lq framing\rq} factors
that cancel out in the final result when computing instanton partition functions, 
conformal blocks, {\it etc.}

\subsubsection{The refined topological vertex of Awata and Kanno}
In \cite{awata.kanno.01, awata.kanno.02}, Awata and Kanno introduced a refined 
topological vertex, that depends on two parameters, as a building block of refined 
topological string partition functions. These partition functions are equal to 
instanton partition functions in 5D 
$\cN \! = \! 2$ supersymmetric Yang-Mills theories with Nekrasov parameters such 
that $\eps_1 + \eps_2 \neq 0$ \cite{nekrasov, awata.yamada.01, awata.yamada.02}. 
This construction is not based on vertex operators and will not be used in this 
work
\footnote{\,
We further comment on the refined topological vertex of Awata and Kanno in 
{\bf \ref{comparison.with.awata.kanno}}.
}.
	
\subsubsection{The refined topological vertex of Iqbal, Kozcaz and Vafa}
In \cite{iqbal.kozcaz.vafa}, Iqbal, Kozcaz and Vafa constructed a refined topological vertex 
using the same vertex operators, but in the final expressions, the parameters 
$x_i$ of $\Gamma_{-} \ll x_i \rr$ are set to powers of a parameter $x$, while 
those of $\Gamma_{+} \ll x^\prime_i \rr$ are set to powers of a different parameter $y \neq x$
\footnote{\
In \cite{iqbal.kozcaz.vafa}, the box-weight parameters are $\ll q, t \rr$. In the present work, 
we use $\ll x, y\rr$ and, following \cite{vuletic.02}, we reserve $\ll q, t\rr$ for 
the Macdonald deformation parameters.
}.  
In the limit $y \rightarrow x$, the refined topological vertex reduces to the original
topological vertex of \cite{aganagic.klemm.marino.vafa}. 
	
The refined topological vertex is also a partition function that counts the number of 
plane partitions bounded by three Young diagrams $Y_1$, $Y_2$ and $Y_3$. It depends on 
two parameters $x$ and $y$ since the weight of a box in a plane partition is $x$ or $y$ 
depending on its position in the plane partition.

\subsubsection{Remark} In \cite{awata.feigin.shiraishi}, Awata, Feigin and Shiraishi 
showed that the refined topological vertex of Awata and Kanno \cite{awata.kanno.01, 
awata.kanno.02}, and that of Iqbal {\it et al.} \cite{iqbal.kozcaz.vafa} are the same 
object, expressed in two different symmetric function bases. In the sequel, we use 
{\it \lq the refined topological vertex\lq\,} to refer to that of Iqbal {\it et al.} 
	
\subsection{MacMahon's partition function and its refinements}

\subsubsection{MacMahon's partition function}
Choosing the Young diagrams that label the boundaries of the topological vertex to be 
empty, $Y_1 = Y_2 = Y_3 = \emptyset$, where $\emptyset$ is the trivial Young diagram 
with no cells, the topological vertex reduces to MacMahon's partition function of the
set of random plane partitions $\pi \in \cP$, 
	
\begin{equation}
\sum_{\pi\, \in\, \cP} x^{\, |\, \pi\, |\,} = 
\prod_{i = 1}^\infty \ll \frac{1}{1\, -\, x^i} \rr^i
\label{x.macmahon.function}
\end{equation}
	
\subsubsection{The $y$-refined MacMahon partition function}
Setting the Young diagrams that label the boundaries of the refined topological vertex 
to be empty, $Y_1 = Y_2 = Y_3 = \emptyset$, the refined topological vertex reduces to 
a one-parameter $y$-refinement of MacMahon's partition function, 
	
\begin{equation}
\sum_{\pi\, \in\, \cP} 
x^{\,\sum_{i=1}^\infty |\, Y_{ -i}\, |}\, 
y^{\,\sum_{j=1}^\infty |\, Y_{j-1}\, |}
= 
\prod_{i, j = 1}^\infty 
\ll 
\frac{1}{1\, -\, x^{i}\, y^{j-1}}
\rr, 
\label{x.y.macmahon.function}
\end{equation}

\noindent where $|\, Y_i\,|$ is the number of cells in the Young diagram $Y_i$ 
at the $i$-diagonal slice of the plane partition $\pi$ in the set of random 
plane partitions $\cP$. In the limit $y \rightarrow x$, the right hand side of 
Equation \ref{x.y.macmahon.function} reduces to that of 
Equation \ref{x.macmahon.function}.

\subsubsection{The $q t$-MacMahon partition functions}
In \cite{vuletic.02}, Vuleti\'c introduced a Macdonald $\ll q, t \rr$-refinement 
of MacMahon's plane-\-partition partition function, 

\begin{equation}
\sum_{\pi\, \in\, \cP} 
F^{\, q \, t}_{\pi}\, x^{\,|\, \pi\, |} 
=
\prod_{i, \, \ll n + 1 \rr \,  = \, 1}^\infty 
\ll 
\frac{1\, -\, x^i\, q^{\, n}\,   t}
     {1\, -\, x^i\, q^{\, n}\, \pt}
\rr^i, 
\label{x.q.t.macmahon.function}
\end{equation}
	
\noindent where $F^{\, q \, t}_{\pi}$ is a function of $q$ and $t$ that specifies 
the dependence of a plane partition $\pi$ on the parameters $\ll q, t \rr$. 
The precise form of $F^{\, q \, t}_{\pi}$ is not needed here, and can be found 
in \cite{vuletic.02}. 
In the limit $q \rightarrow t$, the right hand side of 
Equation \ref{x.q.t.macmahon.function} reduces to that of 
Equation \ref{x.macmahon.function}. 
	
\subsubsection{Remarks on the literature}
In \cite{foda.wheeler.02, cai}, the $q t$-vertex operators of Shiraishi 
{\it et al.} \cite{shiraishi.01} were used to derive the right hand side of
Equation \ref{x.q.t.macmahon.function}. A proof of the left hand side of Equation 
\ref{x.q.t.macmahon.function} requires identities that involve Macdonald functions. 
A proof of these identities was not attempted in \cite{foda.wheeler.02}, but relevant 
identities were derived in \cite{cai}. 
A version of these operators was discussed in Section {\bf 6.2} of \cite{kozcaz.thesis}, 
but was not used to build a topological vertex of the type discussed in the present work.
In the limit $t \rightarrow -1$, Equation \ref{x.q.t.macmahon.function} reduces to 
the partition function of strict plane partitions of \cite{foda.wheeler.01, vuletic.01}, 
a vertex-operator derivation of which was obtained in \cite{foda.wheeler.02}.
In this special case, the vertex operators are based on neutral free-fermions. 
A topological vertex that reduces to this $\ll t = -1\rr$-weighted partition function 
was derived in \cite{drissi}.
	
\subsection{The present work}
	
\subsubsection{Combining the $y$-refinement and the $q t$-refinement}
	
It is natural to expect that one can combine the $y$-refinement of Iqbal {\it et al.} 
and the $q t$-refinement of Vuleti\'c. In the present work, we show 
that this is straightforward to do, and that the answer is, 
	
\begin{equation}
\sum_{\pi\, \in\, \cP} 
F^{\, q \, t}_{\pi}\, 
x^{\,\, \sum_{i = 1}^\infty |\, Y_{  - i}\, |}\,
y^{\,\sum_{j = 1}^\infty |\, Y_{j - 1}\, |}
= 
\prod_{i,\,  j, \, \ll n + 1 \rr \, = \, 1}^\infty 
\ll 
\frac{1\, -\, x^{ i}\, y^{j-1}\, q^{\, n}\,   t}     
     {1\, -\, x^{ i}\, y^{j-1}\, q^{\, n}\, \pt}
\rr, 
\label{x.y.q.t.macmahon.function.01}
\end{equation} 
	
\noindent where the left hand side is written in a way that makes it clear that 
the $y$-deformation of Iqbal {\it et al.} and the $q t$-deformation 
of Vuleti\'c are independent
\footnote{\,
The observation that the $y$-refinement of Iqbal {\it et al.} \cite{iqbal.kozcaz.vafa}, 
and the $q t$-deformation of Vuleti\'c are {\it \lq orthogonal\rq}, 
in the sense that they change the weights of the plane partitions in 
independent ways, is the starting point of this work. 
}.

\subsubsection{The Macdonald kernel function}	
\label{ding.iohara.miki.algebra}
The product expression on the right hand side of Equation \ref{x.y.q.t.macmahon.function.01} 
is a specialization of the kernel function of the Macdonald operator 
\cite{saito.01, saito.02, saito.03}.

\subsubsection{Ding-Iohara-Miki algebra}
The free-field realization of the Macdonald operator naturally gives rise to Ding-Iohara-Miki 
algebra \cite{ding.iohara, miki, feigin.01}. This points to a connection between the
$q t$-deformation of the refined topological vertex and Ding-Iohara-Miki algebra.
	
\subsubsection{A Macdonald refined topological vertex}
More generally, it is natural to expect that the Macdonald refined MacMahon function, 
Equation \ref{x.y.q.t.macmahon.function.01} is the 
$Y_1 = Y_2 = Y_3 = \emptyset$ limit of a Macdonald refined topological vertex
\footnote{\,
One can also think of this object as {\it a doubly-refined topological vertex}.
However, as we will see, the Macdonald parameters $\ll q, t\rr$ appear in a distinct 
way from the $y$-refinement parameter of Iqbal {\it et al.}
}.
In the present work is the derivation of this object. 

\subsubsection{Notation}
We use 
$\cO_{Y_1 Y_2 Y_3} \ll x \rr$ for the original topological vertex of Aganagic {\it et al.},
$\cR_{Y_1 Y_2 Y_3} \ll x, y \rr$ for the refined topological vertex of Iqbal {\it et al.}, 
$\cR^{\, AK}_{Y_1 Y_2 Y_3} \ll x, y \rr$ for the refined topological vertex of Awata and Kanno, and 
$\cM^{\, q \, t}_{Y_1 Y_2 Y_3} \ll x, y \rr$ for the Macdonald refined topological vertex derived in 
the present work, which we refer to as {\it \lq the Macdonald vertex\rq}.

\subsubsection{Limits of the Macdonald vertex} 
$\cM^{\, q \, t}_{Y_1 Y_2 Y_3} \ll x, y \rr$ depends on $\ll x, y, q, t \rr$. 
In the limit 
$q \rightarrow t$, we recover $\cR_{Y_1 Y_2 Y_3} \ll x, y \rr$.
In the limit 
$x \rightarrow y$, we obtain a $q t$-version 
of $\cO_{Y_1 Y_2 Y_3} \ll x \rr$.
In the limit 
$q \rightarrow 0$, 
we obtain a Hall-Littlewood version of $\cR_{Y_1 Y_2 Y_3} \ll x, y \rr$, 
then taking $t \rightarrow -1$, we obtain the vertex of Drissi {\it et al.} 
\cite{drissi}.
There are many obvious variations on the above limit, including that obtained 
by taking
$q \rightarrow t^{\, \alpha}$, 
then $t \rightarrow 1$ to obtain a Jack version of $\cR_{Y_1 Y_2 Y_3} \ll x, y \rr$, 
then $y \rightarrow x$ to obtain a Jack version of $\cO_{Y_1 Y_2 Y_3} \ll x \rr$, 
and so on.
	
\subsubsection{4D limits of the Macdonald vertex}
	
All of the above topological vertices can be glued to form instanton partition functions 
of 5D supersymmetric Yang-Mills theories. It is possible to take a 4D limit of each of 
these topological vertices. We consider the 4D of $\cM^{\, q \, t}_{Y_1 Y_2 Y_3} \ll x, y \rr$ in 
Section {\bf \ref{7}}.
	
\subsection{On the choice of parameters}
In \cite{okounkov.reshetikhin.vafa}, the parameter that counts the number of boxes in a plane partition is $q$. 
In \cite{iqbal.kozcaz.vafa}, the parameter that refines this counting is $t$.
In \cite{vuletic.02}, the parameter that counts the number of boxes is $s$, 
there is no refinement in the sense of \cite{iqbal.kozcaz.vafa}, but there are Macdonald-type 
deformation parameters $\ll q, t\rr$. 
In the present work, the parameter that counts the number of boxes is $x$, 
its refinement in the sense of Iqbal {\it et al.} \cite{iqbal.kozcaz.vafa} is $y$.
We find this choice of variables natural in the sense that our $\ll x, y\rr$ 
are related to the arguments of the vertex operators, which are position variables. 
Our deformation parameters $\ll q, t\rr$ are those of \cite{vuletic.02}. 
They are Macdonald-type parameters in the sense that our $q t$-bosons are 
in bijection with the power-sum symmetric functions that the Macdonald functions 
are expanded in, our $q t$-Heisenberg commutation relations are in agreement 
with the power-sum inner products, and in the limit $q \rightarrow t$, both 
parameters drop out of all expressions.
	
\subsection{Outline of contents}
In Section 
\textbf{\ref{2}}, we recall basic facts related to Macdonald functions, in 
\textbf{\ref{3}}, we do the same for free-bosons and vertex operators, and in 
\textbf{\ref{4}}, we recall the isomorphism of power-sum symmetric functions and 
the generators of the Heisenberg algebra which are free-boson mode operators. 
In \textbf{\ref{5}}, we construct the Macdonald refined topological vertex, and 
in \textbf{\ref{6}}, we make remarks on its structure.
In \textbf{\ref{7}}, we glue four Macdonald refined topological vertices to construct 
the Macdonald refined $U \! \ll 2 \rr$ \lq strip\rq\, partition function, that is the 
building block of the Macdonald refined topological string partition functions, then 
in \textbf{\ref{8}}, we compute the 4D limits of the objects computed in \textbf{\ref{6}}.
In \textbf{\ref{9}}, we glue two strips to obtain 
the $\cN = 2$ $U\, \ll 2 \rr$ instanton partition function 
that is equal to a 4-point $q t$-conformal block on a sphere, and take its 4D limit, then 
in \textbf{\ref{10}}, we compute 
the $\cN = 2^\star$ $U\, \ll 2 \rr$ instanton partition function 
that is equal to a 1-point $q t$-conformal block on a torus, and take its 4D limit. 
In \textbf{\ref{11}}, we conclude with a number of remarks.
	
\section{Macdonald symmetric functions}
\label{2}
{\it Our starting point is Macdonald function theory. We list basic definitions related 
to Macdonald symmetric functions that are used 
in subsequent sections. We give detailed references to \cite{macdonald.book}, and refer 
to it for a complete presentation. The aim of this section is to outline the derivation 
of the Cauchy identities in 
\ref{cauchy.identity.skew.01},
\ref{cauchy.identity.skew.02}, and 
\ref{cauchy.identity.skew.03}}
\smallskip 
	
\subsection{Notation}
Let $Y$ be a Young diagram that consists of $m$ non-zero parts, 
$Y =   \ll y_1, \cdots, y_m \rr$, and $\bx$ a set of $n$ variables, 
$\bx = \ll x_1, \cdots, x_n \rr$, such that $m \leq n$. We use the notation, 
	
\begin{equation}
\bx_{\bi}^Y = \ll x_{\iota_1}^{y_1}, \, \cdots, \, x_{\iota_m}^{y_m} \rr, 
\label{sum}
\end{equation}
	
\noindent where $\bi = \ll \iota_1, \cdots, \iota_m \rr$ is defined as 
follows. We start from the set of $n$ integers, ${\pmb n} = \ll 1, \cdots, n \rr$, 
for example, ${\pmb n} = \ll 1, 2, 3, 4 \rr$, 
choose a subset of $m$ integers ${\pmb m}$, such that ${\pmb m} \subseteq {\pmb n}$, 
for example, ${\pmb m} = \ll 1, 2, 4 \rr$, 
and consider a permutation $\bi$ of ${\pmb m}$, 
for example, $\bi = \ll 2, 4, 1 \rr$.
The sum on
the right hand side of Equation \ref{sum} is over {\it all  distinct} 
permutations $\bi$, of {\it all  distinct} subsets ${\pmb m} \subseteq {\pmb n}$
\footnote{\,
In the special case of ${\pmb m} = {\pmb n}$, $\bi$ is a permutation of ${\pmb m}$, 
and the sum over $\bi$ is a sum over all distinct permutations of ${\pmb m}$.
}.

\subsection{The monomial symmetric functions}
The monomial symmetric function $m_Y \ll \bx \rr$ is
\footnote{\, 
Ch. I, p. 18, Equation 2.1, in \cite{macdonald.book}
}, 
	
\begin{equation}
m_Y \ll \bx \rr = \sum_{\bi} x_{\bi}^Y,  
\end{equation}

\noindent where the sum runs over all  distinct permutations of the set $\bi$. 
For example, 

\begin{equation}
m_\emptyset \ll \bx \rr = 1,            \quad	
m_1         \ll \bx \rr = \sum_i x_i,   \quad
m_2         \ll \bx \rr = \sum_i x_i^2, \quad 
m_{443}     \ll x_1, \cdots, x_n \rr = \sum_{\bi} x_i^4 \, x_j^4 \, x_k^3, 	
\end{equation}

\noindent where the sum in the last example is over all distinct permutations $\bi$, 
of all  distinct subsets ${\pmb m} \subseteq {\pmb n} = \ll 1, 2, \cdots, n \rr$, 
of cardinality $| \, {\pmb m} \, | \geq 3$, and $i \neq j \neq k \in {\pmb m}$
	
\subsection{The power-sum symmetric functions}
Given a possibly-infinite set of variables 
$\bx = \ll x_1, x_2, \cdots \rr$, the power-sum function $p_n \ll \bx \rr$, 
$n \in \ll 0, 1, \cdots \rr$, is
\footnote{\,
Ch. I, p. 23, in \cite{macdonald.book}
}, 
	
\begin{equation}
p_0 \ll \bx \rr = 1,
\quad  
p_n \ll \bx \rr = \sum_i x_i^{\, n} = m_n \ll \bx \rr,\, n \in \ll 1, 2, \cdots \rr,  
\end{equation}
	
\noindent and the power-sum function $p_Y \ll \bx \rr$ indexed by 
$Y = \ll y_1, y_2, \cdots \rr$ is
\footnote{\,
Ch. I, p. 24, in \cite{macdonald.book}
}, 
	
\begin{equation}
    p_Y     \ll \bx \rr\, =\, 
    p_{\, Y_1} \ll \bx \rr\, 
    p_{\, Y_2} \ll \bx \rr\, \cdots
\end{equation}

\subsection{The $q t$-inner product of the power-sum functions}
\label{power.sum.inner.product.macdonald.basis}
Consider the ring of symmetric functions in a set of variables $\bx = \ll x_1, x_2 \cdots \rr$, 
with coefficients in the field of rational functions in two variables $\ll q, t\rr$. In this case, 
the power-sum functions are orthogonal in the sense of the inner product
\footnote{\,
Ch. VI, p. 225, Equation 4.11 in \cite{macdonald.book}
}, 
	
\begin{equation}
\langle\, p_{\, Y_1} \ll \bx \rr\, |\, p_{\, Y_2} \ll \bx \rr \rangle_{\, q \, t} = 
z^{\, q \, t}_{\, Y_1}\, \delta_{Y_1 Y_2},
\quad
z^{\, q \, t}_Y 
=
\ll
1^{n_1} \ll n_1 ! \rr
2^{n_2} \ll n_2 ! \rr 
\cdots
\rr
\prod_{i=1}^{y_1^{\, \prime}} 
\ll
\frac{1 - q^{\, y_i}}
     {1 - t^{\, y_i}}
\rr 
\label{young.diagram.power.sum.inner.product}
\end{equation} 
 
\subsubsection{Remark}
The inner product in Equation \ref{young.diagram.power.sum.inner.product} can be understood
as follows
\footnote{\, 
Ch. I, p. 75--76, in \cite{macdonald.book}
}. For every power-sum 
function $p_Y \ll \bx \rr$, there is a dual differential operator 
$D_Y \ll \bx \rr$ in $\bx = \ll x_1, x_2, \cdots \rr$, such that acting with 
$D_Y \ll \bx \rr$ on $p_Y \ll \bx \rr$, then setting 
$x_1 = x_2 = \cdots = 0$, one obtains the right hand side of Equation 
\ref{young.diagram.power.sum.inner.product}.

\subsection{An identity}
The power-sum functions $p_n \ll \bx \rr$ satisfy the identity
\footnote{\,
Ch. VI, p. 310, in \cite{macdonald.book}}, 

\begin{equation}
\prod_{n \, =\, 1}^\infty 
\exp 
\ll
\frac{1}{n} 
\ll
\frac{1 - t^{\, n}}{1 - q^{\, n}}
\rr 
p_n \ll \bx \rr 
p_n \ll \by \rr
\rr
=
\prod_{i,\, j, \, \ll n + 1 \rr \, = \, 1}^\infty
\ll 
\frac{1\, -\, x_i\, y_j\, q^{\, n}\,t}
     {1\, -\, x_i\, y_j\, q^{\, n}\, \pt}
\rr, 
\label{an.exponential.is.a.product}
\end{equation}

\subsubsection{Remark} The right hand side of \ref{an.exponential.is.a.product} is the Macdonald 
kernel. It specializes to the right hand side of Equation \ref{x.y.q.t.macmahon.function.01}.

\subsection{The Macdonald function} $P^{\, q \, t}_Y \ll\, \bx\, \rr$,
$Y = \ll y_1, y_2, \cdots \rr$, 
is the unique symmetric function in $\bx = \ll x_1, x_2, \cdots \rr$, where the cardinality 
$| \, \bx \, | \geq y_1$, that can be written as
\footnote{\,
Ch. VI, p. 322, in \cite{macdonald.book}
}, 
	
\begin{equation}
P^{\, q \, t}_{\, Y_1} \ll\, \bx\, \rr = 
\sum_{Y_\1 \succeq Y_\2} u^{\, q \, t}_{\, Y_1\, Y_2}\, m_{\, Y_2} \ll \bx \rr, 
\quad 
u^{\, q \, t}_{\, Y_1\, Y_1} = 1, 
\end{equation}
	
\noindent where $m_Y \ll \bx \rr$ is the monomial symmetric function labelled by $Y$,
and satisfies the orthogonality relation, 

\begin{equation}
\langle 
P^{\, q \, t}_{\, Y_1} \ll\, \bx\, \rr\, |\, 
P^{\, q \, t}_{\, Y_2} \ll\, \bx\, \rr 
\rangle = 0, 
\quad
\textit{for}
\quad
Y_1 \neq Y_2
\label{macdonald.orthogonality}
\end{equation}

\subsubsection{Remarks}
The inner product in Equation \ref{macdonald.orthogonality} can be understood as follows.
Expanding $P^{\, q \, t}_Y \ll\, \bx\, \rr$ in terms of power-sum functions $p_n$, 
$n \in \ll 0, 1, \cdots \rr$, Equation \ref{macdonald.orthogonality} follows from Equation 
\ref{young.diagram.power.sum.inner.product}

\subsection{The dual Macdonald function} $Q_Y^{\, q \, t} \ll\, \bx\, \rr$ is 
defined in terms of $P^{\, q \, t}_Y \ll\, \bx\, \rr$ as
\footnote{\,
Ch. VI, p. 323, Equation 4.12, in \cite{macdonald.book}
}
\footnote{\,
Ch. VI, p. 339, Equation 6.19, in \cite{macdonald.book}
},
	
\begin{equation}
Q_Y^{\, q \, t} \ll\, \bx\, \rr  = 
b^{\, q \, t}_Y   
\, 
P^{\, q \, t}_Y \ll\, \bx\, \rr, 
\quad  
b^{\, q \, t}_Y = 
\prod_{\wsq\, \in\, Y}
\ll 
\frac{
1\, -\,q^{A_{\wsq, Y}  }\, t^{L_{\wsq, Y}^+}
}{
1\, -\,q^{A_{\wsq, Y}^+}\, t^{L_{\wsq, Y}  }
}
\rr
\label{dual.macdonald}
\end{equation}
	
\subsection{The $q t$-involution}
There is a $q t$-operator $\omega_{q t}$ that acts on power-sum functions, 
$p_n \ll \bx \rr$ as
\footnote{\, 
Ch VI, p. 312, Equation 2.14 in \cite{macdonald.book}
}, 

\begin{equation}
\omega_{\, q \, t}  \ll p_n \ll \bx \rr \rr  = 
\ll\, -\, \rr^{n-1} \ll \frac{1-q^{\, n}}{1-t^{\, n}} \rr\, p_n \ll \bx \rr, 
\label{involution.on.power.sum}
\end{equation} 

\noindent which is a $q t$-involution in the sense that the action of $\omega_{\, q \, t}$ 
is followed by the action of $\omega_{\, t\, q}$, and {\it vice versa}, so that, 

\begin{equation}
\omega_{\, t\, q}\, \ll\, \omega_{\, q \, t} \ll\, p_n \ll \bx \rr \, \rr \rr = p_n \ll \bx \rr,  
\end{equation}

\noindent and consequently, $\omega_{\, q \, t}$ acts on the Macdonald functions as
\footnote{\,
Ch. VI, p. 327, in \cite{macdonald.book}
}, 
     
\begin{equation}
\omega_{\, q \, t} 
\ll
\, P^{\, q \, t}_Y          \ll \bx \rr 
\rr = 
\, Q^{\, t\, q}_{Y^{\, \prime}} \ll \bx \rr, 
\quad 
\omega_{\, q \, t}\,
\ll 
Q^{\, q \, t}_Y            \ll \bx \rr\,
\rr  
= 
P^{\, t\, q}_{Y^{\, \prime}}   \ll \bx \rr
\label{involution.on.macdonald}
\end{equation}
	
\subsubsection{Remark} $\omega_{q t}$ exchanges 
$P^{\, q \, t}_Y               \ll \bx \rr \rightleftharpoons 
 Q^{\, t\, q}_{Y^{\, \prime}} \ll \bx \rr$, 
which includes 
$Y \rightleftharpoons Y^{\, \prime}$, and $q \rightleftharpoons t$. 
	
\subsection{The $q t$-inner product of the Macdonald functions} 
The Macdonald functions satisfy the $q t$-inner product
\footnote{\,
Ch. VI, p. 324, in \cite{macdonald.book}
}, 
	
\begin{equation}
\langle\,
P^{\, q \, t}_{\, Y_1}  \ll\, \bx\, \rr
\, |\, 
Q^{\, q \, t}_{\, Y_2}  \ll\, \bx\, \rr 
\rangle = 
\delta_{Y_1 Y_2}, 
\label{macdonald.inner.product}
\end{equation}
	
\noindent that is, 
$P^{\, q \, t}_{\, Y_1}  \ll\, \bx\, \rr$ and 
$Q^{\, q \, t}_{\, Y_2}  \ll\, \bx\, \rr$ span adjoint bases
with the respect to the $q t$-inner product, Equation \ref{macdonald.inner.product}.
Following \cite{macdonald.book}, 
$P_Y^{\, q \, t} \ll\, \bx\, \rr$ form a          complete orthogonal basis, and 
$Q_Y^{\, q \, t} \ll\, \bx\, \rr$ form an adjoint complete orthogonal basis, 
in the ring of symmetric functions in $\bx$, with coefficients in the field of 
rational functions in $\ll q, t\rr$.
	
\subsection{Cauchy identities for the Macdonald and dual Macdonald functions}
$P^{\, q \, t}_Y \ll\, \bx\, \rr$ and 
$Q^{\, q \, t}_Y \ll\, \bx\, \rr$ satisfy the Cauchy identity
\footnote{\,
Ch. VI, p. 324, Equation 4.13, in \cite{macdonald.book}
}, 
	
\begin{equation}
\sum_Y 
P^{\, q \, t}_Y \ll\, \bx\, \rr\, 
Q^{\, q \, t}_Y \ll\, \by\, \rr  = 
\prod_{i,\, j,\, \ll n + 1 \rr \, = \, 1}^\infty
\ll 
\frac{1\, -\, x_i\, y_j\, q^{\, n}\,   t}
     {1\, -\, x_i\, y_j\, q^{\, n}\, \pt}
\rr
\label{macdonald.cauchy.identity}
\end{equation}

\noindent Using the involution $\omega_{q t}$, 
Equation \ref{involution.on.macdonald}, on the Macdonald functions in the $\bx$-variables 
in Equation \ref{macdonald.cauchy.identity}, then in the $\by$-variables
\footnote{\,
Ch. VI, p. 329, Equation 5.4, in \cite{macdonald.book}
}, 

\begin{equation}
\sum_Y
Q^{\, t\, q}_{Y^{\, \prime}} \ll\, \bx\, \rr
Q^{\, q \, t}_Y          \ll\, \by\, \rr\, 
=
\sum_Y 
P^{\, q \, t}_Y          \ll\, \bx\, \rr
P^{\, t\, q}_{Y^{\, \prime}} \ll\, \by\, \rr\, 
= 
\prod_{i,\, j\, = \, 1}^\infty \ll 1 + x_i\, y_j \rr
\label{more.cauchy.identities}
\end{equation}

\subsubsection{Remark} From Equations 
\ref{macdonald.cauchy.identity} and 
\ref{more.cauchy.identities}
\footnote{\,
Ch. VI, p. 313, Equation 2.18, in \cite{macdonald.book}
}, 

\begin{equation}
\omega_{q t}\,
\ll 
\prod_{i,\, j,\, \ll n + 1 \rr \, = \, 1}^\infty
\ll 
\frac{1\, -\, x_i\, y_j\, q^{\, n}\,   t}
     {1\, -\, x_i\, y_j\, q^{\, n}\, \pt}
\rr
\rr 
= 
\prod_{i, j = 1}^\infty \ll 1 + x_i\, y_j \rr
\end{equation}

\subsection{Notation}
In the sequel, we use 
\footnote{\,
Ch. VI, p. 309, Equation 2.5, in \cite{macdonald.book}
}
\footnote{\,
Ch. VI, p. 352, in \cite{macdonald.book}
},

\begin{equation}
\Pi^{\, q \, t} \ll \bx, \by \rr 
=  
\prod_{i,\, j,\, \ll n + 1 \rr \, = \, 1}^\infty
\ll 
\frac{1\, -\, x_i\, y_j\, q^{\, n}\,   t}
     {1\, -\, x_i\, y_j\, q^{\, n}\, \pt}
\rr, 
\quad 
\Pi_0 \ll \bx, \by \rr\, =\,
\prod_{i, j = 1}^\infty \ll 1 + x_i\, y_j \rr
\label{PI.01}
\end{equation}

\subsection{Structure constants}
The product of two Macdonald functions can be expanded in the form
\footnote{\,
Ch. VI, p. 343, Equation $7.1^{\, \prime}$, in \cite{macdonald.book}
}, 
	
\begin{equation}
P^{\, q \, t}_{\, Y_1} \ll\, \bx\, \rr \, 
P^{\, q \, t}_{\, Y_2} \ll\, \bx\, \rr  = 
\sum_{\, Y_3} 
f^{\, q t,\, Y_3}_{\, Y_1\, Y_2}\, 
P^{\, q \, t}_{\, Y_3}          \, \ll\, \bx\, \rr, 
\label{product.01}
\end{equation}

\noindent which can be used as a definition of the structure constants 
$f^{\, q t,\, Y_3}_{\, Y_1\, Y_2}$. 
Using the involution $\omega_{\, q \, t}$, the product of two dual Macdonald 
functions, $Q_{\, Y_1}$ and $Q_{\, Y_2}$, can be expanded
\footnote{\,
Ch. VI, p. 344, in \cite{macdonald.book} 
}, 

\begin{equation}
Q^{\, q \, t}_{\, Y_1} \ll\, \bx\, \rr\, 
Q^{\, q \, t}_{\, Y_2} \ll\, \bx\, \rr = 
\sum_{\, Y_3} 
f^{\, q t,\, Y_3^{\, \prime}}_{\, Y^{\, \prime}\, Y_2^{\, \prime}}\, 
Q^{\, q \, t}_{\, Y_3} \ll\, \bx\, \rr
\label{product.02}
\end{equation}

\noindent From Equations \ref{product.01} and \ref{product.02}
\footnote{\,
Ch. VI, p. 344, Equation 7.3, in \cite{macdonald.book}
}, 

\begin{equation}
f^{\, q t,\, Y_3^{\, \prime}}_{\, Y_1^{\, \prime}\, Y_2^{\, \prime}} =
\ll
\frac{ 
b^{\, q \, t}_{\, Y_3}
}{
b^{\, q \, t}_{\, Y_1} \, 
b^{\, q \, t}_{\, Y_2} 
} 
\rr\,
f^{\, q t,\, Y_3}_{\, Y_1\, Y_2},  
\end{equation}
	
\noindent where $b^{\, q \, t}_Y$ is defined in Equation \ref{dual.macdonald}.
The structure constant $f^{\, q t,\, Y_3}_{\, Y_1\, Y_2}$ 
can be written as an inner product
\footnote{\,
Ch. VI, p. 343, Equation 7.1, in \cite{macdonald.book}
}, 
	
\begin{equation}
f^{\, q t,\, Y_3}_{\, Y_1\, Y_2} = 
\langle\,
Q^{\, q \, t}_{\, Y_3} \ll\, \bx\, \rr \, |\,
P^{\, q \, t}_{\, Y_1} \ll\, \bx\, \rr \, 
P^{\, q \, t}_{\, Y_2} \ll\, \bx\, \rr \,
\rangle
\end{equation}
	
\subsection{The skew Macdonald and skew dual Macdonald functions}
$P^{\, q \, t}_{Y_1 / Y_2} \ll\, \bx\, \rr$ is defined as
\footnote{\,
Ch. VI, p. 344, Equation $7.6^{\, \prime}$, in \cite{macdonald.book}
}, 
	
\begin{equation}
P^{\, q \, t}_{Y_1 / Y_2}       \ll\, \bx\, \rr  =
\ll \frac{ b^{\, q \, t}_{\, Y_2}}
         { b^{\, q \, t}_{\, Y_1}} 
\rr 
Q^{\, q \, t}_{Y_1 / Y_2}       \ll\, \bx\, \rr, 
\end{equation}
	
\noindent and $Q^{\, q \, t}_{Y_1 / Y_2} \ll\, \bx\, \rr$ 
is defined as
\footnote{\,
Ch. VI, p. 344, Equation 7.5, in \cite{macdonald.book}
}, 
	
\begin{equation}
Q^{\, q \, t}_{Y_1 / Y_2}                 \ll\, \bx\, \rr  =
\sum_{\, Y_3} f^{\, q t,\, Y_1}_{\, Y_2\, Y_3}  \, 
Q^{\, q \, t}_{\, Y_3}                 \ll\, \bx\, \rr 
\end{equation}

\subsection{Cauchy identities for skew Macdonald and dual skew Macdonald functions} 
The skew Macdonald and dual skew Macdonald functions satisfy the Cauchy identity
\footnote{\,
Ch. VI, p. 352, in \cite{macdonald.book}
}, 
	
\begin{equation}
\prod_{i,\, j,\, \ll n + 1 \rr \, = \, 1}^\infty
\ll 
\frac{1\, -\, x_i\, y_j\, q^{\, n}\,   t}
     {1\, -\, x_i\, y_j\, q^{\, n}\, \pt}
\rr
\sum_Y  
 \, 
P^{\, q \, t}_{Y_1 / Y} \ll\,\bx\, \rr\, 
Q^{\, q \, t}_{Y_2 / Y} \ll\,\by\, \rr
=
\sum_Y 
P^{\, q \, t}_{Y    /   Y_2} \ll\,\bx\, \rr \, 
Q^{\, q \, t}_{Y    /   Y_1} \ll\,\by\, \rr
\label{cauchy.identity.skew.01}
\end{equation}

\noindent Applying the involution, Equation \ref{involution.on.macdonald}, to 
the $\bx$-symmetric functions in Equation \ref{cauchy.identity.skew.01} 
\footnote{\,
Ch. VI, p. 352, in \cite{macdonald.book}
}, 

\begin{equation}
\prod_{i, j = 1}^\infty \ll 1 + x_i\, y_j \rr
\sum_Y 
Q^{\, t\, q}_{Y_1^{\, \prime} / Y^{\, \prime}} \ll\,\bx\,\rr\,
Q^{\, q \, t}_{Y_2        / Y  } \ll\,\by\,\rr 
= 
\sum_Y 
Q^{\, t\, q}_{Y^{\, \prime} / Y_2^{\, \prime}} \ll\,\bx\, \rr 
Q^{\, q \, t}_{Y        / Y_1  } \ll\,\by\, \rr\,
\label{cauchy.identity.skew.02}
\end{equation}

\noindent Repeating the exercise on the $\by$-symmetric functions,

\begin{equation}
\prod_{i, j = 1}^\infty \ll 1 + x_i\, y_j \rr
\sum_Y 
P^{\, q \, t}_{Y_1              / Y            }  \ll\, \bx\, \rr\, 
P^{\, t \, q}_{Y_2^{\, \prime}  / Y^{\, \prime}}  \ll\, \by\, \rr  
=\sum_Y 
P^{\, q \, t}_{Y             / Y_2            } \ll\, \bx\, \rr\,
P^{\, t \, q}_{Y^{\, \prime} / Y_1^{\, \prime}} \ll\, \by\, \rr 
\label{cauchy.identity.skew.03}
\end{equation}
	
\subsubsection{Remark} 
In each of Equations 
\ref{cauchy.identity.skew.02} and 
\ref{cauchy.identity.skew.03},
a Macdonald or dual Macdonald function that depends on $\ll q, t \rr$, 
is multiplied by a similar function that depends on $\ll t, q\rr$. 
These identities are important in the sequel. In fact, one can, in 
a sense, trace all constructions in this work to the Cauchy identity 
in Equation \ref{cauchy.identity.skew.01}

\section{$q t$-free bosons and $q t$-vertex operators}
\label{3}
{\it Our next point is the theory of $q t$-Heisenberg algebra and $q t$-vertex 
operators which are simple deformations of free bosons and vertex operators. 
We recall basic definitions related to these objects, which are necessary to obtain 
the correct Macdonald kernel function, a specialization of which is the Macdonald 
refined Macmahon partition function on the right hand side of Equation 
\ref{x.y.q.t.refined.macmahon.function.02}  
}
\smallskip 
	
\subsection{Remarks on earlier works}
All calculations in \cite{okounkov.reshetikhin, okounkov.reshetikhin.vafa, iqbal.kozcaz.vafa} 
use vertex operators based on 2D charged  
free-fermions, without deformation parameters. In \cite{foda.wheeler.02}, vertex 
operators based on $t$-free-fermions were used to generate $t$-plane partitions 
related to Hall-Littlewood symmetric functions
\footnote{\,
To simplify the presentation, we say 
$t$-free fermion,           $t$-plane partitions,          {\it etc.} for 
$t$-deformed free fermions, $t$-weighted plane partitions, {\it etc.}
}. 
These $t$-plane partitions are the $q \rightarrow 0$ limit of the $q t$-plane 
partitions introduced in \cite{vuletic.02}, the same way that Hall-Littlewood functions 
are $q \rightarrow 0$ limits of Macdonald functions. The computations were tedious, 
and an extension using vertex operators based on $q t$-free-fermions to 
generate the $q t$-plane partitions related to Macdonald polynomials, of 
the type introduced in \cite{vuletic.02}, was not attempted. In the present work, 
we choose to work exclusively in terms of 2D free-bosons. 
	
\subsection{$q t$-free bosons}

Following Shiraishi, Kubo, Awata and Odake \cite{shiraishi.01},
given two variables $q$ and $t$, $|\, q \, |<1$ and $|\, t \, |<1$, we introduce
the $q t$-free boson operators that satisfy the $q t$-Heisenberg
algebra, 

\begin{equation}
[a_{ m}^{\, q \, t},a_{ n}^{\, q  t} ] = 
n \ll \frac{ 1\, -\, q^{\, | n |}}{ 1\, -\, t^{\, | n |}} \rr \delta_{ m + n , 0} 
\label{q.t.heisenberg.commutation.relations}
\end{equation}

\noindent Working in terms of the $q t$-free boson operators, left-state
$ \langle\, a_{ Y}\, |$, labelled by a Young diagram $ Y = \ll y_1, y_2, \cdots \rr$, 
is generated from the left vacuum state,

\begin{equation}
\langle\, \ba^{\,q \, t}_Y\, | =  \langle\, 0\, | \, 
a_{\, Y_1}^{\, q \, t}\, 
a_{\, Y_2}^{\, q \, t}\, 
\cdots,
\end{equation}

\noindent while the right-state $ |\, \ba_Y\, \rangle$ labelled by the same Young 
diagram $Y$ is generated from the right vacuum state,

\begin{equation}
|\, \ba^{\, q \, t}_Y\, \rangle = 
a_{ - y_1}^{\, q \, t}\, 
a_{ - y_2}^{\, q \, t}\, 
\cdots |\, 0\, \rangle
\end{equation} 

\noindent Using the $q t$-Heisenberg commutation relations,
the inner product of 
$ \langle\, \ba_{\, Y_1}^{\, q \, t}\, |$ 
and 
$ |\, \ba_{\, Y_2}^{\, q \, t}\,\rangle$ 
is, 

\begin{equation}
\langle\, a_{\, Y_1}^{\, q \, t}\, |\, a_{\, Y_2}^{\, q \, t}\,\rangle = 
z^{\, q \, t}_{ Y_1}\, \delta_{ Y_1 Y_2}
\label{inner.product.right.left.states}
\end{equation}

\subsection{$q t$-vertex operators}

Following Shiraishi \textit{et al.} \cite{shiraishi.01}, we define the 
$q t$-vertex operators,

\begin{equation}
\Gamma^{\, q \, t}_+ \ll x \rr =  
\exp \ll\, -\,\sum_{n = 1}^\infty \frac{x^{-n}}{n} 
\ll 
\frac{ 1\, -\, t^{\, n}}{1\, -\, q^{\, n}}\rr a_n^{\, q \, t}
\rr,
\quad
\Gamma^{\, q \, t}_{-} \ll x \rr =  
\exp \ll\, -\, \sum_{n = 1}^\infty \frac{x^{\, n}}{n} \ll 
\frac{1\, -\, t^{\, n}}{1\, -\, q^{\, n}} \rr a_{- n}^{\, q \, t} \rr
\label{q.t.vertex.operators}
\end{equation}

\subsubsection{Remark}
$\Gamma^{\, q \, t}_{+} \ll x \rr$ depends on the inverse variables 
$x^{\, \prime}, \ll x^2 \rr^{\, \prime}, \cdots$, 
while
$\Gamma^{\, q \, t}_{-} \ll x \rr$ depends on $x, x^2, \cdots$ 
We use this convention to produce known formulas related to 
symmetric functions and to plane partitions without modification.

\subsection{$q t$-vertex operator commutation relations}

\noindent From the $q t$-Heisenberg commutation relations,
Equation \ref{q.t.heisenberg.commutation.relations}, and the dependence 
of $\Gamma^{\, q \, t}_{\pm}$ on their variables, 
Equation \ref{q.t.vertex.operators}, 
we obtain the commutation relations,

\begin{equation}
\Gamma^{\, q \, t}_{+} \ll x^{\, \prime} \rr\,
\Gamma^{\, q \, t}_{-} \ll y             \rr 
=
\prod_{n \, = \, 0}^\infty 
\ll
\frac{1\, -\, x \, y\,q^{\, n}\,  t}
     {1\, -\, x \, y\,q^{\, n}\,\pt} 
     \rr
\Gamma^{\, q \, t}_{-} \ll y             \rr\,
\Gamma^{\, q \, t}_{+} \ll x^{\, \prime} \rr
\label{q.t.vertex.operator.commutation.relation}
\end{equation}

\subsection{On the relation to Ding-Iohara-Miki algebra}
	
The $q t$-operators and Heisenberg algebra in 
Equation \ref{q.t.heisenberg.commutation.relations}, and 
the $q t$-vertex operators in 
Equation \ref{q.t.vertex.operator.commutation.relation}, and 
related operators, are identical to those that appear in 
free-field realisation of the Macdonald operator 
\cite{
shiraishi.01,
feigin.01,
shiraishi.02,
shiraishi.03,
shiraishi.04}, 
and of Ding-Iohara-Miki algebra \cite{saito.01, saito.02, saito.03}.
The vertex operators 
$\Gamma^{\, q \, t}_{+} \ll x \rr$ and
$\Gamma^{\, q \, t}_{-} \ll x \rr$ are the vertex operators 
$\phi^{\star}           \ll z \rr$ and 
$\phi                   \ll z \rr$ in Equation 2.12 in \cite{saito.01}, respectively.
	
\subsection{From $q t$-vertex operators to the $q t$-MacMahon function}
From the $q t$-vertex operator commutation 
relations, Equation \ref{q.t.vertex.operator.commutation.relation}, 
	
\begin{equation}
\langle\, 0\, |\, 
\prod_{i=1}^\infty \Gamma^{\, q \, t}_{+} \ll x^{ -i}  \rr
\prod_{j=1}^\infty \Gamma^{\, q \, t}_{-} \ll y^{j-1} \rr\, 
|\, 0\, \rangle  
=
\prod_{i,\, j,\, \ll n + 1 \rr = 1}^\infty 
\ll 
\frac{1\, -\, x^i\, y^{j-1}\, q^{\, n}\,t}
     {1\, -\, x^i\, y^{j-1}\, q^{\, n}\, \pt} 
\rr, 
\label{x.y.q.t.refined.macmahon.function.02}
\end{equation}
	
\noindent which is the result in Equation \ref{x.y.q.t.macmahon.function.01}. 
The right hand side of Equation \ref{x.y.q.t.refined.macmahon.function.02} is 
a specialization of that in Remark {\bf 2.7} in \cite{saito.01} on the Macdonald 
kernel function. 
	
\section{The power-sum/Heisenberg correspondence}
\label{4}
{\it From identities that involve Macdonald functions, we obtain identities 
that involve operator-valued Macdonald functions that act on states labelled 
by Macdonald functions. The aim of this section is to derive Equation \ref{two.identities},
which are used in Section \ref{5} to derive the Macdonald vertex.
}
\smallskip 
	
\subsection{An isomorphism}
	
Comparing 
{\bf 1.} the inner product of power-sum functions in the Macdonald basis, 
Equation \ref{young.diagram.power.sum.inner.product}, and 
{\bf 2.} the inner product of the right and left-states, 
Equation \ref{inner.product.right.left.states}, 
we deduce that the symmetric power-sum function basis is isomorphic 
to the Fock space spanned by the left-states $\langle\,a_Y\, |$, 
as well as that spanned by the right-states $|\, a_Y\, \rangle$,
where $Y$ is an arbitrary partition. More precisely, we have 
the correspondence, 
	
\begin{equation}
p_n \ll \bx \rr \rightleftharpoons - \, a_{ n}, \quad n \geq 1,  
\label{power.sum.heisenberg.correspondence}
\end{equation}

\noindent and also the correspondence, 

\begin{equation}
p_n \ll \bx \rr \rightleftharpoons - \, a_{-n}, \quad n \geq 1,  
\label{power.sum.heisenberg.correspondence.dual}
\end{equation}
	
Since the power-sum functions form a complete basis, we can think of 
the Macdonald functions as functions of the power-sum functions, then 
formally replace the latter with Heisenberg generators to obtain 
operator-valued Macdonald functions that act on left and right 
vacuum states to produce left and right Macdonald states. We use 
this formal substitution to start from known Macdonald symmetric 
function identities and obtain identities that involve operator-valued 
Macdonald functions acting on left and right-states labelled by
Macdonald functions.
	
\subsection{The action of operator-valued Macdonald functions on states}
We start from the Cauchy identity for skew Macdonald and skew dual Macdonald 
functions, Equation \ref{cauchy.identity.skew.01}, which we recall here, 
	
\begin{equation} 
\prod_{i,\, j,\, \ll n + 1 \rr \, = \, 1}^\infty
\ll 
\frac{1\, -\, x_i\, y_j\, q^{\, n}\,   t}
     {1\, -\, x_i\, y_j\, q^{\, n}\, \pt}
\rr
\sum_Y \, 
P^{\, q \, t}_{Y_1 / Y} \ll\,\bx\, \rr\, 
Q^{\, q \, t}_{Y_2 / Y} \ll\,\by\, \rr
=
\sum_Y 
P^{\, q \, t}_{Y    /   Y_2} \ll\,\bx\, \rr \, 
Q^{\, q \, t}_{Y    /   Y_1} \ll\,\by\, \rr
\label{step.01}
\end{equation}
	
\noindent Using Equations \ref{PI.01} and \ref{an.exponential.is.a.product},
	
\begin{multline}
\exp 
\ll \sum_{n=1}^\infty 
\frac{1}{n} 
\ll \frac{1\, -\,t^{\, n}}{1\, -\,q^{\, n}} \rr 
p_n \ll\, \bx\, \rr\, 
p_n \ll\, \by\, \rr\,
\rr
\sum_Y
P^{\, q \, t}_{Y_1 / Y} \ll\, \bx     \, \rr\, 
Q^{\, q \, t}_{Y_2 / Y} \ll\, \by     \, \rr
\\
= \sum_Y
P^{\, q \, t}_{Y / Y_2} \ll\, \bx     \, \rr\, 
Q^{\, q \, t}_{Y / Y_1} \ll\, \by     \, \rr
\label{step.02}
\end{multline}

\subsubsection{The action of $\Gamma^{\, q \, t}_{+}$ on a left-state}	
Using the power-sum/Heisenberg correspondence, Equation \ref{power.sum.heisenberg.correspondence.dual}, 
on $p_n \ll \bx \rr$, on the right hand side of Equation \ref{step.02}, we introduce free-boson mode 
operators that act as creation operators on a left-state, to obtain the operator-valued Macdonald Cauchy 
identity,

\begin{multline}
\exp 
\ll 
\sum_{n=1}^\infty 
- \frac{1}{n} 
\ll \frac{1\, -\,t^{\, n}}{1\, -\,q^{\, n}} \rr
a_n\, 
p_n \ll\, \by\, \rr\, 
\rr
\sum_Y
P^{\, q \, t}_{Y_1 / Y} \ll\, \ba_{+}\, \rr\, 
Q^{\, q \, t}_{Y_2 / Y} \ll\, \by    \, \rr
\\
= \sum_Y
P^{\, q \, t}_{Y / Y_2} \ll\, \ba_{+}\, \rr\, 
Q^{\, q \, t}_{Y / Y_1} \ll\, \by    \, \rr, 
\label{step.03}
\end{multline}

\noindent where 
$P^{\, q \, t}_{Y_1 / Y} \ll\, \ba_{+}\, \rr$ and 
$P^{\, q \, t}_{Y / Y_2} \ll\, \ba_{+}\, \rr$ 
are obtained by expanding 
$P^{\, q \, t}_{Y_1 / Y} \ll\, \bx     \, \rr$ and
$P^{\, q \, t}_{Y / Y_2} \ll\, \bx     \, \rr$ 
in the power-sum basis, 
then using the power-sum/Heisenberg correspondence, 
Equation  \ref{power.sum.heisenberg.correspondence}. 
From the definition of the $\Gamma^{\, q \, t}_{+}$ vertex operators, 
Equation \ref{q.t.vertex.operators}, 
	
\begin{equation}
\prod_{i \, =\, 1}^\infty \Gamma^{\, q \, t}_{+} \ll y^{\, \prime}_i \rr 
\sum_Y
P^{\, q \, t}_{Y_1 / Y} \ll\, \ba_{+}     \, \rr\, 
Q^{\, q \, t}_{Y_2 / Y} \ll\, \by         \, \rr
= \sum_Y
p^{\, q \, t}_{Y / Y_2} \ll\, \ba_{+}     \, \rr\, 
Q^{\, q \, t}_{Y / Y_1} \ll\, \by         \, \rr
\label{step.04}
\end{equation}

\noindent Using the $q t$-Heisenberg commutation relation, Equation
\ref{q.t.heisenberg.commutation.relations}, and acting with each side 
of Equation \ref{step.04.repeated} on a left vacuum state, 
	
\begin{equation}
\sum_Y
\langle\, 
P^{\, q \, t}_{Y_1 / Y}\, |\,
Q^{\, q \, t}_{Y_2 / Y} \ll\, \by         \, \rr
\prod_{i \, =\, 1}^\infty \Gamma^{\, q \, t}_{+} \ll y^{\, \prime}_i \rr            
= \sum_Y
\langle\, 
P^{\, q \, t}_{Y / Y_2}\, |\, 
Q^{\, q \, t}_{Y / Y_1} \ll\, \by\, \rr, 
\label{step.05}
\end{equation}

\noindent where $\langle\, P^{\, q \, t}_{Y_1 / Y_2}\,|$ is a state in 
the free-boson Fock space obtained by the action of the operator-valued 
Macdonald function labelled by the skew Young diagram $Y_1 / Y_2$,
	
\begin{equation}
\langle\, \emptyset\, |\, 
P^{\, q \, t}_{Y_1 / Y_2}  \ll\, \ba_{+}\, \rr\, 
= 
\langle\, P^{\, q \, t}_{Y_1 / Y_2}\, | 
\label{macdonald.action.01}
\end{equation}

\noindent Setting $Y_2 = \emptyset$ in Equation \ref{step.05}, we force 
$Y = \emptyset$ in the sum on the left hand side,
	
\begin{equation}
\langle\, 
P^{\, q \, t}_{\, Y_1}\, |\,
\prod_{i \, =\, 1}^\infty \Gamma^{\, q \, t}_+ \ll y^{\, \prime}_i \rr            
= \sum_Y
\langle\, 
P^{\, q \, t}_Y\, |\, 
Q^{\, q \, t}_{Y / Y_1} \ll\, \by\, \rr
\label{step.06}
\end{equation}

\subsubsection{The action of\, $\Gamma^{\, q \, t}_{-}$\, on a right-state}
Using the power-sum/Heisenberg correspondence, 
Equation \ref{power.sum.heisenberg.correspondence.dual}, on $p_n \ll \by \rr$, to 
introduce free-boson mode operators that act as creation operators on a right-state, 
to obtain the operator-valued Macdonald Cauchy identity,
	
\begin{multline}
\exp 
\ll \sum_{n=1}^\infty 
- \frac{1}{n} 
\ll \frac{1\, -\,t^{\, n}}{1\, -\,q^{\, n}} \rr 
p_n \ll\, \bx\, \rr\, a_{-n}
\rr
\sum_Y
P^{\, q \, t}_{Y_1 / Y} \ll\, \bx     \, \rr\, 
Q^{\, q \, t}_{Y_2 / Y} \ll\, \ba_{-}\, \rr
\\
= \sum_Y
P^{\, q \, t}_{Y / Y_2} \ll\, \bx     \, \rr\, 
Q^{\, q \, t}_{Y / Y_1} \ll\, \ba_{-}\, \rr
\label{step.03.repeated}
\end{multline}
	
\noindent From the definition of the $\Gamma^{\, q \, t}_{-}$ vertex operators, 
Equation \ref{q.t.vertex.operators}, 
	
\begin{equation}
\prod_{i \, =\, 1}^\infty \Gamma^{\, q \, t}_{-} \ll x_i \rr 
\sum_Y
P^{\, q \, t}_{Y_1 / Y} \ll\, \bx     \, \rr\, 
Q^{\, q \, t}_{Y_2 / Y} \ll\, \ba_{-}\, \rr
= \sum_Y
P^{\, q \, t}_{Y / Y_2} \ll\, \bx     \, \rr\, 
Q^{\, q \, t}_{Y / Y_1} \ll\, \ba_{-}\, \rr
\label{step.04.repeated}
\end{equation}
	
\noindent Acting with each side of Equation \ref{step.04.repeated} on a right vacuum state, 
	
\begin{equation}
\prod_{i \, =\, 1}^\infty \Gamma^{\, q \, t}_{-} \ll x_i \rr 
\sum_Y
P^{\, q \, t}_{Y_1 / Y} \ll\, \bx     \, \rr\, 
|\, 
Q^{\, q \, t}_{Y_2 / Y} \, \rangle 
= \sum_Y
P^{\, q \, t}_{Y / Y_2} \ll\, \bx     \, \rr\, 
|\, 
Q^{\, q \, t}_{Y / Y_1}\, \rangle, 
\label{step.05.repeated}
\end{equation}
		
\noindent where $|\, Q^{\, q \, t}_{Y_1 / Y_2} \, \rangle$ is a state in 
the free boson Fock space obtained by the action of the operator-valued 
Macdonald function labelled by the skew Young diagram $Y_1 / Y_2$,
	
\begin{equation}
Q^{\, q \, t}_{Y_1 / Y_2}  \ll\, \ba_{-}\, \rr\, |\, \emptyset\, \rangle = 
|\, Q^{\, q \, t}_{Y_1 / Y_2} \, \rangle 
\label{macdonald.action.02}
\end{equation}
	
\noindent Setting $Y_1 = \emptyset$ in Equation \ref{step.05.repeated}, we force 
$Y = \emptyset$ in the sum on the left hand side, 
	
\begin{equation}
\prod_{i \, =\, 1}^\infty \Gamma^{\, q \, t}_{-} \ll x_i \rr 
|\, Q^{\, q \, t}_{\, Y_2} \, \rangle 
= \sum_Y
P^{\, q \, t}_{Y/Y_2} \ll\, \bx     \, \rr\, 
|\, 
Q^{\, q \, t}_Y\, \rangle 
\label{step.06.repeated}
\end{equation}

\subsubsection{
The action of 
\, $\Gamma^{\, q \, t}_{-}$ on a left-state, and
\, $\Gamma^{\, q \, t}_{+}$ on a right-state}	
Using Equations 
\ref{step.06} and  
\ref{step.06.repeated}, then Equation 
\ref{macdonald.cauchy.identity},
	
\begin{multline}
\langle\, P^{\, q \, t}_{\, Y_1}\, |\, 
\prod_{i \, =\, 1}^\infty\, \Gamma^{\, q \, t}_{+} \ll\, x^{\, \prime}_i \rr
\prod_{j \, =\, 1}^\infty\, \Gamma^{\, q \, t}_{-} \ll\, y_j \rr\, 
|\,       Q^{\, q \, t}_{Y_2            }\, \rangle = 
\\
=\, \sum_Y 
Q^{\, q \, t}_{Y / Y_1}\, \ll \bx \rr\, 
P^{\, q \, t}_{Y / Y_2            }\, \ll \by \rr 
= 
\prod_{i,\, j,\, \ll n + 1 \rr \, = \, 1}^\infty
\ll 
\frac{1\, -\, x_i\, y_j\, q^{\, n}\,   t}
     {1\, -\, x_i\, y_j\, q^{\, n}\, \pt}
\rr
\sum_Y 
Q^{\, q \, t}_{Y_2 / Y} \ll \bx \rr\, 
P^{\, q \, t}_{Y_1 / Y} \ll \by \rr
\label{version.01}
\end{multline}
	
\noindent On the other hand, using the $q t$-vertex operator commutation relation, 
Equation \ref{q.t.vertex.operator.commutation.relation}, then 
inserting a complete set of orthonormal states, the left 
hand side of Equation \ref{version.01} can be re-written as, 
	
\begin{multline} 
\langle\, 
P^{\, q \, t}_{\, Y_1}  \, |\, 
\prod_{i \, =\, 1}^\infty \Gamma^{\, q \, t}_{+} \ll x^{\, \prime}_i \rr 
\prod_{j \, =\, 1}^\infty \Gamma^{\, q \, t}_{-} \ll y_j \rr\, 
|\, 
Q^{\, q \, t}_{\, Y_2}\, \rangle = 
\\ 
\prod_{i,\, j,\, \ll n + 1 \rr =\, 1}^\infty
\ll 
\frac{1\, -\, x_i\, y_j\, q^{\, n}\,   t}
     {1\, -\, x_i\, y_j\, q^{\, n}\, \pt}
\rr
\langle\, 
P^{\, q \, t}_{\, Y_1}  \, |\, 
\prod_{j \, =\, 1}^\infty \Gamma^{\, q \, t}_{-} \ll y_j \rr  
\prod_{i \, =\, 1}^\infty \Gamma^{\, q \, t}_{+} \ll x^{\, \prime}_i \rr\, 
\, | \, 
Q^{\, q \, t}_{\, Y_2}\, \rangle = 
\\ 
\prod_{i,\, j,\, \ll n + 1 \rr \, = \, 1}^\infty
\ll 
\frac{1\, -\, x_i\, y_j\, q^{\, n}\,   t}
     {1\, -\, x_i\, y_j\, q^{\, n}\, \pt}
\rr
\sum_Y
\langle\, P^{\, q \, t}_{\, Y_1}\, |\, 
\prod_{j = 1}^\infty  
\Gamma^{\, q \, t}_{-} \ll y_j \rr 
\, | \,   Q^{\, q \, t}_Y\, \rangle\, 
\langle\, P^{\, q \, t}_Y\, |\, 
\prod_{i\, = \, 1}^\infty  
\Gamma^{\, q \, t}_+ \ll x^{\, \prime}_i \rr \, |\, Q^{\, q \, t}_{\, Y_2}\, \rangle
\label{version.02}
\end{multline}
	
\noindent Comparing Equations \ref{version.01} and \ref{version.02},  

\begin{multline}
\ll 
\langle\, P^{\, q \, t}_{\, Y_1}\, | 
\, 
\prod_{i \, =\, 1}^\infty 
\Gamma^{\, q \, t}_{-} \ll y_j \rr\,
\rr\, 
|\, Q^{\, q \, t}_{\, Y_2} \, \rangle 
=   
P^{\, q \, t}_{Y_1 / Y_2} \ll \by \rr, 
\\ 
\langle \, P^{\, q \, t}_{\, Y_1} \, |\,
\ll 
\prod_{i \, =\, 1}^\infty\,
\Gamma^{\, q \, t}_+  \ll x^{\, \prime}_i \rr\,
|\, Q^{\, q \, t}_{\, Y_2}\, \rangle\,
\rr 
=  
Q^{\, q \, t}_{Y_2 / Y_1} \ll\, \bx \,\rr
\label{two.preliminary.identities}
\end{multline}

\noindent Since the states        $\langle\, P^{\, q \, t}_{\, Y_1}\, |$ 
form a basis of left-states,  and $|\, Q^{\, q \, t}_{\, Y_2}\, \rangle$
form a basis of right-states, and given the $q t$-inner product 
Equation \ref{macdonald.inner.product}, 

\begin{multline}
\ll 
\langle\, P^{\, q \, t}_{\, Y_1}\, | 
\, 
\prod_{i \, = \, 1}^\infty 
\Gamma^{\, q \, t}_{-} \ll y_j \rr\,
\rr\, = 
\sum_Y 
\langle\, P^{\, q \, t}_Y\, |\, \alpha_Y \ll \by \rr, 
\\ 
\ll 
\prod_{i \, =\, 1}^\infty\,
\Gamma^{\, q \, t}_+  \ll x^{\, \prime}_i \rr\,
|\, Q^{\, q \, t}_{\, Y_2}\, \rangle\,
\rr 
=
\sum_Y\, 
\beta_Y \ll \bx \rr\, 
|\, Q^{\, q \, t}_{\, Y_2}\, \rangle, 
\label{expanding}
\end{multline}

\noindent where $\alpha_Y \ll \by \rr$ and $\beta_Y \ll \bx \rr$ 
are expansion coefficients that carry the dependence on the variables
$\ll \bx \rr$ and $\ll \by \rr$, while the expansion is in the set of Young 
diagrams $Y$. Using Equation \ref{two.preliminary.identities}, we determine 
$\alpha_Y \ll \by \rr$ and 
$ \beta_Y \ll \bx \rr$, 

\begin{multline}
\langle\, P^{\, q \, t}_{\, Y_1}\, |\, 
\prod_{i \, =\, 1}^\infty 
\Gamma^{\, q \, t}_{-} \ll y_j \rr  =  
\sum_Y 
\langle\, P^{\, q \, t}_Y  \, |\, 
P^{\, q \, t}_{Y_1 / Y} \ll \by \rr, 
\quad 
\prod_{i \, =\, 1}^\infty 
\Gamma^{\, q \, t}_+  \ll x^{\, \prime}_i \rr\, |\, Q^{\, q \, t}_{\, Y_1}\, \rangle =  
\sum_Y  
Q^{\, q \, t}_{Y_1 / Y} \ll\, \bx \,\rr\, 
|\, Q^{\, q \, t}_Y \rangle, 
\label{two.identities}
\end{multline}
	
\noindent which we need in the derivation of the Macdonald topological vertex 
in Section {\bf \ref{5}}.
	
\section{The Macdonald vertex}
\label{5}
{\it Following \cite{okounkov.reshetikhin.vafa, iqbal.kozcaz.vafa}, we use the operator-valued Macdonald function identities 
derived in Section \ref{4} to construct the Macdonald vertex that represents the matrix 
element of 
a       state labelled by a Young diagram $Y_3$, between 
a left-state labelled by a      Macdonald function $P^{\, q \, t}_{\, Y_1} \ll \bx \rr$, 
a right-state labelled by a dual Macdonald function $Q^{\, q \, t}_{\, Y_2} \ll \by \rr$.}
\smallskip 
	
\subsection{Four steps}
We construct the Macdonald vertex $\cM^{\, q \, t}_{Y_1 Y_2 Y_3} \ll x, y \rr$, 
that is labelled by three finite Young diagrams $\ll Y_1, Y_2 \right.$, 
$\left. Y_3 \rr$, and depends 
on two parameters $\ll x, y \rr$, as well two Macdonald parameters $\ll q, t \rr$, 
in four steps. 
\1 We consider the infinitely-extended profile of $Y_3$, 
and construct an infinite sequence of vertex operators, 
$\ll \prod_{Maya \ll Y_3 \rr} \Gamma^{\, q \, t}_{\pm} \rr$,
that encodes this profile, 
\2 We commute the $q t$-vertex operators in 
$\ll \prod_{Maya \ll Y_3 \rr} \Gamma^{\, q \, t}_{\pm} \rr$ so that all 
$\Gamma^{\, q \, t}_{-}$ vertex operators are on the left, and all
$\Gamma^{\, q \, t}_{+}$ vertex operators are on the right, and in 
the process, we pick up a product of $\ll x, y, q, t \rr$-dependent factors, 
\3 we compute the expectation value of the resulting configuration of vertex operators
between a Macdonald state labelled by $Y_1$ on the left,
and     another           labelled by $Y_2$ on the right, 
using the operator-valued Macdonald function identities, and finally, 
\4 we normalize the result so that $\cM^{\, q \, t}_{\emptyset \emptyset \emptyset} \ll x, y \rr= 1$. 
But before we do that, we need to recall simple correspondences between Young diagrams,
Maya diagrams, and sequences of vertex operators, and make a number of remarks on the 
choice of arguments in sequences of vertex operators. 
    
\subsection{A Young diagram/Maya diagram/vertex operator correspondence}
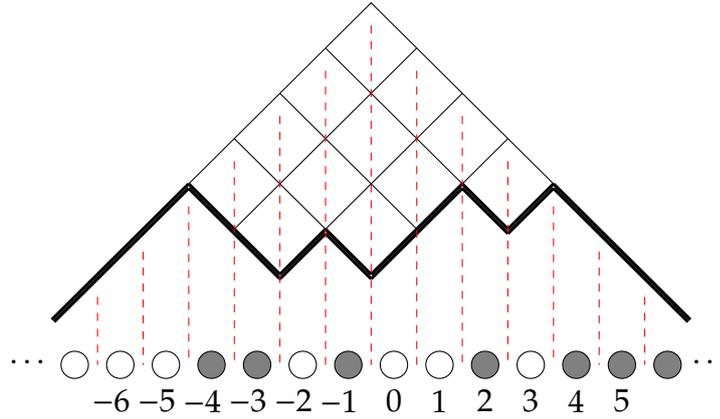
\begin{figure}
\begin{center}
\begin{tikzpicture}[scale=.6]
\draw [thin] (-1.5,1.0)--(5.5,8.0)--(12.5,1.0);
\draw [thin] ( 2.5,3.0)--(6.5,7.0);
\draw [thin] ( 3.5,2.0)--(7.5,6.0);
\draw [thin] ( 5.5,2.0)--(8.5,5.0);
\draw [thin] ( 8.5,3.0)--(9.5,4.0);
\draw [thin] ( 1.5,4.0)--(3.5,2.0);
\draw [thin] ( 2.5,5.0)--(5.5,2.0);
\draw [thin] ( 3.5,6.0)--(6.5,3.0);
\draw [thin] ( 4.5,7.0)--(8.5,3.0);

\draw [very thick] (-1.5,1.0)--(1.5,4.0);
\draw [very thick] (-1.45,0.95)--(1.5,3.9);

\draw [very thick] ( 1.5,4.0)--(3.5,2.0);
\draw [very thick] ( 1.5,3.9)--(3.5,1.9);

\draw [very thick] ( 3.5,2.0)--(4.5,3.0);
\draw [very thick] ( 3.5,1.9)--(4.5,2.9);

\draw [very thick] ( 4.5,3.0)--(5.5,2.0);
\draw [very thick] ( 4.5,2.9)--(5.5,1.9);

\draw [very thick] ( 5.5,2.0)--(7.5,4.0);
\draw [very thick] ( 5.5,1.9)--(7.5,3.9);

\draw [very thick] ( 7.5,4.0)--(8.5,3.0);
\draw [very thick] ( 7.5,3.9)--(8.5,2.9);

\draw [very thick] ( 8.5,3.0)--(9.5,4.0);
\draw [very thick] ( 8.5,2.9)--(9.5,3.9);

\draw [very thick] ( 9.5,4.0)--(12.5,1.0);
\draw [very thick] ( 9.5,3.9)--(12.45,0.95);

\foreach \i in {0,...,6} 
{
\draw [dashed, red] (\i - .5, \i + 1.5)--(\i - .5, 0);
}
\foreach \j in {7,...,12} 
{
\draw [dashed, red](\j -.5, 13.5 -\j)--(\j - .5, 0);
}
\node [left]  at (-1.3,0) {$\cdots$};
\node [right] at (12.3,0) {$\cdots$};
\foreach \x in {0,...,2}
{
\draw                 (\x- 1,0) circle (0.3);
\draw [fill=black!50] (\x+10,0) circle (0.3);
}
\draw [fill=black!50] (2,0) circle (0.3);
\draw [fill=black!50] (3,0) circle (0.3);
\draw (4,0) circle (0.3);
\draw [fill=black!50] (5,0) circle (0.3);
\draw (6,0) circle (0.3);
\draw  (7,0) circle (0.3);
\draw [fill=black!50] (8,0) circle (0.3);
\draw (9,0) circle (0.3);

\foreach \a in {6, 5, ..., 1} 
{
\node [below] at (- \a + 5.8, -.3) {$- \a$};
}
\foreach \a in {0, 1, ..., 5} 
{
\node [below] at (  \a + 6.0, -.3) {$  \a$};
}				
\end{tikzpicture}
\end{center}
\caption{ \it The Young diagram/Maya diagram correspondence for 
$Y = \ll 4, 3, 3, 2 \rr$. The infinite profile of the Young diagram 
is indicated with a heavy line. The integer below a stone is its position in the Maya diagram.
The apex of the inverted Young diagram is located between positions $-1$ and $0$.
}
\label{young.maya.correspondence}
\end{figure}

Given a Young diagram $Y$, we position it as in Figure \ref{young.maya.correspondence}, 
consider its infinite profile, map the segments of the profile to black and white stones 
as follows,  

\begin{equation}
\diagup   \rightleftharpoons \Circle, 
\quad
\quad
\diagdown \rightleftharpoons \CIRCLE,   
\end{equation} 

\noindent to obtain a Maya diagram \cite{blue.book}. From the Maya diagram, obtain 
an infinite sequence of $q t$-vertex operators, by mapping the black and white 
stones to vertex operators, so that we end up with the correspondences,

\begin{equation}
\diagup   \rightleftharpoons \Circle \rightleftharpoons \Gamma^{\, q \, t}_{+}, 
\quad
\quad
\diagdown \rightleftharpoons \CIRCLE \rightleftharpoons \Gamma^{\, q \, t}_{-}  
\end{equation} 

%FIGURE.5.1   
	
\subsection{The choice of arguments in sequences of vertex operators}
To reproduce the refined topological vertex of Iqbal {\it et al.} from 
the Macdonald (refined) topological vertex in the limit $q \rightarrow t$, 
we choose the arguments in the $q t$-vertex operators as in \cite{iqbal.kozcaz.vafa},

\begin{equation}
\Gamma_{+}^{\,q \, t} \ll x^{  - i}\, y^{  y_{3, i}            }   \rr, \quad 
\Gamma_{-}^{\,q \, t} \ll y^{j - 1}\, x^{- y_{3, j}^{\, \prime}}   \rr, 
\label{full.arguments}
\end{equation}

\noindent where 
$y_{3, i}$ is the length of the $i$-row of the Young diagram $Y_3$ that labels the 
upper boundary of the vertex, and 
$y^{\, \prime}_{3, j}$ is the length of the $j$-row of the transpose Young diagram 
$Y^{\, \prime}_3$.

\subsubsection{Remarks on the arguments in the vertex operators with reference to Figure 
\ref{young.maya.correspondence}} 
The arguments in the infinite sequences of vertex operators, that we need to construct,
can be explained, with reference to Figure \ref{young.maya.correspondence} as follows.
\1 For $Y_3 = \emptyset$, the assignment of arguments simplify, 

\begin{equation}
\Gamma_{+}^{\,q \, t} \ll x^i \rr, i \in \ll \cdots, -2, -1 \rr, 
\quad 
\Gamma_{-}^{\,q \, t} \ll y^j \rr, j \in \ll 0, 1, \cdots \rr 
\label{simple.arguments}
\end{equation}

\noindent 
\2 Consider the infinitely-extended profile of $Y_3 = \emptyset$. 
One can think of 
the  left-half of this profile as the negative $x$-axis 
in a Cartesian coordinate system in $\RR^2$, and of 
the right-half                 as the positive $y$-axis. 
\3 
Split the negative $x$-axis into $x$-segments of unit length each, and label them 
with $i \in \ll \cdots, -2, -1 \rr$, and  
split the positive $y$-axis into $y$-segments of unit length each, and label them 
with $j \in \ll 0, 1, \cdots \rr$
\4 
Consider the infinite extended profile of $Y_3 \neq \emptyset$, as in 
Figure \ref{young.maya.correspondence}. 
Finitely-many $x$-segments and the $y$-segments are shifted away from their 
original positions, but they remain parallel to the latter. 
\5 
The $x$-segments correspond to white stones, which in turn correspond to 
$\Gamma^{\, q \, t}_{+}$ vertex operators.  
The $y$-segments correspond to black stones, which in turn correspond to 
$\Gamma^{\, q \, t}_{-}$ vertex operators.
\6  
The argument of $\Gamma^{\, q \, t}_{+}$ that corresponds to an $x$-segment whose 
$\ll Y_3 = \emptyset \rr$-position is $i \in \ll \cdots, -2, -1 \rr$,  
on the negative $x$-axis, and was shifted by $y_{3, i}$ away from that position is 
$\ll x^{- i}\, y^{y_{3, i}} \rr$
\7 
The argument of $\Gamma^{\, q \, t}_{-}$ that corresponds to a  $y$-segment whose  
$\ll Y_3 = \emptyset \rr$-position is  
$\ll j - 1 \rr$, $j \in  \ll 1, 2, \cdots \rr$,  
on the positive $y$-axis, and was shifted by $y^{\, \prime}_{3, j}$ away from that position is 
$\ll y^{j-1}\, x^{- y^{\, \prime}_{3, i}} \rr$
\8 
We label the positions of the $y$-segments with $\ll j-1 \rr, j \in \ll 1, 2, \cdots \rr$, 
to conform with the conventions common in the literature.

\subsubsection{Example} The Young diagram/Maya diagram correspondence in 
Figure \ref{young.maya.correspondence} leads to the vertex-operator sequence, 

\begin{multline}
\ll 
\prod_{Maya \ll Y_3 \rr} \Gamma^{\, q \, t}_{\pm}
\rr 
=
\cdots 
\Gamma_{+}^{\,q \, t}\ll x^{-5}     \rr
\Gamma_{-}^{\,q \, t}\ll x^{-4}     \rr
\Gamma_{-}^{\,q \, t}\ll y     \, x^{-4}    \rr
\Gamma_{+}^{\,q \, t}\ll x^{-4}\, y^{ 2}    \rr
\Gamma_{-}^{\,q \, t}\ll y^2   \, x^{-3}    \rr
\\
\Gamma_{+}^{\,q \, t}\ll x^{-3}\, y^3\rr
\Gamma_{+}^{\,q \, t}\ll x^{-2}\, y^3\rr
\Gamma_{-}^{\,q \, t}\ll y^3   \, x^{-1} \rr
\Gamma_{+}^{\,q \, t}\ll x^{-1}\, y^4 \rr
\Gamma_{-}^{\,q \, t}\ll           y^4 \rr
\cdots
\label{vertex.operator.sequence}
\end{multline}

\noindent To obtain the \lq unnormalized\rq\, Macdonald vertex, 
$\cM^{\, q t,\, unnorm}_{\, Y_1\,Y_2\,Y_3} \ll x, y \rr$,
we need to evaluate the sequence $\ll \prod_{Maya \ll Y_3 \rr} \Gamma^{\, q \, t}_{\pm} \rr$ 
between a left-state labelled by $Y_1$ and a right-state labelled by $Y_2$,

\begin{equation}
\cM^{\, q t,\, unnorm}_{\, Y_1\, Y_2\, Y_3} \ll x, y \rr 
=
\langle\, 
P^{\, q \, t}_{\, Y_1} 
\, |\,
\ll 
\prod_{ {\it Maya} \ll Y_3 \rr}
\Gamma_{\pm}^{\,q \, t}
\rr 
\, |\, 
Q_{\, Y_2}^{\, q \, t} \rangle
\label{unnormalized.macdonald.vertex}
\end{equation} 

\noindent The final, normalised Macdonald vertex 
$\cM^{\, q \, t}_{\, Y_1\,Y_2\, Y_3} \ll x, y \rr$ is,

\begin{equation}
\cM^{\, q \, t}_{\, Y_1\,Y_2\, Y_3} \ll x, y \rr 
=
\frac{
\cM^{\, unnorm}_{Y_1     \,Y_2     \,Y_3} \ll x, y \rr
}{
\cM^{\, unnorm}_{\emptyset\,\emptyset\,\emptyset} \ll x, y \rr
}, 
\label{normalized.macdonald.vertex}
\end{equation}

\noindent where 
$\cM^{\, unnorm}_{\emptyset\,\emptyset\,\emptyset} \ll x, y \rr$ 
is easily identified with the $\ll x, y, q, t\rr$-MacMahon partition 
function on the right hand side of Equation \ref{x.y.q.t.macmahon.function.01}
We are now in a position to construct the Macdonald vertex following 
the four steps outlined at the start of this section.

\subsection{Step \1}
We prepare an infinite sequence of $q t$-vertex operators 
$\ll \prod_{Maya \ll Y_3 \rr} \Gamma^{\, q \, t}_{\pm} \rr$ that encodes 
the infinite profile of the finite Young diagram $Y_3$.

\subsection{Step \2}
Given $\ll \prod_{Maya \ll Y_3 \rr} \Gamma^{\, q \, t}_{\pm} \rr$, there are 
two ways to \lq straighten\rq the sequence in preparation for evaluating 
its expectation value between states, we can either
\1 perform a finite number of commutations to put 
all $\Gamma^{\, q \, t}_{-}$ vertex operators on the right, and 
all $\Gamma^{\, q \, t}_{+}$ vertex operators on the left, or 
\2 perform an infinite number of commutations to put 
all $\Gamma^{\, q \, t}_{+}$ vertex operators on the right, and 
all $\Gamma^{\, q \, t}_{-}$ vertex operators on the left.
In the present work, we choose the latter in order to obtain 
expressions that reduce to those in the literature in appropriate 
limits.
From Equation \ref{q.t.vertex.operator.commutation.relation}, 

\begin{multline}
\Gamma^{\, q \, t}_{+} \ll x^{  - i}\, y^{  y_{            3, i}} \rr  \,
\Gamma^{\, q \, t}_{-} \ll y^{j - 1}\, x^{- y^{\, \prime}_{3, j}} \rr =
\\
\prod_{n \, = \, 0}^\infty 
\ll
\frac{1\, -\, x^{- y^{\, \prime}_{3, j} + i}\, y^{- y^+_{3, i} + j}\,q^{\, n}\, t}
     {1\, -\, x^{- y^{\, \prime}_{3, j} + i}\, y^{- y^+_{3, i} + j}\,q^{\, n}\,\pt} 
     \rr
\Gamma^{\, q \, t}_{-} \ll  y^{j - 1}\, x^{- y^{\, \prime}_{3, j}} \rr\,
\Gamma^{\, q \, t}_{+} \ll  x^{  - i}\, y^{  y_{            3, i}} \rr, 
\label{q.t.vertex.operator.commutation.relation.with.arguments}
\end{multline}

\noindent where $y^+_{3,\, i} = y_{3,\, i} + 1$, and $y_{3, i}$ is the length 
of the $i$-row in $Y_3$. In the limit $q \rightarrow t$, the factor on the right 
hand side of Equation \ref{q.t.vertex.operator.commutation.relation.with.arguments} 
reduces to that in the corresponding commutation relation that appears in the derivation 
of the refined topological vertex in \cite{iqbal.kozcaz.vafa}. If we think of 
$\Gamma^{\, q \, t}_{+}$ as attached to a segment 
$\diagup$ in the extended profile of $Y_3$, and 
$\Gamma^{\, q \, t}_{-}$ as attached to an adjacent segment 
$\diagdown$ to the right of the former, 
then the commutation relation, 
Equation \ref{q.t.vertex.operator.commutation.relation.with.arguments}
describes replacing the adjacent pair 
$\diagup   \diagdown$ with the pair 
$\diagdown   \diagup$, thereby adding a cell to $Y_3$, to generate a larger 
Young diagram. 
The exponents that appear in the factor on the right hand side of 
Equation \ref{q.t.vertex.operator.commutation.relation.with.arguments} have simple 
interpretations,

\begin{equation}
y_{3, i}        - j = A_{\wsq}, \quad 
y_{3, j}^{\, \prime} - i = L_{\wsq}, 
\end{equation}

\noindent where $A_{\wsq}$ and $L_{\wsq}$ are the arm-length and the leg-length 
of the cell $\wsq$ that is added to $Y_3$ {\it via} the commutation in Equation
\ref{q.t.vertex.operator.commutation.relation.with.arguments},
to generate a larger Young diagram, 
that is $\wsq \notin Y_3$.
Inserting the sequence $\ll \prod_{Maya \ll Y_3 \rr} \Gamma^{\, q \, t}_{\pm} \rr$ 
between a left-state 
$\langle\, P^{\, q \, t}_{\, Y_1}\, |$, and 
a right-state 
$|\, Q^{\, q \, t}_{\, Y_2}\, \rangle$, then 
commuting the (infinitely-many) $\Gamma^{\, q \, t}_{+}$ vertex operators to the right 
of the  $\Gamma^{\, q \, t}_{+}$ vertex operators, 

\begin{multline}
\langle\, P^{\, q \, t}_{\, Y_1}\, |\, 
\ll \prod_{i=1}^\infty \Gamma^{\, q \, t}_{+} \ll x^{  - i}\, y^{  y_{            3, i}} \rr  \rr \,
\ll \prod_{j=1}^\infty \Gamma^{\, q \, t}_{-} \ll y^{j - 1}\, x^{- y^{\, \prime}_{3, j}} \rr  \rr \, 
|\, Q^{\, q \, t}_{\, Y_2}\, \rangle\, 
=
\\
\prod_{n \, = \, 0}^\infty 
\ll 
\prod_{\wsq \notin Y_3} 
\frac{1\, -\, x^{- L_\wsq}\, y^{- A^+_\wsq}\,q^{\, n}\, t}
     {1\, -\, x^{- L_\wsq}\, y^{- A^+_\wsq}\,q^{\, n}\,\pt} 
\rr\, 
\langle\, P^{\, q \, t}_{\, Y_1}\, |
\ll \prod_{j=1}^\infty \Gamma^{\, q \, t}_{-} \ll  y^{j - 1}\, x^{- y^{\, \prime}_{3, j}} \rr \rr\,
\ll \prod_{i=1}^\infty \Gamma^{\, q \, t}_{+} \ll  x^{  - i}\, y^{  y_{            3, i}} \rr \rr
|\, Q^{\, q \, t}_{\, Y_2}\, \rangle
\label{macdonald.top.vertex.step.01}
\end{multline} 

\subsection{Step \3}
Using Equation \ref{two.identities}, 

\begin{multline}
\langle\, P^{\, q \, t}_{\, Y_1}\, |\, 
\ll \prod_{i=1}^\infty \Gamma^{\, q \, t}_{+} \ll x^{  - i}\, y^{  y_{            3, i}} \rr \rr\,
\ll \prod_{j=1}^\infty \Gamma^{\, q \, t}_{-} \ll y^{j - 1}\, x^{- y^{\, \prime}_{3, j}} \rr \rr\, 
|\, Q^{\, q \, t}_{\, Y_2}\, \rangle\, 
=
\\
\prod_{n \, = \, 0}^\infty 
\ll 
\prod_{\wsq \notin Y_3} 
\frac{1\, -\, x^{- L_\wsq}\, y^{- A^+_\wsq}\,q^{\, n}\,  t}
     {1\, -\, x^{- L_\wsq}\, y^{- A^+_\wsq}\,q^{\, n}\,\pt} 
\rr 
\,
\sum_Y
P_{\, Y_1 / Y}^{\,q \, t} \ll y^{\, \bj - 1}\, x^{\, - Y_3^{\, \prime}} \rr 
Q_{\, Y_2 / Y}^{\,q \, t} \ll x^{\, \bi    }\, y^{\, - Y_3            } \rr, 
\label{macdonald.top.vertex.step.02}
\end{multline} 

\noindent where 
$\bi = \ll 1, 2, \cdots \rr$, 
$\bj = \ll 1, 2, \cdots \rr$, and the arguments in 
$P_{\, Y_1 / Y}^{\,q \, t} \ll y^{\, \bj    - 1}\, x^{\, -Y_3^{\, \prime}} \rr$ and 
$Q_{\, Y_2 / Y}^{\,q \, t} \ll x^{\, \bi       }\, y^{\, -Y_3            } \rr$, 
should be understood in the sense of Section \ref{sequences}

\subsection{Step \4}
To normalize the expression in Equation \ref{macdonald.top.vertex.step.02} such that 
$\cM^{\, q \, t}_{\emptyset\, \emptyset\, \emptyset} = 1$, we divide 
it by the $q t$-Macmahon partition function, the right hand side of 
Equation \ref{x.y.q.t.macmahon.function.01}. Using the identity, 

\begin{equation}
\ll 
\prod_{\wsq \notin Y_3} 
\frac{1\, -\, x^{- L_\wsq}\, y^{- A^+_\wsq}\,q^{\, n}\,  t}
     {1\, -\, x^{- L_\wsq}\, y^{- A^+_\wsq}\,q^{\, n}\,\pt} 
\rr 
\, 
\ll
\prod_{i, j = 1}^\infty 
\ll
\frac{1-x^i\,y^{\,j-1}\,q^{\, n}\,   t}
     {1-x^i\,y^{\,j-1}\,q^{\, n}\, \pt}
\rr
\rr^{-1}
=
\ll 
\prod_{\wsq \in Y_3}
\frac{1 -\,x^{\,L^+_\wsq}\, y^{\,A_\wsq}\, q^{\, n}\,   t}
     {1 -\,x^{\,L^+_\wsq}\, y^{\,A_\wsq}\, q^{\, n}\, \pt}
\rr,  
\end{equation}

\noindent which follows from Equations 2.8 and 2.11 in \cite{awata.kanno.02}. 
The result of the above steps is that the Macdonald vertex is, 

\begin{empheq}[box=\fbox]{equation}
\cM^{\, q \, t}_{\, Y_1 \, Y_2 \, Y_3} \ll x, y \rr 
= 
\prod_{n \, = \, 0}^\infty 
\ll 
\prod_{\wsq \in Y_3}
\frac{1 -\,x^{\,L^+_{\wsq, Y_3}}\, y^{\,A_{\wsq, Y_3}}\,q^{\, n}\,t}
     {1 -\,x^{\,L^+_{\wsq, Y_3}}\, y^{\,A_{\wsq, Y_3}}\,q^{\, n}\, \pt}
\rr 
\sum_Y
P_{\, Y_1 / Y}^{\,q \, t} \ll y^{\, \bj -1}\, x^{-  Y_3^{\, \prime}} \rr 
Q_{\, Y_2 / Y}^{\,q \, t} \ll x^{\, \bi   }\, y^{ - Y_3            } \rr, 
\label{macdonald.vertex}
\end{empheq}

\noindent where $\bi = \ll 1, 2, \cdots \rr$, and the arguments in 
$P_{\, Y_1 / Y}^{\,q \, t} \ll y^{\, \bj -1}\, x^{-Y_3^{\, \prime}} \rr$ and
$Q_{\, Y_2 / Y}^{\,q \, t} \ll x^{\, \bi   }\, y^{-Y_3            } \rr$
are in the sense of Section \ref{sequences}. In the limit $q \rightarrow t$, all 
dependence on $\ll q, t \rr$ drops out and we recover, 
	
\begin{equation}
\cR_{\, Y_1\, Y_2\, Y_3} \ll x, y \rr 
=  
\ll
\prod_{\wsq \in Y_3} 
\frac{1}{1 - x^{\, L^+_{\wsq, Y_3}}\, y^{\, A_{\wsq, Y_3}}}
\rr 
\sum_Y 
s_{Y_1 / Y} \ll y^{\, \bj -1}\, x^{- Y_3^{\, \prime}}  \rr
s_{Y_2 / Y} \ll x^{\, \bi   }\, y^{- Y_3            }  \rr, 
\label{refined.vertex}
\end{equation}
	
\noindent which, in our choice of variables, is the refined topological vertex 
of Iqbal {\it et al.} \cite{iqbal.kozcaz.vafa}, in the {\it \lq diagonal slicing\rq\,} in 
the sense of Section \ref{slicing}
\footnote{\,
Our $\ll x, y \rr$ are $\ll q, t \rr$ in \cite{iqbal.kozcaz.vafa}, and our convention for 
$Y_3$ is the transpose of the corresponding Young diagram in \cite{iqbal.kozcaz.vafa}
}. 
We do not consider the {\it \lq perpendicular slicing\rq\,} in the sense of 
Section \ref{slicing}, since the additional factors involved cancel out when 
we glue topological vertices to form instanton partition functions. 
In the limit $y \rightarrow x$, we recover the corresponding expression for 
the original vertex of \cite{aganagic.klemm.marino.vafa}.

\section{Remarks on the structure of the Macdonald vertex}
\label{6}
{\it We collect a number of remarks on the structure of the Macdonald vertex.}
\smallskip

\subsection{Not all legs are on equal footing}
A topological vertex has three external legs.  
The $x$-leg, labelled by a Young diagram $Y_1$, and associated with the negative 
$x$-coordinates in the plane partition representation, and 
the $y$-leg, labelled by a Young diagram $Y_2$, and associated with the positive 
$y$-coordinates in the plane partition representation, are {\it \lq non-preferred 
legs\rq}. 
The remaining leg, labelled by the Young diagram $Y_3$, which is encoded in the 
sequence of vertex operators used to construct the vertex. 
Given the asymmetric way that the vertex is constructed, not all legs are on equal 
footing, and the third leg labelled by $Y_3$ is usually called {\it \lq the preferred leg\rq}. 

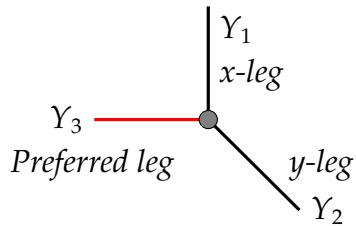
\begin{figure}
\begin{centering}
\begin{tikzpicture}[scale=.6]
\draw [red,   very thick] (1.5, 4.0)--(4.0, 4.0);
\draw [black, very thick] (4.0, 4.0)--(6.0, 2.0);
\draw [black, very thick] (4.0, 6.5)--(4.0, 4.0);
\node [left]  at (1.5, 4.0) {$Y_3$};
\node [left]  at (3.5, 3.0) {\it Preferred leg};
\node [right] at (6.0, 2.0) {$Y_2$};
\node [right] at (5.5, 3.0) {\it $y$-leg};
\node [right] at (4.0, 6.0) {$Y_1$};
\node [right] at (4.0, 5.0) {\it $x$-leg};
\draw [fill=black!50] (4.0,4.0) circle (0.2);
\end{tikzpicture} 
\par\end{centering}
\caption{
\it The Macdonald vertex $\cM^{\, q \, t}_{\, Y_1\, Y_2\, Y_3} \ll x, y \rr$ as a trivalent 
vertex.
The $x$-leg, associated with negative $x$-coordinates in the plane partition 
representation, comes from the top, labelled by $Y_1$.
The $y$-leg, associated with positive $y$-coordinates in the plane partition 
representation, goes to the bottom, labelled by a Young diagram $Y_2$.
The preferred leg is horizontal, labelled by $Y_3$.
}
\label{figure.topological.vertex} 
\end{figure}

\subsection{The planar representation of the vertex}
There are various, equivalent ways to represent the Macdonald vertex, 
$\cM^{\, q \, t}_{Y_1 Y_2 Y_3} \ll x, y \rr$, or any 
topological vertex for that matter, as a trivalent vertex in a diagram. 
For the purposes of Section {\bf \ref{6}}, it is convenient to use the 
conventions in Figure {\bf \ref{figure.topological.vertex}}.  
The $x$-leg, associated with the negative $x$-coordinates in the plane partition 
representation of the Macdonald vertex, comes vertically from top of the diagram,
labelled by $Y_1$.
Turning clockwise around the vertex, one encounters 
the $y$-leg, associated with the positive $y$-coordinates in the plane partition
representation of the Macdonald vertex, going diagonally to the bottom of the diagram,
labelled by $Y_2$.
Continuing clockwise around the vertex, one encounters the preferred leg, represented 
as a horizontal segment, labelled by $Y_3$.

\subsection{The arguments and the diagrams}
The $x$-leg is associated with the      Macdonald function $P^{\, q \, t}_{\, Y_1}$, and
when $Y_3 = \emptyset$, the argument of $P^{\, q \, t}_{\, Y_1}$ is $\ll \bx \rr$. 
The $y$-leg is associated with the dual Macdonald function $Q^{\, q \, t}_{\, Y_2}$, and 
when $Y_3 = \emptyset$, the argument of $Q^{\, q \, t}_{\, Y_2}$ is $\ll \by \rr$.
In this sense, $Y_3$ is a {\it \lq mixing Young diagram\rq}\, that mixes the arguments 
of the Macdonald functions. When $Y_3 \neq \emptyset$, 
the argument of $P^{\, q \, t}_{\, Y_1}$ is no longer purely $\bx$, and 
the argument of $Q^{\, q \, t}_{\, Y_2}$ is no longer purely $\by$. 
In the planar representation of $\cM^{\, q \, t}_{Y_1, Y_2, Y_3} \ll x, y \rr$, 
in Figure \ref{figure.topological.vertex}, 
$Y_1$ labels the vertical leg, 
$Y_2$        the diagonal leg, and
$Y_3$        the horizontal (preferred) leg. 
These remarks be useful to check consistency when we glue Macdonald vertices. 

\subsection{The Macdonald vertex {\it vs.} the refined topological vertex 
of Awata and Kanno \cite{awata.kanno.01, awata.kanno.02}}
\label{comparison.with.awata.kanno}

Using the notation, 

\begin{multline}
| Y | = \ll y_1   + y_2   + \cdots \rr,
\quad 
| Y_1 / Y_2 | = | Y_1 | - | Y_2 |, 
\quad 
|| Y || = \ll y_1^2 + y_2^2 + \cdots \rr, 
\\  
\quad
\rho = \ll - \frac12, - \frac32, \cdots \rr,   
\end{multline}

\noindent the refined topological vertex of Awata and Kanno 
\cite{awata.kanno.01, awata.kanno.02} is,  

\begin{multline}
\cR^{\, AK}_{Y_1 Y_2 Y_3} \ll x, y \rr =
\\
\ll - 1 \rr^{ |Y_2|} 
x^{- \frac12\, ||Y_2            ||}\, 
y^{  \frac12\, ||Y_2^{\, \prime}||}\,
P_{\, Y_3}^{\, x\,y}\, \ll y^{\, \rho} \rr 
\sum_Y 
\ll \frac{y}{x} \rr^{\, \frac12\, |Y_2 / Y|}\,
\ll
\iota_{AK} 
P_{\, Y_1^{\, \prime} / Y^{\, \prime}}^{\,y\,x}\, \ll -\, x^{\, \rho}\, y^{Y_3^{\, \prime}} \rr\, \rr 
P_{\, Y_2             / Y            }^{\,x\,y}\, \ll  \, y^{\, \rho}\, x^{Y_3            } \rr, 
\end{multline}

\noindent where $\iota_{AK}$ is an involution that acts on the power-sum functions as, 

\begin{equation}
\iota_{AK} \ll p_n \rr  = -\, p_n
\end{equation}

\noindent Since the arguments in the Macdonald symmetric functions in 
$\cM^{\, q \, t}_{Y_1 Y_2 Y_3} \ll x, y \rr$ depend on the variables $\ll x,\, y \rr$, 
while Macdonald parameters $\ll q,\, t \rr$ are additional, independent 
parameters, $\cM^{\, q \, t}_{Y_1 Y_2 Y_3} \ll x, y \rr$ is not the same as the refined 
topological vertex of Awata and Kanno.
	
\section{The Macdonald $U \! \ll 2 \rr$ strip partition function}
\label{7}
{\it 
We glue four Macdonald vertices to obtain the Macdonald-analogue of 
the normalized contribution of the bifundamental hypermultiplet to 
the 5D $U \! \ll 2 \rr$ instanton partition function. The rules that 
we use are the same as those used in \cite{iqbal.kozcaz.vafa} plus an additional 
rule specific to Macdonald vertices that requires that we glue 
a vertex that depends on $\ll q, t\rr$ to a vertex that depends on 
$\ll t, q\rr$, and {\it vice versa}.
}  
\smallskip 

\subsection{From topological vertices to partition functions}
Topological vertices are simple combinatorial objects that can be glued to form 
arbitrarily complicated 5D instanton partition functions in 5D supersymmetric gauge 
theories \cite{geometric.engineering.01, geometric.engineering.02}. These in turn 
are equal to off-critical deformations of conformal blocks in 2D conformal field 
theories \cite{nekrasov, awata.yamada.01, awata.yamada.02}. 
In the 4D-limit discussed in Section \ref{8}, one recovers 4D instanton partition 
functions that are equal to critical conformal blocks \cite{alday.gaiotto.tachikawa}. 
Let us focus on instanton partition functions in 
$U_1 \! \ll 2 \rr \times \cdots \times U_{N+1} \! \ll 2 \rr$ gauge theories, 
$N \in \ll 0, 1, \cdots \rr$
In these cases, the 5D instanton partition functions are equal to off-critical 
conformal blocks in diagonal $\textit{Virasoro} \times\, U \! \ll 1 \rr$-conformal 
field theories. These instanton partition functions can be built in two steps.

\subsubsection{Gluing vertices to form a strip}
One starts by gluing four topological vertices to form a partition function 
that, following Iqbal and Kashani-Poor 
\cite{iqbal.kashani.poor.01, iqbal.kashani.poor.02}, 
we call the $U \!\ll 2 \rr$ {\it \lq strip\rq} partition function, 
see Figure \ref{figure.u2.strip.diagram}. This gluing is non-trivial in the 
sense that it requires the use of non-trivial symmetric function identities.
At the level of 2D conformal field theory, the 5D strip is equal to the matrix
element of a primary field between two arbitrary states in an off-critical 
deformation of the conformal field theory. In the 4D-limit discussed in Section \ref{8},  
we recover the corresponding critical expressions.

\subsubsection{Gluing strips to form an instanton partition function}
One glues $\ll N + 1 \rr$ copies of the strip partition function to obtain 
the instanton partition function of a 
$U_1     \! \ll 2 \rr \times$
$\cdots               \times$
$U_{N+1} \! \ll 2 \rr$
5D supersymmetric gauge theory. This gluing is essentially trivial in 
the sense that one proceeds as follows.
\1 Assign K\"ahler parameters to the would-be internal legs, 
\2 Take the product of the strip partition functions, with the same partitions
on the external legs that we wish to glue into internal legs, and
\3 Sum over all partitions that label the new internal legs, weighted 
with the assigned K\"ahler parameters, {\it without further evaluation}.
In this section, we focus on the gluing of four Macdonald vertices to obtain 
the Macdonald analogue of the $U \! \ll 2 \rr$ strip partition function, which 
is the non-trivial step in the computation of the full instanton partition 
functions and conformal blocks.

\subsection{
$\cM_{\, Y_1\, Y_2\, Y_3}^{\,q \, t} \ll x, y \rr$, and  
$\cM_{\, Y_1\, Y_2\, Y_3}^{\,t \, q} \ll y, x \rr$}
In gluing Macdonald vertices, we need
two versions of Macdonald vertices that differ in the way that we assign the variables. 
The first, $\cM_{\, Y_1\, Y_2\, Y_3}^{\,q \, t} \ll x, y \rr$, 
is that defined in Equation \ref{macdonald.vertex}.
The second, $\cM_{\, Y_1\, Y_2\, Y_3}^{\,t\,q} \ll y, x \rr$, 
is obtained from the first by swapping the variable assignments 
$x \rightleftharpoons y$, and 
$q \rightleftharpoons t$.

\subsection{Four vertices make a $U \! \ll 2 \rr$ strip}
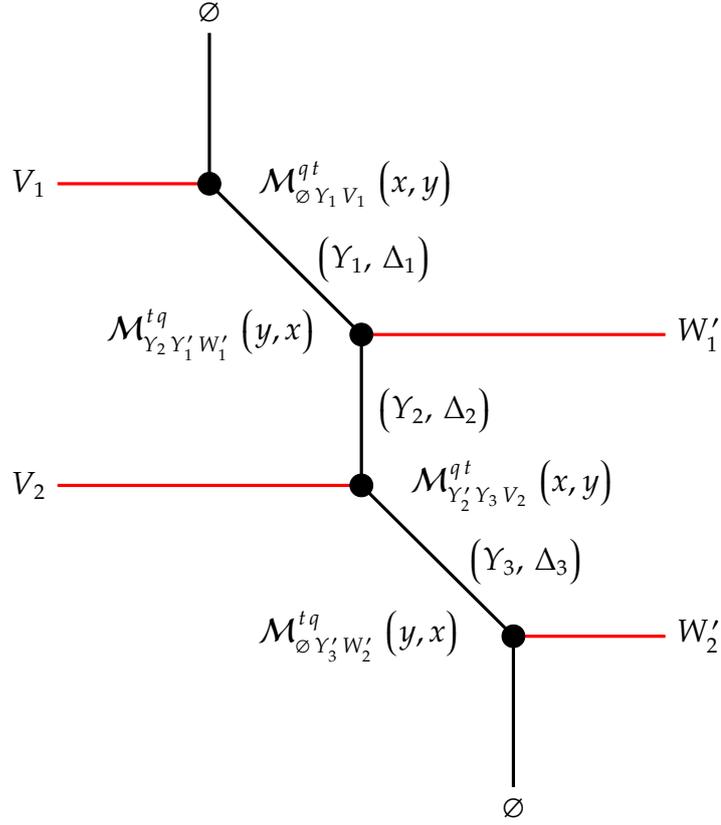
\begin{figure}
\begin{centering}
\begin{tikzpicture}[scale=1.0]
\draw [red, very thick] (0.0, 8.0)--(2.0, 8.0);
\draw [fill=black!100] (2.0,8.0) circle (0.15);
\draw [red, very thick] (0.0, 4.0)--(4.0, 4.0);
\draw [fill=black!100] (4.0,4.0) circle (0.15);
\draw [red, very thick] (6.0, 2.0)--(8.0, 2.0);
\draw [fill=black!100] (6.0,2.0) circle (0.15);
\draw [red, very thick] (4.0, 6.0)--(8.0, 6.0);
\draw [fill=black!100] (4.0,6.0) circle (0.15);

\draw [black, very thick] (2.0,10.0)--(2.0,8.0);
\draw [black, very thick] (6.0, 0.0)--(6.0,2.0);
\draw [black, very thick] (2.0, 8.0)--(4.0,6.0);
\draw [black, very thick] (4.0, 4.0)--(6.0,2.0);
\draw [black, very thick] (4.0, 6.0)--(4.0,4.0);

\node [above] at (2.0,10.0) {$\emptyset$};
\node [left]  at (0.0, 8.0) {$V_1$};
\node [left]  at (0.0, 4.0) {$V_2$};
\node [right] at (8.0, 2.0) {$W_2^{\, \prime}$};
\node [right] at (8.0, 6.0) {$W_1^{\, \prime}$};
\node [below] at (6.0, 0.0) {$\emptyset$};

\node [right] at (3.2,7.0) {$\ll Y_1,\, \Delta_1 \rr$};
\node [right] at (4.0,5.0) {$\ll Y_2,\, \Delta_2 \rr$};
\node [right] at (5.2,3.0) {$\ll Y_3,\, \Delta_3 \rr$};

\node [right] at (2.5,8.0) {$\cM_{\emptyset      \, Y_1            \, V_1 }^{\,q \, t}          \ll x, y \rr$};
\node [right] at (0.5,6.0) {$\cM_{Y_2            \, Y_1^{\, \prime}\, W_1^{\, \prime}}^{\,t\,q} \ll y, x \rr$};
\node [right] at (4.5,4.0) {$\cM_{Y_2^{\, \prime}\, Y_3            \, V_2 }^{\,q \, t}          \ll x, y \rr$};
\node [right] at (2.5,2.0) {$\cM_{\emptyset      \, Y_3^{\, \prime}\, W_2^{\, \prime}}^{\,t\,q} \ll y, x \rr$};
\end{tikzpicture} 
\par\end{centering}
\caption{\it 
The $ U \! \ll 2 \rr$ strip obtained by gluing four topological vertices, 
represented with black circles, along non-preferred legs. Each internal 
leg is assigned a partition $Y$ 
and a K\"ahler parameter $\Delta$. Each external leg is assigned a partition, 
but no K\"ahler parameters. The partitions on the vertical external legs are 
trivial.
}
\label{figure.u2.strip.diagram} 
\end{figure}

The $U \! \ll 2 \rr$ strip diagram is formed by gluing four topological vertices along 
their non-preferred $x$- and $y$-legs, as in Figure \ref{figure.u2.strip.diagram}. 
Each of these vertices is of type 
$\cM_{\, Y_1\, Y_2\, Y_3}^{\,q \, t} \ll x, y \rr$, or 
$\cM_{\, Y_1\, Y_2\, Y_3}^{\,t \, q} \ll y, x \rr$, as follows.
As in Figure \ref{figure.u2.strip.diagram}, there are 
two external legs to the left,  assigned partitions $\ll V_1,       \, V_2        \rr$, 
two external legs to the right, assigned partitions $\ll W_1^{\, \prime},\, W_2^{\, \prime} \rr$. 
The internal lines are assigned exponentiated K\"ahler parameters $Q_i,\, i = 1, 2, 3$, and 
partitions $Y_1, Y_2$, and $Y_3$, from top to bottom. 
The vertical external legs are assigned empty partitions, and no K\"ahler parameters.

%FIGURE.6.1

\subsection{The rules of gluing}
We consider two versions of the Macdonald vertex, 
$\cM_{\, Y_1\, Y_2\, Y_3}^{\,q \, t} \ll x, y \rr$, and
$\cM_{\, Y_1\, Y_2\, Y_3}^{\,t\,q} \ll y, x\rr$, and glue them as follows.
{\1} In gluing vertices along their non-preferred legs, that is the $x$-legs and $y$-legs, 
we can only glue a
$\cM_{\, Y_1\, Y_2\, Y_3}^{\,q \, t} \ll x, y \rr$ vertex and
$\cM_{\, Y_1\, Y_2\, Y_3}^{\,t \, q} \ll y, x \rr$ vertex, 
{\2} An $x$-leg can be only glued to an $x$-leg, and 
     a  $y$-leg can be only glued to a  $y$-leg, 
{\3} In gluing two legs, if one is assigned a partition $Y$, the other must be assigned 
the transpose partition $Y^{\, \prime}$, and finally 
{\4} The gluing is weighted by an exponentiated K\"ahler parameter $Q$, 
which contributes a factor $\ll - Q \rr^{| Y |}$ to the result.

\subsection{The normalized $U \! \ll 2 \rr$ strip
$\cS_{\, \bV\, \bW\, \bD}^{\, norm}$}
We compute the partition function that corresponds to the strip diagram, then normalize 
the result. In the context of the original, and the refined topological vertices, this 
calculation is standard, see for instance \cite{taki, kozcaz.thesis}. In the present work, 
we follow the computation in \cite{foda.wu.01}. We write the normalized $U\ll 2 \rr$ strip 
partition function as,

\begin{equation}
\cS_{\, \bV\, \bW\, \bD}^{\, norm}
=
\frac{
\cS_{\, \bV\, \bW\, \bD} 
}{
\cS_{\, \pmb{\emptyset}\, \pmb{\emptyset}\, \bD} 
}, 
\label{normalized.strip}
\end{equation}

\noindent where $\pmb{\emptyset} = \ll \emptyset, \emptyset \rr$, 
and $\emptyset$ is the empty Young diagram
\footnote{\,
We write 
$\cS_{\, \bV\, \bW\, \bD}^{\, norm}$ rather than
$\cS_{\, \bV\, \bW\, \bD}^{\, norm} \ll x, y\rr$, 
and do the same for the factors of 
$\cS_{\, \bV\, \bW\, \bD}^{\, norm} \ll x, y\rr$ discussed 
below, as that simplifies the notation without leading to 
ambiguities. On the other hand, we write 
$\cM_{\, Y_1\, Y_2\, Y_3}^{\,q \, t} \ll x, y \rr$ to distinguish it from 
$\cM_{\, Y_1\, Y_2\, Y_3}^{\,t \, q} \ll y, x \rr$, as in 
Equation \ref{u2.strip.partition.function.numerator}
}. 
The numerator on the right hand side of Equation \ref{normalized.strip} is, 

\begin{multline}
\cS_{\, \bV\, \bW\, \bD} 
=
\sum_{\, Y_1\, Y_2\, Y_3} 
\ll - Q_1  \rr^{ | Y_1 |}
\ll - Q_2  \rr^{ | Y_2 |}
\ll - Q_3  \rr^{ | Y_3 |}
\\  
\cM_{\, \emptyset        \, Y_1            \, V_1            }^{\, q \, t} \ll  x, y \rr\,
\cM_{\, Y_2              \, Y_1^{\, \prime}\, W_1^{\, \prime}}^{\, t \, q} \ll  y, x \rr\,
\cM_{\, Y_2^{\, \prime}  \, Y_3            \, V_2            }^{\, q \, t} \ll  x, y \rr\,
\cM_{\, \emptyset        \, Y_3^{\, \prime}\, W_2^{\, \prime}}^{\, t \, q} \ll  y, x \rr, 
\label{u2.strip.partition.function.numerator}
\end{multline}

\noindent where $Q_i$, $i=1, 2, 3$ are exponentiated K\"ahler parameters. 
We glue $x$-legs to $x$-legs and $y$-legs to $y$-legs, so the dependence 
of the vertices on $\ll x, y \rr$, and consequently the dependence on 
$\ll q, t \rr$ as well, must alternate along the strip.  
The denominator on the right hand side of Equation \ref{normalized.strip} 
is the partition function of the same strip diagram, but with all external 
partition pairs empty.

\subsection{Evaluating the numerator $\cS_{\, \bV\, \bW\, \bD}$}
Using the notation,

\begin{equation}
Z_{ Y }^{\, q \, t} \ll x,\, y \rr
=
\prod_{n \, = \, 0}^\infty
\ll 
\prod_{\wsq \in Y}
\frac{
1 - x^{\, L^+_{\wsq, \,Y}}\, y^{\, A_{\wsq,\,Y}}\,q^{\, n}\,t
}{
1 - x^{\, L^+_{\wsq, \,Y}}\, y^{\, A_{\wsq,\,Y}}\,q^{\, n}\, \pt}
\rr, 
\label{Z}
\end{equation}

\noindent together with the fact that,  

\begin{equation} 
P_{\, \emptyset / Y}^{\, q \, t} \ll \bx \rr =
P_{\, \emptyset    }^{\, q \, t} \ll \bx \rr = 1,
\quad 
Q_{\, \emptyset / Y}^{\, q \, t} \ll \bx \rr = 
Q_{\, \emptyset    }^{\, q \, t} \ll \bx \rr = 1, 
\end{equation}

\noindent and the definition of the Macdonald vertex in 
Equation \ref{macdonald.vertex}, 
Equation \ref{u2.strip.partition.function.numerator} becomes, 

\begin{multline}
\cS_{\, \bV\, \bW\,\bD}  
=
Z_{V_1            }^{\, q \, t} \ll x, y\rr
Z_{W_1^{\, \prime}}^{\, t\, q} \ll y, x\rr
Z_{V_2            }^{\, q \, t} \ll x, y\rr 
Z_{W_2^{\, \prime}}^{\, t\, q} \ll y, x\rr
\\
\sum_{ Y_1\, Y_2\, Y_3}
\ll - Q_1  \rr^{ | Y_1 |}
\ll - Q_2  \rr^{ | Y_2 |}
\ll - Q_3  \rr^{ | Y_3 |}
\\
Q_{\, Y_1}^{\, q \, t} \ll x^{\, \iota}\, y^{- V_1} \rr
\ll
\sum_{Y_4}
Q_{\, Y_1^{\, \prime} / Y_4}^{\, t\, q} \ll y^{\, \bj    }\, x^{- W_1^{\, \prime}}\,  \rr 
P_{\, Y_2             / Y_4}^{\, t\, q} \ll x^{\, \bi - 1}\, y^{- W_1            }\,  \rr
\rr 
\\
\ll 
\sum_{Y_5}
P_{\, Y_2^{\, \prime}/ Y_5}^{\, q \, t} \ll y^{\, \bj - 1}\, x^{- V_2^{\, \prime}}\,  \rr
Q_{\, Y_3^{\, \prime}/ Y_5}^{\, q \, t} \ll x^{\, \bi    }\, y^{- V_2            }\,  \rr
\rr 
Q_{\,  Y_3                }^{\, t \, q} \ll y^{\, \bj    }\, x^{- W_2^{\, \prime}}\,  \rr, 
\label{numerator}
\end{multline}

\noindent where the arguments in the Macdonald and dual Macdonald functions should
be understood in the sense of Section \ref{sequences}. 

\subsubsection{Remark} The first Macdonald vertex from the top contributes  
$Q_{\, Y_1}^{\, q \, t} \ll x^{\, \bi}\, y^{- V_1} \rr$ to the sum on the right hand side 
of Equation \ref{numerator}, since the Macdonald function associated with the vertical 
$x$-leg of this vertex, which is external, is trivial, and $P_{\emptyset}^{\, q \, t} = 1$. 
Similarly, the fourth Macdonald vertex from the top contributes  
$Q_{\, Y_3}^{\, t\, q} \ll y^{\, \bj}\, x^{- W_2^{\, \prime}} \rr$, 
since the Macdonald function associated with the vertical 
$x$-leg of this vertex, which is also external, is also trivial 
$P_{\emptyset}^{\, q \, t} = 1$. Due to alternating the $\ll x, y \rr$ and $\ll q, t \rr$ 
dependence of adjacent Macdonald vertices, the external legs of both vertices are $x$-legs. 

\subsubsection{The $Y_1$-sum}
From the definition of the skew Macdonald function, 

\begin{equation}
Q^{ | Y_1 / Y_2 |}\, 
P_{Y_1 / Y_2}^{\, q \, t} \ll     \bx \rr
=
P_{Y_1 / Y_2}^{\, q \, t} \ll Q\, \bx \rr, 
\quad
Q^{ | Y_1 / Y_2 |}                              \,
P_{Y_1 / Y_2}^{\, t\, q} \ll      \bx \rr
=
P_{Y_1 / Y_2}^{\, t\, q} \ll Q \, \bx \rr, 
\end{equation}

\noindent and using the Cauchy identity, Equation \ref{cauchy.identity.skew.02}, 

\begin{multline}
\sum_{\, Y_1}
\ll -\, Q_1 \rr^{ |Y_1|}
Q^{\, q\, t}_{\, Y_1                  } \ll x^{\, \bi}\, y^{ -V_1            }\,\rr
Q^{\, t\, q}_{\, Y_1^{\, \prime} / Y_4} \ll y^{\, \bj}\, x^{ -W_1^{\, \prime}}\,\rr 
\\
=
\prod_{i, j = 1}^\infty 
\ll 1 - Q_1\, x^{ - w_{1, j}^{\, \prime} + i} \, y^{- v_{1, i} + j}   \,  \rr\,
Q^{\, q \, t}_{\, Y_4^{\, \prime}} \ll - Q_1\, x^{\, \bi}\, y^{ - V_1}\,  \rr
\label{y1.sum}
\end{multline}

\subsubsection{The $Y_3$-sum}
The $Y_3$-sum is evaluated similarly to the $Y_1$-sum,

\begin{multline}
\sum_{\, Y_3}
\ll - Q_3 \rr^{ |Y_3 |}
Q^{\, t\, q}_{Y_3                  } \ll y^{\, \bj}\, x^{- W_2^{\, \prime}}  \rr
Q^{\, q\, t}_{Y_3^{\, \prime} / Y_5} \ll x^{\, \bi}\, y^{- V_2            }  \rr
\\
=
\prod_{i, j = 1}^\infty 
\ll 1 - Q_3\, x^{ - w_{2, j}^{\, \prime} + i}\, y^{- v_{2, i} + j} \rr\,
Q_{\, Y_5^{\, \prime}}^{\, t\, q} \ll -\, Q_3\,y^{\, \bj}\, x^{-W_2^{\, \prime}} \rr
\label{y3.sum}
\end{multline}

\subsubsection{The $Y_2$-sum}

Using the Cauchy identity, Equation \ref{cauchy.identity.skew.03}, 
we re-write this in terms of a sum over a new set of partition $Y_6$,

\begin{multline}
\sum_{\, Y_2} 
\ll - Q_2 \rr^{ |Y_2|}
P^{\, q\, t}_{ Y_2^{\, \prime} / Y_5} \ll y^{\, \bj - 1}\, x^{ - V_2^{\, \prime}} \rr\,
P^{\, t\, q}_{ Y_2             / Y_4} \ll x^{\, \bi - 1}\, y^{ - W_1            } \rr
\\
=
\ll - Q_2 \rr^{ |Y_4|}
\prod_{i, j = 1}^\infty 
\ll 1 - Q_2\, x^{ - v_{2, j}^{\, \prime\, +} + i}\, y^{- w^+_{1, i} + j} \rr
\\
\sum_{Y_6} 
P^{\, q \, t}_{\,Y_4^{\,\prime} / Y_6^{\,\prime}} \ll         y^{\,\bi - 1}\, x^{- V_2^{\,\prime}}\,\rr
P^{\, t \, q}_{\,Y_5^{\,\prime} / Y_6           } \ll - Q_2\, x^{\,\bi - 1}\, y^{- W_1           }\,\rr,
\label{y2.sum}
\end{multline}

\noindent where
$v_{2, j}^{\, \prime\, +} = v_{2, j}^{\, \prime} + 1$, and
$w^+_{1, i}               = w_{1, i}             + 1$ 
We use this notation below.

\subsubsection{The $Y_4$-sum}
Using the Cauchy identity, Equation \ref{cauchy.identity.skew.01}, 

\begin{multline}
\sum_{Y_4} 
\ll - Q_2 \rr^{ |Y_4|} 
Q^{\,q \, t}_{Y_4^{\,\prime}                 }\ll - Q_1\, x^{\,\bi}    \, y^{- V_1           }\rr
P^{\,q \, t}_{Y_4^{\,\prime} / Y_6^{\,\prime}}\ll         y^{\,\bi - 1}\, x^{- V_2^{\,\prime}}\rr
\\
=
\Pi^{\,q \, t} \ll Q_{\, 12}\, x^{\, \bi} \, y^{- V_1}, y^{\, \bj -1}\, x^{- V_2^{\,\prime}}\rr
Q^{\, q \, t}_{Y_6^{\, \prime}} \ll Q_{\, 12}\, x^{\,\bi}\, y^{- V_1} \rr, 
\label{y4.sum}
\end{multline}

\noindent where $Q_{\, 12} = Q_1\, Q_2$, and the arguments in 
$\Pi^{\,q \, t} \ll Q_{\, 12}\, x^{\, \bi} \, y^{- V_1}, y^{\, \bj -1}\, x^{- V_2^{\,\prime}}\rr$ 
are meant in the following sense. 

\subsubsection{Constructing 
$\Pi^{\,q \, t} \ll Q_{\, 12}\, x^{\, \bi} \, y^{- V_1}, y^{\, \bj -1}\, x^{- V_2^{\,\prime}}\rr$}
\label{constructing.PI}

Consider a scalar $Q_{\, 12}$, 
two sequences 
$\bi       = \ll 1, 2, \cdots \rr$ and
$\bj  = \ll 1, 2, \cdots \rr$, and 
two Young diagrams
$V_1 = \ll v_{1, 1}, v_{1, 2}, \cdots \rr$ and 
$V_2 = \ll v_{2, 1}, v_{2, 2}, \cdots \rr$. 
Consider the transpose 
$V_2^{\, \prime} = \ll v^{\, \prime}_{2, 1}, v^{\, \prime}_{2, 2}, \cdots \rr$, 
and form the product sequences, 

\begin{equation}
Q_{\, 12}\, x^{\, \bi}            \, y^{- V_1} = 
\ll Q_{\, 12} x   \, y^{- v_{1, 1}}, \, Q_{\, 12} x^2 \, y^{- v_{1, 2}}, \, \cdots \rr, 
\quad 
         y^{\, \bj  - 1}  \, x^{- V_2^{\, \prime}} = 
\ll                               \, x^{- v_{2, 1}}, \,        
         y                        \, x^{- v_{2, 2}}, \, \cdots \rr
\label{PI.02}
\end{equation} 

\noindent 
$\Pi^{\,q \, t} \ll Q_{\, 12}\, x^{\, \bi} \, y^{- V_1}, y^{\, \bj -1}\, x^{- V_2^{\,\prime}}\rr$ 
is understood in the sense of Equation \ref{PI.01}, but with the elements of 
$\bx = \ll x_1, x_2, \cdots \rr$ in Equation \ref{PI.01} replaced by the elements of 
$Q_{\, 12}\, x^{\, \bi}            \, y^{- V_1}$ in Equation \ref{PI.02}, and the elements of 
$\by = \ll y_1, y_2, \cdots \rr$ in Equation \ref{PI.01} replaced by the elements of
$y^{\, \bj  - 1}  \, x^{- V_2^{\, \prime}}$ in Equation \ref{PI.02}

\subsubsection{The $Y_5$-sum}

Similarly, 

\begin{multline}
\sum_{Y_5}
Q^{\, t\, q}_{Y^{\, \prime}_5      } \ll - Q_3\, y^{\, \bi    }\, x^{ - W_2^{\, \prime}} \rr\, 
P^{\, t\, q}_{Y^{\, \prime}_5 / Y_6} \ll - Q_2\, x^{\, \bi - 1}\, y^{ - W_1            } \rr
\\
=
\Pi^{\, t\, q} 
\ll 
Q_{\, 23}\, x^{\, \bi   - 1}\, y^{- W_1}, 
         y^{\, \bj }\, x^{- W_2^{\, \prime}} 
\rr\, 
Q^{\,t \,q}_{Y_6} \ll - Q_3\, y^{\,\bi}\, x^{- W_2^{\,\prime}} \rr, 
\label{y5.sum}
\end{multline}

\noindent where $Q_{\, 23} = Q_2\, Q_3$

\subsubsection{The $Y_6$-sum}

Finally, using the Cauchy identity, Equation \ref{more.cauchy.identities}, 
the $Y_6$-sum introduced in Equation \ref{y2.sum} is evaluated,  

\begin{equation}
\sum_{Y_6}       
Q^{\, q \, t}_{Y_6^{\, \prime}} \ll   Q_{\, 12}\, x^{\, \bi}\, y^{- V_1            }\, \rr 
Q^{\, t \, q}_{Y_6            } \ll - Q_3      \, y^{\, \bi}\, x^{- W_2^{\, \prime}}\, \rr
=
\prod_{i, j = 1}^\infty
\ll 
1 - Q_{\, 123}\, x^{- w_{2, j}^{\, \prime} + i}\, y^{- v_{1, i} + j}
\rr, 
\label{y6.sum}
\end{equation}

\noindent where $Q_{\, 123} = Q_1\, Q_2\, Q_3$

\subsubsection{Remark} The factors collected on the right sides of Equations 
\ref{y1.sum}, 
\ref{y3.sum}, 
\ref{y2.sum} and 
\ref{y6.sum} are independent of $\ll q, t\rr$. 

\subsection{The evaluated numerator $\cS_{\,\bV\, \bW\, \bD}$}
From Equation \ref{numerator}, and Equations \ref{y1.sum}--\ref{y6.sum}, the evaluated 
numerator of the $U \! \ll 2 \rr$ strip partition function is, 

\begin{equation}
\cS_{\, \bV\, \bW\, \bD} =
\cA_{\, \bV\, \bW\, \bD}\, 
\cB_{\, \bV\, \bW\, \bD}\, 
\cC_{\, \bV\, \bW\, \bD},
\label{numerator.evaluated.factorized}
\end{equation}

\noindent where,  

\begin{equation}
\cA_{\, \bV\, \bW\, \bD} = 
Z_{V_1 }^{\,q \, t} \ll x, y \rr
Z_{V_2 }^{\,q \, t} \ll x, y \rr
Z_{W_1^{\, \prime}}^{\,t\,q} \ll x,y \rr
Z_{W_2^{\, \prime}}^{\,t\,q} \ll x,y \rr,
\label{A.numerator}
\end{equation}

\begin{multline}
\cB_{\, \bV\, \bW\, \bD} = 
\prod_{i, j = 1}^\infty
\ll 1 - Q_1\, x^{- w_{1, j}^{\, \prime} + i}\,y^{- v_{1, i} + j} \rr
\ll 1 - Q_3\, x^{- w_{2, j}^{\, \prime} + i}\,y^{- v_{2, i} + j} \rr
\\ 
\ll 1 - Q_2         \, x^{- v_{2, j}^{\, \prime} + i - 1}\,y^{- w_{1, i} + j - 1}\, \rr
\ll 1 - Q_{\, 1 2 3}\, x^{- w_{2, j}^{\, \prime} + i    }\,y^{- v_{1, i} + j    }\, \rr, 
\label{B.numerator}
\end{multline}

\begin{equation}
\cC_{\, \bV\, \bW\, \bD} =  
\Pi^{\, q \, t} \ll Q_{\, 1 2}\, x^{\, \bi    }\, y^{- V_1}, y^{\, \bj  -1}\, x^{- V_2^{\,\prime}}\rr
\Pi^{\, t \, q} \ll Q_{\, 2 3}\, x^{\, \bi - 1}\, y^{- W_1}, y^{\, \bj    }\, x^{- W_2^{\,\prime}}\rr
\label{C.numerator}
\end{equation}

\subsection{The evaluated denominator 
$\cS_{\,\pmb{\emptyset}\,\pmb{\emptyset}\, \bD}$}
From Equation \ref{normalized.strip}, 
the evaluated denominator of the $U \! \ll 2 \rr$ strip partition function is, 

\begin{equation}
\cS_{\,\pmb{\emptyset}\,\pmb{\emptyset}\, \bD} = 
\cA_{\pmb{\emptyset}\, \pmb{\emptyset} \, \bD}\, 
\cB_{\pmb{\emptyset}\, \pmb{\emptyset} \, \bD}\, 
\cC_{\pmb{\emptyset}\, \pmb{\emptyset} \, \bD}, 
\end{equation}

\noindent where 
$\cA_{\pmb{\emptyset}\, \pmb{\emptyset} \, \bD}$, 
$\cB_{\pmb{\emptyset}\, \pmb{\emptyset} \, \bD}$ and  
$\cC_{\pmb{\emptyset}\, \pmb{\emptyset} \, \bD}$ are easily read off 
Equations \ref{A.numerator} and \ref{C.numerator} by setting 
$\bV = \pmb{\emptyset}$ and
$\bW = \pmb{\emptyset}$. 
In particular,
$\cA_{\pmb{\emptyset}\, \pmb{\emptyset} \, \bD} = 1$

\subsection{A normalized infinite product}
Given two two Young diagrams, 
$V = \ll v_1, v_2, \cdots \rr$, 
$W = \ll w_1, w_2, \cdots \rr$, we define the normalized infinite product
\footnote{\,
The fact that this product is normalized is indicated by $\checkmark$ at 
the top right corner of the factor. We use this notation in the sequel.
},

\begin{equation}
{\prod_{i, j = 1}^\infty} 
\ll 
1 - Q\, x^{\, i - v_j}\, y^{\, j - 1 - w_i} 
\rr^{\checkmark}
=  
\prod_{i, j = 1}^\infty 
\ll 
\frac{
1 - Q\, x^{\, i            - v_j}\, y^{\, j - 1           - w_i}
}{
1 - Q\, x^{\, _i \phantom{- v_j}}\, y^{\, j - 1  \phantom{- w_i}}
}
\rr, 
\label{normalised.products.01}
\end{equation}

\noindent which is normalized in the sense of the normalized strip partition 
function, Equation \ref{normalized.strip}.

\subsection{From a normalized infinite product to a finite product}
Following \cite{nakajima.yoshioka, awata.kanno.02}
\footnote{\,
Equations {\bf 2.8--2.11} in \cite{awata.kanno.02}.
}, 
we re-write the normalized infinite product in Equation \ref{normalised.products.01} as, 

\begin{multline}
{\prod_{i, j = 1}^{\infty}} 
\ll 1 - Q\, x^{\, - v_j \, + i \, - 1 }\, y^{\, - w_i^{\, \prime} + j } \rr^{\checkmark}
=  
\prod_{\, \wsq \in V} \ll 1 - Q\, x^{\, - A^+_{\wsq, W}}\, y^{\, -L_{\wsq, V}} \rr
\prod_{\, \bsq \in W} \ll 1 - Q\, x^{\,     A_{\bsq, V}}\, y^{\,L^+_{\bsq, W}} \rr
\\
=  
\prod_{\, \bsq \in W} \ll 1 - Q\, x^{\, - A^+_{\bsq, W}}\, y^{\, - L_{\bsq, V}} \rr
\prod_{\, \wsq \in V} \ll 1 - Q\, x^{\,     A_{\wsq, V}}\, y^{\, L^+_{\wsq, W}} \rr, 
\label{nakajima.identity}
\end{multline}

\noindent where, given a cell $\wsq \in V$,
$A_{\wsq, W}$ is the arm-length, and 
$L_{\wsq, W}$ is the leg-length of $\wsq$, measured with respect to the profile 
of a Young diagram $W$, that may be identical to, or different from $V$.
In particular, for $\wsq \notin W$, $A_{\wsq, W}$ and $L_{\wsq, W}$ are negative.

\subsubsection{Remark} 
The product on the left hand side of Equation \ref{nakajima.identity} is normalised 
in the sense of Equation \ref{normalised.products.01}, but the products on the right 
hand side are not. 
Equation \ref{nakajima.identity} expresses the normalized infinite products on 
the left hand side to the non-normalized finite products on the right hand side.

\subsection{The evaluated, normalized $\cS_{\,\bV\,\bW\, \bD}^{\, norm}$}
From Equations \ref{A.numerator}--\ref{C.numerator}, the normalised $U \! \ll 2 \rr$ 
strip partition function is,

\begin{equation}
\cS_{\,\bV\,\bW\, \bD}^{\, norm} = 
\ll \frac{\cA_{\, \bV\, \bW\, \bD}}{\cA_{\pmb{\emptyset}\, \pmb{\emptyset} \, \bD}} \rr \,  
\ll \frac{\cB_{\, \bV\, \bW\, \bD}}{\cB_{\pmb{\emptyset}\, \pmb{\emptyset} \, \bD}} \rr \,  
\ll \frac{\cC_{\, \bV\, \bW\, \bD}}{\cC_{\pmb{\emptyset}\, \pmb{\emptyset} \, \bD}} \rr   
= 
\cA^{\, norm}_{\bV\, \bW\, \bD}\, 
\cB^{\, norm}_{\bV\, \bW\, \bD}\, 
\cC^{\, norm}_{\bV\, \bW\, \bD}, 
\label{S.norm}
\end{equation}

\noindent where the factors are as follows.

\subsubsection{$\cA^{\, norm}_{\bV\, \bW\, \bD}$}
This is the product of four factors that depend on $\ll q, t \rr$, generated 
by normal ordering the $q t$-vertex operators in the Macdonald vertex, 

\begin{equation}
\cA^{\, norm}_{\bV\, \bW\, \bD} =
Z_{V_1            }^{\,q \, t} \ll x, y \rr
Z_{V_2            }^{\,q \, t} \ll x, y \rr
Z_{W_1^{\, \prime}}^{\,t \, q} \ll y, x \rr
Z_{W_2^{\, \prime}}^{\,t \, q} \ll y, x \rr,
\label{A.norm}
\end{equation}

\noindent where $Z_{V}^{\, q \, t} \ll x, y \rr$ is defined 
in Equation \ref{Z}

\subsubsection{$\cB^{\, norm}_{\bV\, \bW\, \bD}$}
This is the product of four factors that do {\it not} depend on $\ll q, t\rr$, 
generated in performing sums over states labelled by partitions, 

\begin{multline}
\cB^{\, norm}_{\bV\, \bW\, \bD} = 
{\prod_{i, j = 1}^\infty} 
\ll 1 - Q_1      \, x^{- w_{1, j}^{\, \prime} + i}    \,y^{- v_{1, i} + j} \rr^{\! \checkmark}
\ll 1 - Q_3      \, x^{- w_{2, j}^{\, \prime} + i}    \,y^{- v_{2, i} + j} \rr^{\! \checkmark}
\\
\ll 1 - Q_2      \, x^{- v_{2, j}^{\, \prime} + i - 1}\,y^{- w_{1, i} + j - 1} \rr^{\! \checkmark}
\ll 1 - Q_{\, 1 2 3}\, x^{- w_{2, j}^{\, \prime} + i}\,y^{- v_{1, i} + j} \rr^{\! \checkmark}
\label{B.norm.01}
\end{multline}

\noindent Using Equation \ref{nakajima.identity},

\begin{multline}
\cB^{\, norm}_{\, \bV\, \bW} =    
\prod_{\wsq \in V_1} \ll 1-Q_1   \, x^{\, -L_{\wsq, W_1}}\, y^{\, -A_{\wsq, V_1}} \rr
\prod_{\bsq \in W_1} \ll 1-Q_1   \, x^{\,L^+_{\bsq, V_1}}\, y^{\,A^+_{\bsq, W_1}} \rr
\\
\prod_{\wsq \in V_2} \ll 1-Q_3   \, x^{\, -L_{\wsq, W_2}}\, y^{\, -A_{\wsq, V_2}} \rr 
\prod_{\bsq \in W_2} \ll 1-Q_3   \, x^{\,L^+_{\bsq, V_2}}\, y^{\,A^+_{\bsq, W_2}} \rr
\\
\prod_{\bsq \in W_1} \ll 1-Q_2\, x^{\, -L^+_{\bsq, V_2}}\,y^{\, -A^+_{\bsq, W_1}} \rr
\prod_{\wsq \in V_2} \ll 1-Q_2\, x^{\,L_{\wsq, W_1}}\,y^{\,A_{\wsq, V_2}} \rr
\\
\prod_{\wsq \in V_1} \ll 1- Q_{\, 123}   \, x^{\, -L_{\wsq, W_2}}\, y^{\, -A_{\wsq, V_1}} \rr 
\prod_{\bsq \in W_2} \ll 1- Q_{\, 123}   \, x^{\,L^+_{\bsq, V_1}}\, y^{\,A^+_{\bsq, W_2}} \rr, 
\label{B.norm.02}
\end{multline}

\subsubsection{$\cC^{\, norm}_{\bV\, \bW\, \bD}$}
This is the product of two factors that depend on $\ll q, t\rr$, generated 
in performing sums over states labelled by partitions,

\begin{equation}
\cC^{\, norm}_{\bV\, \bW\, \bD} =  
\frac{
\Pi^{\, q \, t} 
\ll Q_{\, 1 2}\, x^{\, \bi}    \, y^{- V_1},\, y^{\, \bj -1}\, x^{- V_2^{\, \prime}} \rr
}{
\Pi^{\, q \, t} 
\ll Q_{\, 1 2}\, x^{\, \bi}    \,          ,\, y^{\, \bj -1}\,                       \rr
}\, 
\frac{ 
\Pi^{\, t\, q} \ll Q_{\, 2 3}\, x^{\, \bi -1}\, y^{- W_1},\, y^{\, \bj }\, x^{- W_2^{\,\prime}} \rr
}{
\Pi^{\, t\, q} \ll Q_{\, 2 3}\, x^{\, \bi -1},\,             y^{\, \bj }                        \rr
}, 
\label{C.norm.01}
\end{equation}

\noindent where the arguments of $\Pi^{ q t}$ in Equation \ref{C.norm.01} are explained 
in Section \ref{constructing.PI}. Using Equation \ref{normalised.products.01}, 

\begin{equation}
\frac{
\Pi^{ q t} 
\ll Q\, x^{\, \bi     } \, y^{\,  - W}, 
        y^{\, \bj  - 1} \, x^{\,  - V} \rr
}{
\Pi^{ q t} 
\ll Q\, x^{\, \bi     },       
        y^{\, \bj  - 1}              \rr
}
=
{\prod_{i,\, j,\, n\, =\, 1}^\infty}
\frac{
\ll 1 - Q\, x^{\, i -v_j} \, y^{\, j - 1 - w_i}\, q^{n-1}\,   t \rr^{\checkmark}          
}{
\ll 1 - Q\, x^{\, i -v_j} \, y^{\, j - 1 - w_i}\, q^{n-1}\, \pt \rr^{\checkmark}
}, 
\end{equation}

\noindent Equation \ref{C.norm.01} becomes, 

\begin{multline}
\cC_{\, \bV\, \bW\, \bD}^{\, norm} 
= 
\prod_{i,\, j,\, \ll n + 1 \rr = 1}^\infty
\frac{
\ll 1 - Q_{\, 12}\, x^{ i - v^{\, \prime}_{2, j}}\, y^{j - v^+_{1, i}}\, q^{n-1}\,    t \rr^{\checkmark}
}{
\ll 1 - Q_{\, 12}\, x^{ i - v^{\, \prime}_{2, j}}\, y^{j - v^+_{1, i}}\, q^{n-1}\,  \pt \rr^{\checkmark}
}
\frac{
\ll 1 - Q_{\, 23}\, x^{i - w^{\, \prime\, +}_{2, j}}\, y^{j - w_{1, i}}\, t^{n-1}\,   q \rr^{\checkmark}
}{
\ll 1 - Q_{\, 23}\, x^{i - w^{\, \prime\, +}_{2, j}}\, y^{j - w_{1, i}}\, t^{n-1}\, \pq \rr^{\checkmark}
}, 
\\
= \prod_{n = 0}^\infty
\ll 
\prod_{\wsq \in V_1}
\frac{ 
\ll 1 - Q_{\, 12}\, x^{ - L_{\wsq,\, V_2}}\, y^{- A^+_{\wsq,\, V_1}}\, q^{\, n}\,   t \rr
}{
\ll 1 - Q_{\, 12}\, x^{ - L_{\wsq,\, V_2}}\, y^{- A^+_{\wsq,\, V_1}}\, q^{\, n}\, \pt \rr
}
\,
\prod_{\wsq \in V_2}
\frac{ 
\ll 1 - Q_{\, 12}\, x^{\, L^+_{\wsq,\, V_1}}\, y^{ A_{\wsq,\, V_2}}\, q^{\, n}\,   t \rr
}{
\ll 1 - Q_{\, 12}\, x^{\, L^+_{\wsq,\, V_1}}\, y^{ A_{\wsq,\, V_2}}\, q^{\, n}\, \pt \rr
}     
\right. 
\\
\left. 
\prod_{\bsq \in W_1}
\frac{ 
\ll 1 - Q_{\, 23}\, x^{ - L^+_{\bsq,\, W_2}}\, y^{- A_{\bsq,\, W_1}}\, t^{\, n}\,   q \rr
}{ 
\ll 1 - Q_{\, 23}\, x^{ - L^+_{\bsq,\, W_2}}\, y^{- A_{\bsq,\, W_1}}\, t^{\, n}\, \pq \rr
}
\,
\prod_{\bsq \in W_2}     
\frac{ 
\ll 1 - Q_{\, 23}\, x^{\, L_{\bsq,\, W_1}}\, y^{\, A^+_{\bsq,\, W_2}}\, t^{\, n}\,   q \rr
}{ 
\ll 1 - Q_{\, 23}\, x^{\, L_{\bsq,\, W_1}}\, y^{\, A^+_{\bsq,\, W_2}}\, t^{\, n}\, \pq \rr
}
\rr
\label{C.norm.02}
\end{multline}

\subsection{Factorization of the $q t$-strip into $q t$-independent and $q t$-dependent 
factors} 
\label{factorization}
While $\cM^{\, q \, t}_{Y_1 Y_2 Y_3} \ll x, y\rr$, Equation \ref{macdonald.vertex}, does 
not factorize into $\cR_{Y_1 Y_2 Y_3} \ll x, y\rr$, Equation \ref{refined.vertex}, times 
a $q t$-dependent factor, the strip partition function 
$\cS_{\,\bV\,\bW\, \bD}^{\, norm}$, Equation \ref{S.norm}, does. 

\subsubsection{Factorization of $\cA_{\, \bV\, \bW\, \bD}^{\, norm}$ into $q t$-independent 
and $q t$-dependent factors}
From Equation \ref{A.norm}, $\cA_{\, \bV\, \bW\, \bD}^{\, norm}$ is a product of four copies 
of $Z_Y^{\, q \, t} \ll x,\, y \rr$, defined in Equation \ref{Z}. Each of these factorizes,

\begin{equation}
Z_{ Y }^{\, q \, t} \ll x,\, y \rr
=
\ll 
\prod_{\wsq \in Y}
\frac{
1
}{
1 - x^{\, L^+_{\wsq, \,Y}}\, y^{\, A_{\wsq,\,Y}}}
\rr
\, 
\ll 
\prod_{n \, = \, 1}^\infty\, 
\prod_{\wsq \in Y}
\frac{
1 - x^{\, L^+_{\wsq, \,Y}}\, y^{\, A_{\wsq,\,Y}}\,q^{n          -1 } \,   t
}{
1 - x^{\, L^+_{\wsq, \,Y}}\, y^{\, A_{\wsq,\,Y}}\,q^{n \phantom{-1}} \, \pt}
\rr, 
\label{Z.factorized}
\end{equation}

\noindent where the first factor, on the right hand side of Equation \ref{Z.factorized}, 
appears in Equation \ref{refined.vertex}, and second factor $\rightarrow 1$ in the limit 
$q \rightarrow t$. Thus,  

\begin{equation}
\cA_{\, \bV\, \bW\, \bD}^{\, norm} = 
\ll \cA_{\, \bV\, \bW\, \bD}^{\, norm} \rr_{\ll 0, \, 0 \rr}
\ll \cA_{\, \bV\, \bW\, \bD}^{\, norm} \rr_{\ll q, \, t \rr}, 
\end{equation}

\begin{equation}
\ll \cA_{\, \bV\, \bW\, \bD}^{\, norm} \rr_{\ll 0, \, 0 \rr} = 
\prod_{i=1}^2 \, 
\ll 
\prod_{\wsq \in V_i } 
\frac{1}{1 - x^{\, L^+_{\wsq, \,V_i}}\, y^{\, A_{\wsq,\, V_i}}} 
\rr
\ll 
\prod_{\wsq \in W_i} \frac{1}{1 - x^{\, L_{\wsq, \,W_i}}\, y^{\, A^+_{\wsq,\,W_i}}} 
\rr,
\label{A.norm.00}
\end{equation}

\begin{equation}
\ll \cA_{\, \bV\, \bW\, \bD}^{\, norm} \rr_{\ll q, \, t \rr} = 
\prod_{n \, = \, 1}^\infty
\prod_{i=1}^2 \, 
\ll 
\prod_{\wsq = V_i } \, 
\frac{
1 - x^{\, L^+_{\wsq, \, V_i }}\, y^{\, A_{\wsq,\, V_i }}\,q^{n          -1} \,   t}{
1 - x^{\, L^+_{\wsq, \, V_i }}\, y^{\, A_{\wsq,\, V_i }}\,q^{n \phantom{-1}}\, \pt}
\rr\, 
\ll 
\prod_{\bsq = W_i}
\frac{
1 - x^{\, L_{\bsq, \, W_i}}\, y^{\, A^+_{\bsq,\, W_i}}\,t^{n          -1} \,   q}{
1 - x^{\, L_{\bsq, \, W_i}}\, y^{\, A^+_{\wsq,\, W_i}}\,t^{n \phantom{-1}}\, \pq}
\rr
\label{A.norm.qt}
\end{equation} 

\subsubsection{$\cB_{\, \bV\, \bW\, \bD}^{\, norm}$ has no $q t$-dependent factors}
From Equation \ref{B.norm.02}, $\cB_{\, \bV\, \bW\, \bD}^{\, norm}$ has no $q t$-dependence, 

\begin{equation}
\cB_{\, \bV\, \bW\, \bD}^{\, norm} = 
\ll \cB_{\, \bV\, \bW\, \bD}^{\, norm} \rr_{\ll 0, \, 0 \rr} 
\label{B.norm.00}
\end{equation}

\subsubsection{The factorization of $\cC_{\, \bV\, \bW\, \bD}^{\, norm}$ into $q t$-independent 
and $q t$-dependent factors}
From Equation \ref{C.norm.02},

\begin{equation}
\cC_{\, \bV\, \bW\, \bD}^{\, norm} = 
\ll \cC_{\, \bV\, \bW\, \bD}^{\, norm} \rr_{\ll 0, \, 0 \rr} 
\ll \cC_{\, \bV\, \bW\, \bD}^{\, norm} \rr_{\ll q, \, t \rr}, 
\end{equation}

\begin{multline}
\ll \cC_{\, \bV\, \bW\, \bD}^{\, norm} \rr_{\ll 0, \, 0 \rr} 
= 
\ll 
\prod_{\wsq \in V_1}
\frac{1}{\ll 1 - Q_{\, 12}\, x^{ - L_{\wsq,\, V_2}}\, y^{ - A^+_{\wsq,\, V_1}} \rr}
\,
\prod_{\wsq \in V_2}
\frac{1}{\ll 1 - Q_{\, 12}\, x^{\, L^+_{\wsq,\, V_1}}\, y^{ A_{\wsq,\, V_2}} \rr}     
\right. 
\\
\left. 
\prod_{\bsq \in W_1}
\frac{1}{\ll 1 - Q_{\, 23}\, x^{ - L^+_{\bsq,\, W_2}}\, y^{ - A_{\bsq,\, W_1}} \rr}
\,
\prod_{\bsq \in W_2}     
\frac{1}{\ll 1 - Q_{\, 23}\, x^{\, L_{\bsq,\, W_1}}\,   y^{\, A^+_{\bsq,\, W_2}} \rr}
\rr, 
\label{C.norm.00}
\end{multline}

\noindent which is the factor that appears when using
$\cR_{Y_1 Y_2 Y_3} \ll x, y\rr$ in \cite{foda.wu.01}
\footnote{\,
Up to a change in notation that transposes Young diagrams, exchanging
arm-lengths and leg-lengths.
}, 

\begin{multline}
\ll \cC_{\, \bV\, \bW\, \bD}^{\, norm} \rr_{\ll q, \, t \rr}   
= \prod_{n = 1}^\infty
\ll 
\prod_{\wsq \in V_1}
\frac{ 
\ll 1 - Q_{\, 12}\, x^{ - L_{\wsq,\, V_2}}\, y^{- A^+_{\wsq,\, V_1}}\, q^{n-1}\,   t \rr}{
\ll 1 - Q_{\, 12}\, x^{ - L_{\wsq,\, V_2}}\, y^{- A^+_{\wsq,\, V_1}}\, q^{n  }\, \pt \rr
}
\,
\prod_{\wsq \in V_2}
\frac{ 
\ll 1 - Q_{\, 12}\, x^{\, L^+_{\wsq,\, V_1}}\, y^{ A_{\wsq,\, V_2}}\, q^{n-1}\,   t \rr}{
\ll 1 - Q_{\, 12}\, x^{\, L^+_{\wsq,\, V_1}}\, y^{ A_{\wsq,\, V_2}}\, q^{n  }\, \pt \rr
}     
\right. 
\\
\left. 
\prod_{\bsq \in W_1}
\frac{ 
\ll 1 - Q_{\, 23}\, x^{ - L^+_{\bsq,\, W_2}}\, y^{- A_{\bsq,\, W_1}}\, t^{\, n-1}\,   q \rr}{ 
\ll 1 - Q_{\, 23}\, x^{ - L^+_{\bsq,\, W_2}}\, y^{- A_{\bsq,\, W_1}}\, t^{\, n  }\, \pq \rr
}
\,
\prod_{\bsq \in W_2}     
\frac{ 
\ll 1 - Q_{\, 23}\, x^{\, L_{\bsq,\, W_1}}\, y^{\, A^+_{\bsq,\, W_2}}\, t^{\, n-1}\,   q \rr}{ 
\ll 1 - Q_{\, 23}\, x^{\, L_{\bsq,\, W_1}}\, y^{\, A^+_{\bsq,\, W_2}}\, t^{\, n  }\, \pq \rr
}
\rr, 
\label{C.norm.qt}
\end{multline}

\noindent which $\rightarrow 1$, in the limit $q \rightarrow t$. 

\subsection{Comments on the structure of $\cS_{\,\bV\, \bW\, \bD}^{\, norm}$}
\label{pole.structure.5D}

\subsubsection{The $q t$-independent terms}
\label{5D.00.poles}

\begin{equation}
\ll \cS_{\, \bV\, \bW\, \bD}^{\, norm} \rr_{\ll 0, \, 0 \rr} = 
\ll \cA_{\, \bV\, \bW\, \bD}^{\, norm} \rr_{\ll 0, \, 0 \rr} \, 
\ll \cB_{\, \bV\, \bW\, \bD}^{\, norm} \rr_{\ll 0, \, 0 \rr} \, 
\ll \cC_{\, \bV\, \bW\, \bD}^{\, norm} \rr_{\ll 0, \, 0 \rr}, 
\end{equation}

\noindent is the $q t$-independent factor that we obtain if we use 
$\cR_{Y_1 Y_2 Y_3} \ll x, y \rr$ to compute $\cS_{\, \bV\, \bW\, \bD}^{\, norm}$
\footnote{\,
\label{changes}
See Equations 3.17--3.19 in \cite{foda.wu.01}, but note the differences in 
notation. In particular, 
$Q_i$, $i \in \ll 1, M, 2 \rr$, $A^{++}_{\wsq, \, Y}$ and $L^{++}_{\wsq, \, Y}$
in \cite{foda.wu.01}, become
$Q_i$, $i \in \ll 1, 2, 3 \rr$, $A^+_{\wsq, \, Y}$ and $L^+_{\wsq, \, Y}$ 
in the present work. 
Also $A^+_{\wsq, \, Y}$ and $L^+_{\wsq, \, Y}$ in \cite{foda.wu.01} are
expanded as $A_{\wsq, \, Y} + \frac12$ and $L_{\wsq, \, Y} + \frac12$ 
in the present work, cancelling factors of $x^{\frac12}$ and $y^{\frac12}$
whose analogues appear in \cite{foda.wu.01}
}. 
Since we choose each of the parameters $\ll x, y, q, t\rr$, to be $< 1$, 
all exponents on the right hand sides of Equations \ref{A.norm.00}--\ref{A.norm.qt}
are non-negative, and at least one exponent in each factor is non-zero, 
$\cA_{\, \bV\, \bW\, \bD}^{\, norm}$ has no poles. 
$\ll \cB_{\, \bV\, \bW\, \bD}^{\, norm} \rr_{\ll 0, \, 0 \rr}$, Equation \ref{B.norm.00}, 
has no poles. 
Each pair of exponents in the same factor on the right hand sides of Equations 
\ref{C.norm.00} is such that, one exponent may (or may not) be positive while 
the other is negative. 
This is because in each factor, $\wsq$ is in one diagram $Y_1$, so it has 
positive arm-length and leg-length with respect to the profile of $Y_1$, 
but may (or may not) be outside the other diagram $Y_2$. When $\wsq \notin Y_2$, 
it has a negative arm-length and leg-length with respect to the profile of $Y_2$.
These cases contribute to the right hand side of Equation \ref{C.norm.00}, 
and $\ll \cC_{\, \bV\, \bW\, \bD}^{\, norm} \rr_{\ll 0, \, 0 \rr}$ has poles
\footnote{\,
See detailed discussion in \cite{bershtein.foda}
}.
These are the poles obtained using $\cR_{Y_1 Y_2 Y_3} \ll x, y \rr$. 
When copies of $\ll \cS_{\,\bV\, \bW\, \bD}^{\, norm} \rr_{\ll 0, \, 0\rr}$
are glued to form 5D instanton partition functions,  
the poles in $\ll \cC_{\, \bV\, \bW\, \bD}^{\, norm} \rr_{\ll 0, \, 0 \rr}$ 
correspond to BPS states in a 5D $\cN = 2$ $U \! \ll 2 \rr$ gauge theory. 
In the 2D interpretation of the instanton partition functions, they correspond
to states that flow in the internal channels of off-critical deformations of 
conformal blocks \cite{awata.yamada.01, awata.yamada.02}. 

\subsubsection{The $q t$-dependent terms}
\label{5D.qt.poles}

\begin{equation}
\ll \cS_{\, \bV\, \bW\, \bD}^{\, norm} \rr_{\ll q, \, t \rr} = 
\ll \cA_{\, \bV\, \bW\, \bD}^{\, norm} \rr_{\ll q, \, t \rr} \, 
\ll \cB_{\, \bV\, \bW\, \bD}^{\, norm} \rr_{\ll q, \, t \rr} \, 
\ll \cC_{\, \bV\, \bW\, \bD}^{\, norm} \rr_{\ll q, \, t \rr}, 
\end{equation}

\noindent is the additional $q t$-dependent factor in $\cS_{\, \bV\, \bW\, \bD}^{\, norm}$
that we obtain when we use 
$\cM_{Y_1 Y_2 Y_3}^{\, q \, t} \ll x, y \rr$. 
$\ll \cA_{\, \bV\, \bW\, \bD}^{\, norm} \rr_{\ll q, \, t \rr}$ and
$\ll \cB_{\, \bV\, \bW\, \bD}^{\, norm} \rr_{\ll q, \, t \rr}$ have 
no poles for the same reasons that 
$\ll \cA_{\, \bV\, \bW\, \bD}^{\, norm} \rr_{\ll 0, \, 0 \rr}$ and
$\ll \cB_{\, \bV\, \bW\, \bD}^{\, norm} \rr_{\ll 0, \, 0 \rr}$ do, while 
$\ll \cC_{\, \bV\, \bW\, \bD}^{\, norm} \rr_{\ll q, \, t \rr}$ has 
poles for the same reasons that
$\ll \cC_{\, \bV\, \bW\, \bD}^{\, norm} \rr_{\ll 0, \, 0 \rr}$ does. 
From Equation \ref{C.norm.qt}, one can see that, for generic positive values of $\ll q, t\rr$,
$\ll \cC_{\, \bV\, \bW\, \bD}^{\, norm} \rr_{\ll q, \, t \rr}$ contributes an infinite tower 
of poles for each pole in 
$\ll \cC_{\, \bV\, \bW\, \bD}^{\, norm} \rr_{\ll 0, \, 0 \rr}$.
For non-generic positive choices of $\ll q, t \rr$, one can cancel some of these new poles, 
but infinite towers remain. 
When copies of $\cS_{\,\bV\, \bW\, \bD}^{\, norm}$, with $q t$-dependent factors 
are glued to form 5D $q t$-deformed instanton partition functions, 
the $\ll \cC_{\, \bV\, \bW\, \bD}^{\, norm} \rr_{\ll q, \, t \rr}$ will contribute. 
To have a better look at these contributions, we take the 4D limit of 
$\cS_{\,\bV\, \bW\, \bD}^{\, norm}$.

\section{The 4D limit of the $U \! \ll 2 \rr$ strip partition function}
\label{8} 
{\it 
We take the 4D limit of the $U \! \ll 2 \rr$ strip partition function.
}
\smallskip 

$\cM^{\, q \, t}_{\, Y_1\, Y_2\, Y_3} \ll x, y \rr$ is a $q t$-deformation of 
$\cR_{Y1\, Y_2\, Y_3} \ll x, y\rr$, and as such, it is a building block of 
$q t$-deformed 5D instanton partition functions. 
In the absence of the $\ll q, t\rr$-parameters, one takes the 4D limit by writing
the parameters $\ll x, y \rr$ and the exponentiated K\"ahler parameters 
$Q_i$, $i = 1, 2, 3$ as, 

\begin{equation}
x   = e^{\,  R\, \eps_1}, \, 
y   = e^{\,- R\, \eps_2}, \,
Q_i = e^{\,  R\, \Delta_i},\, i \in \ll 1,\, 2,\, 3 \rr,
\end{equation}

\noindent where $R$ is the circumference of the $M$-theory circle, 
$\ll \eps_1, \eps_2 \rr$ are Nekarsov's equivariant deformation parameters, 

\begin{equation}
\eps_1 < 0 < \eps_2, 
\end{equation}

\noindent and $\Delta_i$, $i \in \ll 1, 2, 3\rr$ are K\"ahler parameters related to mass 
and Coulomb parameters in a 4D supersymmetric gauge theory, that can be related {\it via} 
the AGT correspondence with conformal dimensions of primary fields in a 2D conformal field 
theory \cite{alday.gaiotto.tachikawa}. 
In the presence of $\ll q, t\rr$-parameters, we choose to expand these parameters 
without loss of generality in terms of $R$ as, 

\begin{equation}
q = e^{\, - R\, \eps_3}, \, 
t = e^{\, - R\, \eps_4}, 
\end{equation}

\noindent where 
$\ll \eps_3, \eps_4 \rr$ are real, non-negative parameters of the same dimensions as 
$\ll \eps_1, \eps_2 \rr$ 

\subsubsection{Remark} If there is another parameter, of the same dimension as $R$, 
to expand $\ll q, t\rr$ in, then the difference of the two expansions can be absorbed 
in a redefinition of $\ll \eps_3, \eps_4 \rr$ 

\subsubsection{Using the factorization to obtain a well-defined 4D limit}
To obtain a well-defined 4D limit that reduces to the known 4D limit for $q=t$, we use 
the factorization into 
$q t$-independent factors, and
$q t$-dependent   factors that $\rightarrow 1$, in the limit $\eps_3 \rightarrow \eps_4$, 
as in Section \ref{factorization}

\subsubsection{$\cA_{\, \bV\, \bW\, \bD}^{\, norm, \, 4D}$}
The 4D limit of $Z_Y^{\,q \, t} \ll x, y \rr$ is,

\begin{equation}
Z_{Y}^{\,q \, t} \ll x, y \rr \bigr\vert_{R \rightarrow 0} = R^{-|Y|}
\,
\ll \prod_{\wsq\in Y} \frac{1}{- \eps_1 L^+_{\wsq, Y} + \eps_2 A_{\wsq, Y}} \rr\,
\ll 
\prod_{n \, = \, 0}^\infty
\prod_{\wsq \in Y} 
\frac{
n\, \eps_3 +\, \eps_4 - \eps_1 L^+_{\wsq, Y} + \eps_2 A_{\wsq, Y}
}{
\ll n + 1 \rr \eps_3  - \eps_1 L^+_{\wsq, Y} + \eps_2 A_{\wsq, Y}
}
\rr, 
\label{z.q.t.4d}
\end{equation}

\noindent where all factors in the double-product on the right hand side 
$\rightarrow 1$, in the limit $\eps_3 \rightarrow \eps_4$ 
Similarly, the 4D limit of $Z_{Y^{\, \prime}}^{\,t\,q} \ll y, x \rr$ is, 

\begin{equation}
Z_{Y^{\, \prime }}^{\,t\,q} \ll y, x \rr \biggr\vert_{R \rightarrow 0} = 
R^{- | Y |}
\ll 
\prod_{\wsq \in Y} 
\frac{
1
}{
- \eps_1 L_{\wsq, Y}  + \eps_2 A^+_{\wsq, Y}
} 
\rr\, 
\ll 
\prod_{n \, = \, 0}^\infty
\prod_{\wsq \in Y}
\frac{         
n\, \eps_4 +\, \eps_3 - \eps_1 L_{\wsq, Y} + \eps_2 A^+_{\wsq, Y}
}{  
\ll n + 1 \rr        \eps_4 - \eps_1 L_{\wsq, Y} + \eps_2 A^+_{\wsq, Y}
}
\rr  
\label{z.t.q.4d}
\end{equation}

\noindent From Equations \ref{z.q.t.4d} and \ref{z.t.q.4d}, 

\begin{equation}
\cA_{\, \bV\, \bW\, \bD}^{\,norm, \, 4D} = 
\ll \cA_{\, \bV\, \bW\, \bD}^{\,norm, \, 4D} \rr_{\ll 0, 0 \rr} \, 
\ll \cA_{\, \bV\, \bW\, \bD}^{\,norm, \, 4D} \rr_{\ll q, t \rr}, 
\label{A.norm.4D}
\end{equation}

\begin{multline}
\ll \cA_{\, \bV\, \bW\, \bD}^{\,norm, \, 4D} \rr_{\ll 0, 0 \rr} = 
\\
R^{-\ll | V_1 | + | V_2 | + | W_1 | + | W_2 | \rr}
\prod_{i=1}^2
\ll 
\prod_{\wsq \in V_i} \frac{1}{ - \eps_1\, L^+_{\wsq, V_i} + \eps_2\,   A_{\wsq, V_i}}\,
\prod_{\bsq \in W_i} \frac{1}{ - \eps_1\,   L_{\bsq, W_i} + \eps_2\, A^+_{\bsq, W_i}}
\rr, 
\label{A.norm.00.4D}
\end{multline}

\begin{multline}
\ll \cA_{\, \bV\, \bW\, \bD}^{\,norm, \, 4D} \rr_{\ll q, \, t\rr} =  
\\
\prod_{n = 0}^\infty  
\prod_{i=1}^2
\ll 
\prod_{\wsq \in V_i} 
\frac{
n\, \eps_3 +\, \eps_4\, -\, \eps_1\, L^+_{\wsq, V_i} +\, \eps_2\, A_{\wsq, V_i}
}{
\ll 
n + 1 \rr       \, \eps_3\, -\, \eps_1\, L^+_{\wsq, V_i} +\, \eps_2\, A_{\wsq, V_i}} 
\, 
\prod_{\bsq \in W_i}
\frac{
n\, \eps_4  +\, \eps_3\, -\, \eps_1\, L_{\bsq, W_i}  +\, \eps_2\, A^+_{\bsq, W_i}
}{ 
\ll 
n + 1 \rr\,\eps_4\, -\, \eps_1\, L_{\bsq, W_i} +\, \eps_2\, A^+_{\bsq, W_i}
}
\rr 
\label{A.4D}
\end{multline}

\subsubsection{$\cB_{\, \bV\, \bW\, \bD}^{\, norm, \, 4D}$}
In the limit $R \rightarrow 0$,  

\begin{multline} 
{\prod_{i, j = 1}^{\infty}} 
\ll 1 - Q\, x^{\, i - 1 - v^\prime_j}\, y^{\, j - w_i} 
\rr^{\checkmark} \biggr\vert_{R \rightarrow 0} 
= \\ 
R^{|V| + |W|}\, 
\prod_{\wsq \in V} \ll - \Delta + \eps_1\, L^+_{\wsq, W} - \eps_2\, A_{\wsq, V}   \rr
\prod_{\bsq \in W} \ll - \Delta - \eps_1\,   L_{\bsq, W} + \eps_2\, A_{\bsq, V}^+ \rr
\label{nakajima.4D.identity}
\end{multline}

\noindent Consequently, 

\begin{multline}
\cB_{\, \bV\, \bW\, \bD}^{\, norm, \, 4D}  =
\ll \cB_{\, \bV\, \bW\, \bD}^{\, norm, \, 4D} \rr_{\ll 0, \, 0\rr}  =
R^{2 \ll |V_1| + |V_2| + |W_1| + |W_2| \rr} 
\\
\prod_{\wsq \in V_1} \ll \Delta_1 - \eps_1\, L_{\wsq, W_1}   + \eps_2\, A_{\wsq, V_1}   \rr
\prod_{\bsq \in W_1} \ll \Delta_1 + \eps_1\, L_{\bsq, V_1}^+ - \eps_2\, A_{\bsq, W_1}^+ \rr
\\
\prod_{\wsq \in V_2} \ll \Delta_2 + \eps_1\, L_{\wsq, W_1}   - \eps_2\, A_{\wsq, V_2}   \rr
\prod_{\bsq \in W_1} \ll \Delta_2 - \eps_1\, L_{\bsq, V_2}^+ + \eps_2\, A_{\bsq, W_1}^+ \rr
\\
\prod_{\wsq \in V_2} \ll \Delta_3 - \eps_1\, L_{\wsq, W_2}   + \eps_2\, A_{\wsq, V_2}   \rr
\prod_{\bsq \in W_2} \ll \Delta_3 + \eps_1\, L_{\bsq, V_2}^+ - \eps_2\, A_{\bsq, W_2}^+ \rr
\\
\prod_{\wsq \in V_1} \ll \Delta_{1 2 3} - \eps_1\, L_{\wsq, W_2}   + \eps_2\, A_{\wsq, V_1}   \rr
\prod_{\bsq \in W_2} \ll \Delta_{1 2 3} + \eps_1\, L_{\bsq, V_1}^+ - \eps_2\, A_{\bsq, W_2}^+ \rr, 
\label{B.norm.00.4D}
\end{multline}

\noindent where $\Delta_{123} = \Delta_1 + \Delta_2 + \Delta_3$

\subsubsection{$\cC_{\, \bV\, \bW\, \bD}^{\, norm, \, 4D}$}
Finally, the 4D limit of $\cC_{\, \bV\, \bW\, \bD}$ is, 

\begin{equation}
\cC^{\, norm, \, 4D}_{\bV\, \bW} = 
\ll \cC^{\, norm, \, 4D}_{\bV\, \bW} \rr_{\ll 0, \, 0\rr} \,
\ll \cC^{\, norm, \, 4D}_{\bV\, \bW} \rr_{\ll q, \, t\rr} \,
\label{C.norm.4D}
\end{equation}

\begin{multline}
\ll \cC^{\, norm, \, 4D}_{\bV\, \bW} \rr_{\ll 0, \, 0\rr} \, = 
R^{- \ll | V_1 | + | V_2 | + | W_1 | + | W_2 | \rr} 
\\
\ll 
\prod_{\wsq \in V_1} 
\frac{1}{ - \Delta_{1 2} - \eps_1 \, L_{\wsq, V_2}^+ + \eps_2\, A_{\wsq, V_1}  }
\rr 
\ll 
\prod_{\wsq \in V_2}
\frac{1}{ - \Delta_{1 2} + \eps_1 \, L_{\wsq, V_1}   - \eps_2\, A_{\wsq, V_2}^+}
\rr 
\\
\ll 
\prod_{\bsq \in W_1}
\frac{1}{- \Delta_{23} - \eps_1\, L_{\bsq, W_2}   + \eps_2\, A_{\bsq, W_1}^+}
\rr 
\ll 
\prod_{\bsq \in W_2}
\frac{1}{- \Delta_{23} + \eps_1\, L_{\bsq, W_1}^+ - \eps_2\, A_{\bsq, W_2}  }
\rr,  
\label{C.norm.00.4D}
\end{multline}

\noindent where 
$\Delta_{12} = \Delta_1 + \Delta_2$, and
$\Delta_{23} = \Delta_2 + \Delta_3$, 

\begin{multline}
\ll \cC^{\, norm, \, 4D}_{\bV\, \bW} \rr_{\ll q, \, t\rr} \, = 
\\ 
\prod_{n = 0}^\infty 
\ll
\prod_{\wsq \in V_1}
\ll 
\frac{\Delta_{1 2} -     n \eps_3 - \eps_4 + \eps_1\, L^+_{\wsq, V_2} - \eps_2\, A_{\wsq, V_1}}{
      \Delta_{1 2} - \ll n + 1 \rr  \eps_3 + \eps_1\, L^+_{\wsq, V_2} - \eps_2\, A_{\wsq, V_1}}
\rr 
\, 
\prod_{\wsq \in V_2}
\ll 
\frac{\Delta_{1 2} -     n \eps_3 - \eps_4 - \eps_1\, L_{\wsq, V_1} + \eps_2\, A^+_{\wsq, V_2}}{
      \Delta_{1 2} - \ll n + 1 \rr  \eps_3 - \eps_1\, L_{\wsq, V_1} + \eps_2\, A^+_{\wsq, V_2}}
\rr 
\right.
\\
\left. 
\prod_{\bsq \in W_1}
\ll
\frac{\Delta_{2 3} -     n \eps_4 - \eps_3 + \eps_1\, L_{\bsq, W_2} - \eps_2\, A^+_{\bsq, W_1}}{
      \Delta_{2 3} - \ll n + 1 \rr  \eps_4 + \eps_1\, L_{\bsq, W_2} - \eps_2\, A^+_{\bsq, W_1}}
\rr\, 
\prod_{\bsq \in W_2}
\ll 
\frac{\Delta_{2 3} -     n \eps_4 - \eps_3 - \eps_1\, L^+_{\bsq, W_1} + \eps_2\, A_{\bsq, W_2}}{
      \Delta_{2 3} - \ll n + 1 \rr  \eps_4 - \eps_1\, L^+_{\bsq, W_1} + \eps_2\, A_{\bsq, W_2}}
\rr
\rr
\label{C.norm.qt.4D}
\end{multline}

\subsubsection{Comparing notation}
The expressions in Equations \ref{A.norm.00.4D}, \ref{B.norm.00.4D} and \ref{C.norm.00.4D} 
in the present work, were computed in Equations 4.3 and 4.4 in \cite{foda.wu.01}, using 
$\cR_{Y_1 Y_2 Y_3} \ll x, y\rr$, and matched in the 4D limit with the Nekrasov partition 
functions, as in \cite{bershtein.foda}.  
The notation used in the present work translate to that used in \cite{foda.wu.01}, 
denoted with {\it \lq old\rq}, as follows,

\begin{multline}
\eps_1 =  \eps_2^{\, old}, \quad
\eps_2 =  \eps_1^{\, old}, \quad
V_i        = V_i^{\, \prime, \, old}, \quad  
W_i        = W_i^{\, \prime, \, old}, \quad
i \in \ll 1, 2\rr, 
\\ 
\Delta_1 = - \, \Delta^{\, old}_1 - \frac12 \, \eps_2^{\, old} + \frac12 \, \eps_1^{\, old}, \quad 
\Delta_2 = - \, \Delta^{\, old}_M + \frac12 \, \eps_2^{\, old} - \frac12 \, \eps_1^{\, old}, \quad
\Delta_3 = - \, \Delta^{\, old}_2 - \frac12 \, \eps_2^{\, old} + \frac12 \, \eps_1^{\, old}
\end{multline} 

\subsubsection{The 4D limit is well-defined}
Putting Equations \ref{A.norm.4D}--\ref{C.norm.4D} together, all factors 
of $R$ cancel out and we obtain a well-defined 4D limit,  

\begin{equation}
\cS_{\, \bV\,\bW\,\bD}^{\,norm, 4D} \ll x, y \rr =
\cA_{\, \bV\, \bW\, \bD}^{\,norm, \, 4D}\, 
\cB_{\, \bV\, \bW\, \bD}^{\,norm, \, 4D}\, 
\cC_{\, \bV\, \bW\, \bD}^{\,norm, \, 4D}
\label{S.norm.4D}
\end{equation}

\subsection{Comments on the structure of $\cS_{\,\bV\, \bW\, \bD}^{\, norm}$ 
in the 4D limit}
\label{pole.structure.4D}

\subsubsection{The $q t$-independent terms}
\label{4D.00.poles}
The $q t$-independent product,  

\begin{equation}
\ll \cS_{\, \bV\, \bW\, \bD}^{\, norm, 4D} \rr_{\ll 0, \, 0\rr} = 
\ll \cA_{\, \bV\, \bW\, \bD}^{\, norm, 4D} \rr_{\ll 0, \, 0 \rr} \, 
\ll \cB_{\, \bV\, \bW\, \bD}^{\, norm, 4D} \rr_{\ll 0, \, 0 \rr} \, 
\ll \cC_{\, \bV\, \bW\, \bD}^{\, norm, 4D} \rr_{\ll 0, \, 0 \rr},
\label{product.00.4D}
\end{equation}

\noindent is the result obtained using $\cR_{Y_1 Y_2 Y_3} \ll x, y \rr$, 
as in Equations 4.2--4.4 in \cite{foda.wu.01}
\footnote{\,
As in the 5D case, in Equations 4.2--4.4 in \cite{foda.wu.01}, slightly 
different notation was used. In particular, 
$\Delta_i$, $i \in \ll 1, M, 2 \rr$ in \cite{foda.wu.01}, become
$\delta_i$, $i \in \ll 1, 2, 3 \rr$ in the present work. 
Other changes are discussed in Footnote \ref{changes}. 
}. 
The first thing to note is that the overall factors of $R$ in Equations 
\ref{A.norm.00.4D}, \ref{B.norm.00.4D} and \ref{C.norm.00.4D}, cancel 
when we compute $\cS_{\, \bV\, \bW\, \bD}^{\, norm}$ as in Equation
\ref{product.00.4D}
Aside from that, $\cA_{\, \bV\, \bW\, \bD}^{\, norm}$, Equation \ref{A.norm.00.4D}, 
has no poles since
$\eps_1 < 0 < \eps_i$, $i \in \ll 2, 3, 4\rr$, 
$A_{\wsq, \, Y} \geq 0$  and  
$L_{\wsq, \, Y} \geq 0$, for $\wsq \in Y$. 
$\cB_{\, \bV\, \bW\, \bD}^{\, norm}$, Equation \ref{B.norm.00.4D}, 
has no poles, and  
$\ll \cC^{\, norm, \, 4D}_{\bV\, \bW} \rr_{\ll 0, \, 0\rr}$, 
Equation \ref{C.norm.00.4D}, has the same poles obtained using 
$\cR_{Y_1 Y_2 Y_3} \ll x, y \rr$
\footnote{\, 
The pole structure of rational expressions as in Equation \ref{C.norm.00.4D} 
were discussed in detail in \cite{bershtein.foda}. 
}.

\subsubsection{The $q t$-dependent terms}
\label{4D.qt.poles}

\begin{equation}
\ll \cS_{\, \bV\, \bW\, \bD}^{\, norm, \, 4D} \rr_{\ll q, \, t \rr} = 
\ll \cA_{\, \bV\, \bW\, \bD}^{\, norm, \, 4D} \rr_{\ll q, \, t \rr} \, 
\ll \cB_{\, \bV\, \bW\, \bD}^{\, norm, \, 4D} \rr_{\ll q, \, t \rr} \, 
\ll \cC_{\, \bV\, \bW\, \bD}^{\, norm, \, 4D} \rr_{\ll q, \, t \rr}, 
\label{product.qt.4D}
\end{equation}

\noindent is the additional $q t$-dependent factor in 
$\cS_{\, \bV\, \bW\, \bD}^{\, norm}$ that we obtain when we use
$\cM_{Y_1 Y_2 Y_3}^{\, q \, t} \ll x, y \rr$, then take the 4D limit. 
$\ll \cA_{\, \bV\, \bW\, \bD}^{\, norm, \, 4D} \rr_{\ll 0, \, 0 \rr}$, 
Equation \ref{A.norm.00.4D}, has no poles for the same reasons as in 5D.
$\ll \cB_{\, \bV\, \bW\, \bD}^{\, norm} \rr_{\ll 0, \, 0 \rr}$, Equation 
\ref{B.norm.00.4D}, has no poles, and  
$\ll \cC^{\, norm, \, 4D}_{\bV\, \bW} \rr_{\ll q, \, t\rr}$, 
Equation \ref{C.norm.qt.4D}, 
has an infinite tower of poles for every pole in 
$\ll \cC^{\, norm, \, 4D}_{\bV\, \bW} \rr_{\ll 0, \, 0\rr}$, 
for generic positive values of $\eps_3$ and $\eps_4$. 
A fraction of these infinite towers cancel for non-generic values of 
$\eps_3$ and $\eps_4$, but not all.

\subsubsection{Remark} At the level of the strip, the additional $q t$-dependent 
terms, containing infinitely-many new poles, appear in overall, multiplicative factors. 
As we show, in Section \ref{9}, when we glue strips to form instanton partition functions, 
these terms are no longer overall multiplicative, but give different weights to the different 
terms summed over to form the instanton partition functions. 

\section{The $U \! \ll 2 \rr$ instanton partition function and its 4D limit}
\label{9}
{\it We glue two strips along non-preferred legs to obtain a 5D $q t$-instanton
partition function, then take its 4D limit.}
\smallskip 

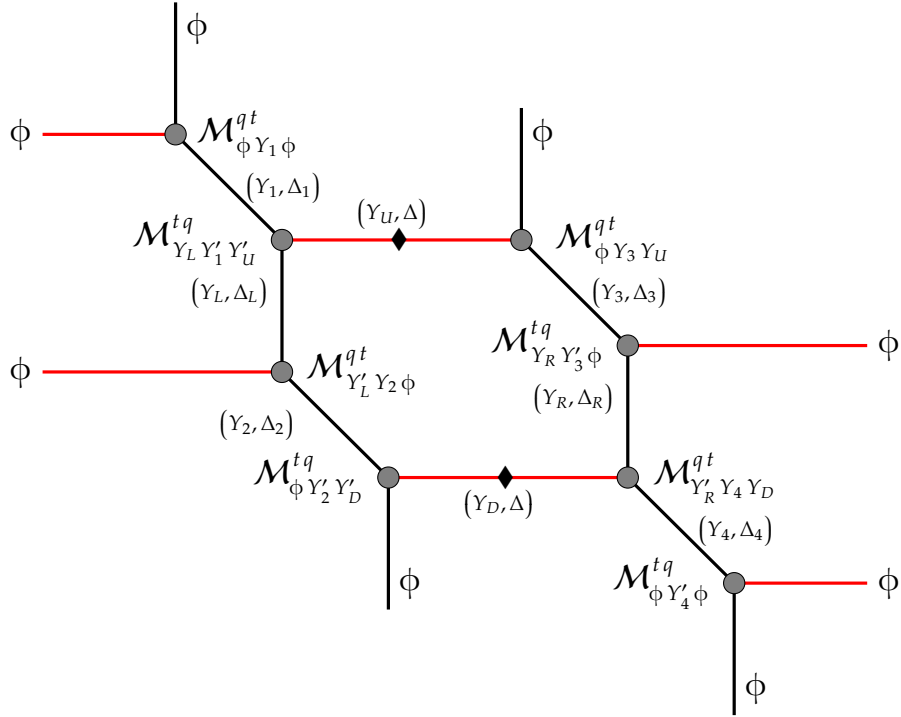
\begin{figure}
\begin{centering}
\begin{tikzpicture}[scale=.7]

\draw [red,   very thick] (01.5,  4.0)--(04.0,  4.0);
\draw [black, very thick] (04.0,  4.0)--(06.0,  2.0);
\draw [black, very thick] (04.0,  6.5)--(04.0,  4.0);
\draw [red,   very thick] (01.5, -0.5)--(06.0, -0.5);
\draw [black, very thick] (06.0,  2.0)--(06.0, -0.5);
\draw [red,   very thick] (06.0,  2.0)--(10.5,  2.0);
\draw [black, very thick] (06.0, -0.5)--(08.0, -2.5);
\draw [red,   very thick] (08.0, -2.5)--(12.5, -2.5);
\draw [black, very thick] (08.0, -2.5)--(08.0, -5.0);
\draw [black, very thick] (10.5,  4.5)--(10.5,  2.0);
\draw [black, very thick] (10.5,  2.0)--(12.5,  0.0);
\draw [black, very thick] (12.5,  0.0)--(12.5, -2.5);
\draw [red,   very thick] (12.5,  0.0)--(17.0,  0.0);
\draw [black, very thick] (12.5, -2.5)--(14.5, -4.5);
\draw [black, very thick] (14.5, -4.5)--(14.5, -7.0);
\draw [red,   very thick] (14.5, -4.5)--(17.0, -4.5);

\draw [fill=black!50] (04.0,  4.0) circle (0.2);
\draw [fill=black!50] (06.0,  2.0) circle (0.2);
\draw [fill=black!50] (06.0, -0.5) circle (0.2);
\draw [fill=black!50] (08.0, -2.5) circle (0.2);
\draw [fill=black!50] (10.5,  2.0) circle (0.2);
\draw [fill=black!50] (12.5,  0.0) circle (0.2);
\draw [fill=black!50] (12.5, -2.5) circle (0.2);
\draw [fill=black!50] (14.5, -4.5) circle (0.2);

\node [left]  at (01.5,  4.0) {$\phiup$};		
\node [left]  at (01.5, -0.5) {$\phiup$};		
\node [right] at (04.0,  6.0) {$\phiup$};
\node [right] at (08.0, -4.5) {$\phiup$};
\node [right] at (10.5,  4.0) {$\phiup$};
\node [right] at (14.5, -6.5) {$\phiup$};
\node [right] at (17.0,  0.0) {$\phiup$};
\node [right] at (17.0, -4.5) {$\phiup$};

\node [left] at (07.0,  3.0){$\scriptstyle{\ll Y_1,\, \Delta_1 \rr}$};
\node [left] at (09.0,  2.5){$\scriptstyle{\ll Y_U,\, \Delta   \rr}$};
\node [left] at (13.5,  1.0){$\scriptstyle{\ll Y_3,\, \Delta_3 \rr}$};
\node [left] at (06.0,  1.0){$\scriptstyle{\ll Y_L,\, \Delta_L \rr}$};
\node [left] at (12.5, -1.0){$\scriptstyle{\ll Y_R,\, \Delta_R \rr}$};
\node [left] at (06.5, -1.5){$\scriptstyle{\ll Y_2,\, \Delta_2 \rr}$};
\node [left] at (11.0, -3.0){$\scriptstyle{\ll Y_D,\, \Delta   \rr}$};
\node [left] at (15.5, -3.5){$\scriptstyle{\ll Y_4,\, \Delta_4 \rr}$};	

\node [left] at (06.50,  4.00){$\cM_{\, \phiup    \, Y_1       \, \phiup    }^{\, q \, t}$};
\node [left] at (13.50,  2.00){$\cM_{\, \phiup    \, Y_3       \, Y_U       }^{\, q \, t}$};
\node [left] at (05.75,  2.00){$\cM_{\, Y_L       \, Y_1^\prime\, Y_U^\prime}^{\, t\, q}$};
\node [left] at (12.25,  0.00){$\cM_{\, Y_R       \, Y_3^\prime\,\phiup     }^{\, t\, q}$};
\node [left] at (08.75, -0.50){$\cM_{\, Y_L^\prime\, Y_2       \, \phiup    }^{\, q \, t}$};
\node [left] at (07.75, -2.50){$\cM_{\phiup       \, Y_2^\prime\, Y_D^\prime}^{\, t\, q}$};
\node [left] at (15.50, -2.50){$\cM_{Y_R^\prime   \, Y_4       \, Y_D       }^{\, q \, t}$};
\node [left] at (14.25, -4.50){$\cM_{\phiup       \, Y_4^\prime\, \phiup    }^{\, t\, q}$};

\node at (08.2,  2.0) {$\blacklozenge$};
\node at (10.2, -2.5) {$\blacklozenge$};

\end{tikzpicture} 
\par\end{centering}
\caption{\it 
The web diagram of the 5D $U \! \ll 2 \rr$ instanton partition function, 
and the 2D $q t$-conformal block}
\label{4.point}
\end{figure}

\subsection{Two strips}
Following the conventions used Figure \ref{4.point}, 
the left-strip has a partition pair $\ll \emptyset, \,\emptyset  \rr$ on the left,   
$\ll Y_U^\prime,\,Y_D^\prime \rr$ on the right, and contributes, 

\begin{equation}
\cS_{\,\pmb{\emptyset}\,\bY\,\pmb{\Delta_{\, L}}}^{\, norm} 
= 
\cA^{\, norm}_{\pmb{\emptyset}\, \bY\, \pmb{\Delta_{\, L}}}\, 
\cB^{\, norm}_{\pmb{\emptyset}\, \bY\, \pmb{\Delta_{\, L}}}\, 
\cC^{\, norm}_{\pmb{\emptyset}\, \bY\, \pmb{\Delta_{\, L}}}, 
\end{equation}

\noindent where $\pmb{\Delta_{\, L}} = \ll \Delta_1,\, \Delta_L,\, \Delta_2 \rr$. 
The right-strip has a partition pair $\ll Y_U,\, Y_D \rr$ on the left,   
$\ll \emptyset,\, \emptyset \rr$ on the right, and contributes,
 
\begin{equation}
\cS_{\, \bY\,\pmb{\emptyset}\, \pmb{\Delta_{\, R}}}^{\, norm} = 
\cA^{\, norm}_{\bY\,\pmb{\emptyset}\, \pmb{\Delta_{\, R}}}\, 
\cB^{\, norm}_{\bY\,\pmb{\emptyset}\, \pmb{\Delta_{\, R}}}\, 
\cC^{\, norm}_{\bY\,\pmb{\emptyset}\, \pmb{\Delta_{\, R}}}, 
\end{equation}

\noindent where $\pmb{\Delta_{\, R}} = \ll \Delta_3,\, \Delta_R,\, \Delta_4\rr$
	
\subsection{The $U \! \ll 2 \rr$ $q t$-instanton partition function}
Gluing the left-strip and the right-strip in Figure \ref{4.point} along 
non-preferred legs that share a common partition pair $\ll Y_U ,\, Y_D\rr $, 
weighted by $\ll - Q_C \rr^{|Y_U|+|Y_D|}$, where $Q_C$ is an exponentiated K\"ahler 
parameter, we obtain the $U\! \ll 2 \rr$ $q t$-instanton partition function, 

\begin{equation}
\cW_{\, \pmb{\Delta_{\, L}}, \, \pmb{\Delta_{\, R}}} 
= 
\sum_{ Y_U,\, Y_D}
\ll - Q_C \rr^{ | Y_U | + | Y_D |}
\cS_{\, \pmb{\emptyset}\, \bY\,\pmb{\Delta_{\, L}}}^{\, norm}
\cS_{\, \bY \, \pmb{\emptyset}\,\pmb{\Delta_{\, R}}}^{\, norm}
\label{4.point.function}
\end{equation}

\noindent It is useful to split the summand 
$\cS_{\, \pmb{\emptyset}\, \bY            \,\pmb{\Delta_{\, L}}}^{\, norm}
 \cS_{\, \bY            \, \pmb{\emptyset}\,\pmb{\Delta_{\, R}}}^{\, norm}$, 
into a factor
$\ll
\cS_{\, \pmb{\emptyset}\, \bY            \,\pmb{\Delta_{\, L}}}^{\, norm}
\cS_{\, \bY            \, \pmb{\emptyset}\,\pmb{\Delta_{\, R}}}^{\, norm}
\rr_{\ll 0, \, 0 \rr}
$,
that does not depend on $\ll q, \, t\rr$, and a factor 
$\ll
\cS_{\, \pmb{\emptyset}\, \bY            \,\pmb{\Delta_{\, L}}}^{\, norm}
\cS_{\, \bY            \, \pmb{\emptyset}\,\pmb{\Delta_{\, R}}}^{\, norm}
\rr_{\ll q, \, t\rr}$,
that depends on $\ll q, \, t\rr$, and $\rightarrow 1$ 
in the limit $q \rightarrow t$, 

\begin{equation}
\ll
\cS_{\, \pmb{\emptyset}\, \bY            \,\pmb{\Delta_{\, L}}}^{\, norm}
\cS_{\, \bY            \, \pmb{\emptyset}\,\pmb{\Delta_{\, R}}}^{\, norm}
\rr_{\ll 0, \, 0\rr} 
=
\prod_{\begin{subarray}{c} \wsq \in Y_U \\ \bsq \in Y_D \end{subarray}}
\frac{
\ll 
\begin{array}{l}
\ll 1 - Q_1  \, x^{\,  L^+_{\wsq, \emptyset}}\, y^{\,  A^+_{\wsq, Y_U}} \rr
\ll 1 - Q_3  \, x^{\, -  L_{\wsq, \emptyset}}\, y^{\, -  A_{\wsq, Y_U}} \rr
\\
\ll 1 - Q_L  \, x^{\, -L^+_{\wsq, \emptyset}}\, y^{\, -A^+_{\wsq, Y_U}} \rr
\ll 1 - Q_R  \, x^{\,    L_{\bsq, \emptyset}}\, y^{\,    A_{\bsq, Y_D}} \rr
\\
\ll 1 - Q_2  \, x^{\,  L^+_{\bsq, \emptyset}}\, y^{\,  A^+_{\bsq, Y_D}} \rr
\ll 1 - Q_4  \, x^{\, -  L_{\bsq, \emptyset}}\, y^{\, -  A_{\bsq, Y_D}} \rr
\\
\ll 1 - Q_{1 L 2} \, x^{\,  L^+_{\bsq, \emptyset}}\, y^{\,  A^+_{\bsq, Y_D}} \rr
\ll 1 - Q_{3 R 4} \, x^{\, -  L_{\wsq, \emptyset}}\, y^{\, -  A_{\wsq, Y_U}} \rr
\end{array}
\rr
}{ 
\ll
\begin{array}{l}
\ll 1 - x^{\,   L_{\wsq,\, Y_U}}\, y^{ A^+_{\wsq,\, Y_U}} \rr
\ll 1 - x^{\, L_{\wsq,\, Y_U}^+}\, y^{   A_{\wsq,\, Y_U}} \rr
\\
\ll 1 - x^{\,   L_{\bsq,\, Y_D}}\, y^{ A^+_{\bsq,\, Y_D}} \rr
\ll 1 - x^{\, L_{\bsq,\, Y_D}^+}\, y^{   A_{\bsq,\, Y_D}} \rr
\\
\ll 1 - Q_{\, 2 L}\, x^{     L_{\bsq,\, Y_U}}\, y^{   A^+_{\bsq,\, Y_D}} \rr
\ll 1 - Q_{\, 3 R}\, x^{   L^+_{\bsq,\, Y_U}}\, y^{     A_{\bsq,\, Y_D}} \rr
\\
\ll 1 - Q_{\, 2 L}\, x^{ - L^+_{\wsq,\, Y_D}}\, y^{ -   A_{\wsq,\, Y_U}} \rr
\ll 1 - Q_{\, 3 R}\, x^{ -   L_{\wsq,\, Y_D}}\, y^{ - A^+_{\wsq,\, Y_U}} \rr
\end{array}
\rr
}
\label{4.point.5D.00}
\end{equation}

\begin{equation} 
\ll
\cS_{\, \pmb{\emptyset}\, \bY\,\pmb{\Delta_{\, L}}}^{\, norm}
\cS_{\, \bY \, \pmb{\emptyset}\,\pmb{\Delta_{\, R}}}^{\, norm}
\rr_{\ll q, \, t\rr} 
=
\prod_{n \, = \, 0}^\infty
\prod_{\begin{subarray}{c} \wsq \in Y_U \\ \bsq \in Y_D \end{subarray}}\, 
\frac{
\ll 
\begin{array}{r} 
\begin{array}{l}  
\ll 1 - x^{\, L_{\wsq,\, Y_U}^+}\, y^{\, A_{\wsq,\, Y_U}}\, q^{\, n}\, t \rr
\ll 1 - x^{\, A_{\wsq,\, Y_U}^+}\, y^{\, L_{\wsq,\, Y_U}}\, t^{\, n}\, q \rr
\\
\ll 1 - x^{\, L_{\bsq,\, Y_D}^+}\, y^{\, A_{\bsq,\, Y_D}}\, q^{\, n}\, t \rr
\ll 1 - x^{\, A_{\bsq,\, Y_D}^+}\, y^{\, L_{\bsq,\, Y_D}}\, t^{\, n}\, q \rr
\\
\ll 1 - Q_{\, 2 L}\, x^{ - L^+_{\wsq,\, Y_D}}\, y^{ -   A_{\wsq,\, Y_U}}\, t^{\, n}\, q \rr
\ll 1 - Q_{\, 3 R}\, x^{ -   L_{\wsq,\, Y_D}}\, y^{ - A^+_{\wsq,\, Y_U}}\, q^{\, n}\, t \rr
\end{array}
\\
\begin{array}{r}  
\ll 1 - Q_{\, 2 L}\, x^{     L_{\bsq,\, Y_U}}\, y^{   A^+_{\bsq,\, Y_D}}\, t^{\, n}\, q \rr
\ll 1 - Q_{\, 3 R}\, x^{   L^+_{\bsq,\, Y_U}}\, y^{     A_{\bsq,\, Y_D}}\, q^{\, n}\, t \rr
\end{array}
\end{array} 
\rr
}{ 
\ll
\begin{array}{r}  
\begin{array}{l}
\ll 1 - x^{\, L_{\wsq,\, Y_U}^+}\, y^{   A_{\wsq,\, Y_U}}\, q^{n+1}  \rr
\ll 1 - x^{\, L_{\wsq,\, Y_U}  }\, y^{ A^+_{\wsq,\, Y_U}}\, t^{n+1}  \rr
\\
\ll 1 - x^{\, L_{\bsq,\, Y_D}^+}\, y^{   A_{\bsq,\, Y_D}}\, q^{n+1}  \rr
\ll 1 - x^{\, L_{\bsq,\, Y_D}  }\, y^{ A^+_{\bsq,\, Y_D}}\, t^{n+1}  \rr
\\
\ll 1 - Q_{\, 2 L}\, x^{     L_{\bsq,\, Y_U}}\, y^{   A^+_{\bsq,\, Y_D}}\, t^{n+1}  \rr
\ll 1 - Q_{\, 3 R}\, x^{   L^+_{\bsq,\, Y_U}}\, y^{     A_{\bsq,\, Y_D}}\, q^{n+1}  \rr
\end{array}  
\\
\begin{array}{r} 
\ll 1 - Q_{\, 2 L}\, x^{ - L^+_{\wsq,\, Y_D}}\, y^{ -   A_{\wsq,\, Y_U}}\, t^{n+1}  \rr
\ll 1 - Q_{\, 3 R}\, x^{ -   L_{\wsq,\, Y_D}}\, y^{ - A^+_{\wsq,\, Y_U}}\, q^{n+1} \rr
\end{array}
\end{array} 
\rr
}
\label{4.point.5D.qt}
\end{equation}

\subsection{The 4D limit of the $U \! \ll 2 \rr$ instanton partition function}
To take the 4D limit of the instanton partition function, we expand the variables, 
	
\begin{equation}
x=e^{ R\, \eps_1},\, 
y=e^{-R\, \eps_2},\,
q=e^{ R\, \eps_3},\,
t=e^{ R\, \eps_4},\,
Q_i = e^{R\, \Delta_i},\, i = \ll 1,\, 2,\, 3,\, 4,\, L,\, R,\, U,\, D \rr, 
\end{equation}

\noindent and readily find, 

\begin{equation}
\ll 
\cS_{\, \pmb{\emptyset}\, \bY\,\pmb{\Delta_{\, L}}}^{\, norm, \, 4D}
\cS_{\, \bY \, \pmb{\emptyset}\,\pmb{\Delta_{\, R}}}^{\, norm, \, 4D}
\rr_{\ll 0, \, 0 \rr} 
= 
\prod_{\begin{subarray}{c} \wsq \in Y_U \\ \bsq \in Y_D \end{subarray}}
\frac{
\ll
\begin{array}{c}
\ll \Delta_1       + \eps_1\, L_{\wsq, \,\emptyset}^+ - \eps_2\, A_{\wsq, \, Y_U}^+ \rr
\ll \Delta_3       - \eps_1\, L_{\wsq, \,\emptyset}   + \eps_2\, A_{\wsq, \,Y_U}    \rr
\\
\ll \Delta_L       - \eps_1\, L_{\wsq, \,\emptyset}^+ + \eps_2\, A_{\wsq, \, Y_U}^+ \rr
\ll \Delta_R       + \eps_1\, L_{\bsq, \,\emptyset}   - \eps_2\, A_{\bsq, \,Y_D}    \rr
\\
\ll \Delta_2       + \eps_1\, L_{\bsq, \,\emptyset}^+ - \eps_2\, A_{\bsq, \, Y_D}^+ \rr
\ll \Delta_4       - \eps_1\, L_{\bsq, \,\emptyset}   + \eps_2\, A_{\bsq, \,Y_D}    \rr
\\
\ll \Delta_{1 L 2} - \eps_1\, L_{\bsq, \,\emptyset}^+ + \eps_2\, A_{\bsq, \, Y_D}^+ \rr
\ll \Delta_{3 R 4} + \eps_1\, L_{\wsq, \,\emptyset}   - \eps_2\, A_{\wsq, \,Y_U}    \rr 
\end{array}
\rr
}{  
\ll
\begin{array}{c} 
\ll \eps_1\, L^+_{\wsq,\, Y_U} - \eps_2\,   A_{\wsq,\, Y_U} \rr
\ll \eps_1\,   L_{\wsq,\, Y_U} - \eps_2\, A^+_{\wsq,\, Y_U} \rr
\\
\ll \eps_1\, L^+_{\bsq,\, Y_D} - \eps_2\,   A_{\bsq,\, Y_D} \rr
\ll \eps_1\,   L_{\bsq,\, Y_D} - \eps_2\, A^+_{\bsq,\, Y_D} \rr
\\
\ll \Delta_{2 L} - \eps_1\, L^+_{\wsq,\, Y_D}   + \eps_2\,   A_{\wsq,\, Y_U} \rr
\ll \Delta_{3 R} - \eps_1\,   L_{\wsq,\, Y_D}   + \eps_2\, A^+_{\wsq,\, Y_U} \rr
\\
\ll \Delta_{2 L} + \eps_1\,   L_{\bsq,\, Y_U} - \eps_2\, A^+_{\bsq,\, Y_D} \rr
\ll \Delta_{3 R} + \eps_1\, L^+_{\bsq,\, Y_U} - \eps_2\,   A_{\bsq,\, Y_D} \rr
\end{array}
\rr
},  
\label{4.point.4D.00}
\end{equation}

\begin{multline}
\ll 
\cS_{\, \pmb{\emptyset}\, \bY\,\pmb{\Delta_{\, L}}}^{\, norm, \, 4D}
\cS_{\, \bY \, \pmb{\emptyset}\,\pmb{\Delta_{\, R}}}^{\, norm, \, 4D}
\rr_{\, \ll q, \, t\rr} 
= 
\\
\prod_{n \, = \, 0}^\infty
\prod_{\begin{subarray}{c} \wsq \in Y_U \\ \bsq \in Y_D \end{subarray}}
\frac{
\ll\begin{array}{c}
\ll  n        \eps_4 + \eps_3 - \eps_1\,   L_{\wsq, \,Y_U} + \eps_2\, A^+_{\wsq, \,Y_U} \rr
\ll  n        \eps_3 + \eps_4 - \eps_1\, L^+_{\wsq, \,Y_U} + \eps_2\,   A_{\wsq, \,Y_U} \rr
\\
\ll  n        \eps_4 + \eps_3 - \eps_1\,   L_{\bsq, \,Y_D} + \eps_2\, A^+_{\bsq, \,Y_D} \rr
\ll  n        \eps_3 + \eps_4 - \eps_1\, L^+_{\bsq, \,Y_D} + \eps_2\,   A_{\bsq, \,Y_D} \rr
\\
\ll \Delta_{2 L}  - n \eps_4 - \eps_3 - \eps_1\, L^+_{\wsq, \,Y_D}  - \eps_2\,   A_{\wsq, \,Y_U} \rr
\ll \Delta_{3 R}  - n \eps_3 - \eps_4 - \eps_1\,   L_{\wsq, \,Y_D}  + \eps_2\, A^+_{\wsq, \,Y_U} \rr
\\
\ll \Delta_{2 L}  - n \eps_4 - \eps_3 + \eps_1\,   L_{\bsq, \,Y_U}  - \eps_2\, A_{\bsq, \,Y_D}^+ \rr
\ll \Delta_{3 R}  - n \eps_3 - \eps_4 + \eps_1\, L^+_{\bsq, \,Y_U}  - \eps_2\, A_{\bsq, \,Y_D}   \rr
\end{array}
\rr
}{  
\ll
\begin{array}{c}
\ll  \ll n + 1 \rr \eps_4 - \eps_1\, L_{\wsq, \,Y_U}   + \eps_2\, A^+_{\wsq, \,Y_U}       \rr
\ll  \ll n + 1 \rr \eps_3 - \eps_1\, L_{\wsq, \,Y_U}^+ + \eps_2\,   A_{\wsq, \,Y_U}       \rr
\\
\ll  \ll n + 1 \rr \eps_4 - \eps_1\, L_{\bsq, \,Y_D}   + \eps_2\, A^+_{\bsq, \,Y_D}       \rr
\ll  \ll n + 1 \rr \eps_3 - \eps_1\, L_{\bsq, \,Y_D}^+ + \eps_2\,   A_{\bsq, \,Y_D}       \rr
\\
\ll \Delta_{2 L} - \ll n + 1\rr \, \eps_4 - \eps_1\, L^+_{\wsq, \,Y_D} + \eps_2\,   A_{\wsq, \,Y_U} \rr
\ll \Delta_{3 R} - \ll n + 1\rr \, \eps_3 - \eps_1\,   L_{\wsq, \,Y_D} = \eps_2\, A^+_{\wsq, \,Y_U} \rr
\\
\ll \Delta_{2 L} - \ll n + 1\rr \, \eps_4 + \eps_1\, L_{\bsq, \,Y_U}   - \eps_2\, A_{\bsq, \,Y_D}^+ \rr
\ll \Delta_{3 R} - \ll n + 1\rr \, \eps_3 + \eps_1\, L_{\bsq, \,Y_U}^+ - \eps_2\, A_{\bsq, \,Y_D}   \rr
\end{array}
\rr
} 
\label{4.point.4D.qt}
\end{multline}

\subsection{Comments on the structure of 
$\cW_{\, \pmb{\Delta_{\, L}}, \, \pmb{\Delta_{\, R}}}$}
Equation \ref{4.point.4D.00} is identical to that obtained using the 4D Nekrasov 
instanton partition functions, such as in \cite{bershtein.foda}.
Equations \ref{4.point.5D.qt} and \ref{4.point.4D.qt} show that, for generic
values of $\ll q, t\rr$, the $q t$-dependent contributions to instanton 
partition function bring in new poles, that are also present in the 4D limit.
In Section \ref{10}, we consider these poles in the context 
of the $\cN = 2^\star$ instanton partition function, or equivalently the 1-point 
function on a torus, in a 2D Virasoro conformal field theory. 

\section{The $\cN = 2^\star$ instanton partition function and its 4D limit}
\label{10}
{\it 
We compute the $\cN = 2^\star $ instanton partition function, or equivalently 
the 1-point $q t$-conformal block of a Virasoro conformal field theory.
}
\smallskip 

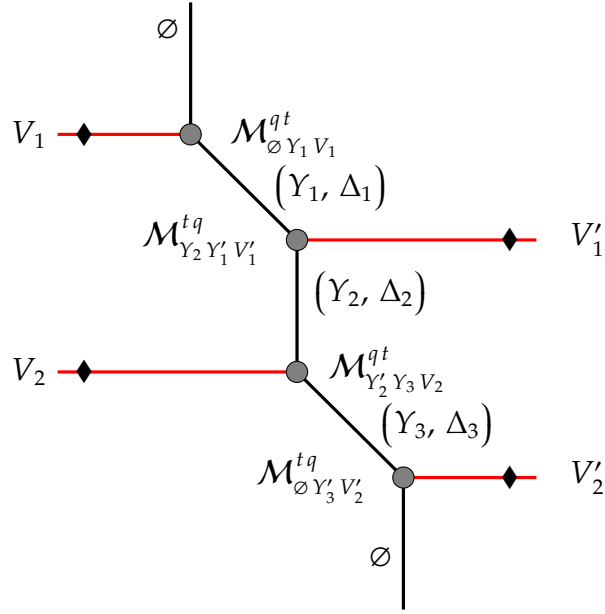
\begin{figure}
\begin{centering}
\begin{tikzpicture}[scale=.7]
\draw [red,   very thick] (1.5, 4.0)--(4.0,   4.0);
\draw [black, very thick] (4.0, 4.0)--(6.0,   2.0);
\draw [black, very thick] (4.0, 6.5)--(4.0,   4.0);
\draw [red,   very thick] (1.5, -.5)--(6.0,   -.5);
\draw [black, very thick] (6.0, 2.0)--(6.0,   -.5);
\draw [red,   very thick] (6.0, 2.0)--(10.5,  2.0);
\draw [black, very thick] (6.0, -.5)--(8.0,  -2.5);
\draw [red,   very thick] (8.0,-2.5)--(10.5, -2.5);
\draw [black, very thick] (8.0,-2.5)--(8.0,  -5.0);

\node [left] at ( 4.0,  6.0) {$\emptyset$};
\node [left] at ( 1.5,  4.0) {$V_1$};		
\node [left] at (12.0,  2.0) {$V_1^\prime$};
\node [left] at ( 1.5, -0.5) {$V_2$};		
\node [left] at (12.0, -2.5) {$V_2^\prime$};
\node [left] at ( 8.0, -4.0) {$\emptyset$};
		
\draw [fill=black!50] (4.0, 4.0) circle (0.2);
\draw [fill=black!50] (6.0, 2.0) circle (0.2);
\draw [fill=black!50] (6.0, -.5) circle (0.2);
\draw [fill=black!50] (8.0,-2.5) circle (0.2);

\node at ( 2.0, 4.0) {$\blacklozenge$};
\node at (10.0, 2.0) {$\blacklozenge$};
\node at ( 2.0,-0.5) {$\blacklozenge$};
\node at (10.0,-2.5) {$\blacklozenge$};
		
\node [left] at ( 8.00, 3.0){$\ll Y_1,\, \Delta_1 \rr$};
\node [left] at ( 8.75, 1.0){$\ll Y_2,\, \Delta_2 \rr$};
\node [left] at (10.00,-1.5){$\ll Y_3,\, \Delta_3 \rr$};
		
\node [left] at (7.0,  4.0){$\cM_{\emptyset\,  Y_1\,        V_1       }^{\,q \, t}$};
\node [left] at (5.5,  2.0){$\cM_{Y_2\,        Y_1^\prime\, V_1^\prime}^{\,t\,q}$};
\node [left] at (9.0, -0.5){$\cM_{Y_2^\prime\, Y_3\,        V_2       }^{\,q \, t}$};
\node [left] at (7.5, -2.5){$\cM_{\emptyset\,  Y_3^\prime\, V_2^\prime}^{\,t\,q}$};

\end{tikzpicture} 
\par\end{centering}
\caption{\it 
The web diagram that corresponds to the $\cN  = 2^\star$ $U \! \ll 2 \rr$ instanton 
partition function, or equivalently a 2D 1-point conformal block on a torus. 
}
\label{toric.web.diagram} 
\end{figure}

To compute the 1-point conformal block on the torus, we start from the strip partition 
function $\cS_{\, \bV\, \bW\, \bD}^{\, norm}$, set $\bV = \bW$, introduce a gluing 
exponentiated K\"ahler parameter $Q$ that contributes a factor $\ll - Q \rr^{|V_1|+|V_2|}$, 
and sum over $\bV = \ll V_1, V_2\rr$,
	
\begin{equation}
W^{\, \cN = 2^\star}_{\bD}
= 
\sum_{\bV}
\ll - Q \rr^{ | V_1 | + | V_2 |}
\cS_{\bV\, \bV\, \bD}
=
\sum_{\bV}
\ll - Q \rr^{ | V_1 | + | V_2 |}
\ll \cS_{\bV\, \bV\, \bD} \rr_{\ll 0, \, 0\rr} \, 
\ll \cS_{\bV\, \bV\, \bD} \rr_{\ll q, \, t\rr}, 
\label{1.point}
\end{equation}

\begin{multline}  
\ll \cS_{\bV\, \bV\, \bD} \rr_{\ll 0, \, 0\rr} = 
\\
\prod_{\wsq \in V_1} 
\frac{
\ll 1 - Q_1         \, x^{\, -  L_{\wsq, \, V_1}}\, y^{\, -  A_{\wsq, \, V_1}} \rr
\ll 1 - Q_1         \, x^{\,  L^+_{\wsq, \, V_1}}\, y^{\,  A^+_{\wsq, \, V_1}} \rr
\ll 1 - Q_2         \, x^{\, -L^+_{\wsq, \, V_2}}\, y^{\, -A^+_{\wsq, \, V_1}} \rr
\ll 1 - Q_{\, 1 2 3}\, x^{\, -  L_{\wsq, \, V_2}}\, y^{\, -  A_{\wsq, \, V_1}} \rr 
}{
\ll 1 -              x^{\,   L^+_{\wsq, \, V_1}}\, y^{\,     A_{\wsq, \, V_1}}\rr
\ll 1 -              x^{\,     L_{\wsq, \, V_1}}\, y^{\,   A^+_{\wsq, \, V_1}}\rr
\ll 1 - Q_{\, 1 2}\, x^{\, -    L_{\wsq,\, V_2}}\, y^{\, - A^+_{\wsq, \, V_1}}\rr
\ll 1 - Q_{\, 2 3}\, x^{\, -  L^+_{\wsq,\, V_2}}\, y^{\, -   A_{\wsq, \, V_1}}\rr	
}
\\
\prod_{\bsq \in V_2} 
\frac{
\ll 1- Q_3         \, x^{\, - L_{\bsq, V_2}}\, y^{\, - A_{\bsq, V_2}} \rr 
\ll 1- Q_3         \, x^{\, L^+_{\bsq, V_2}}\, y^{\, A^+_{\bsq, V_2}} \rr 
\ll 1- Q_2         \, x^{\,   L_{\bsq, V_1}}\, y^{\,   A_{\bsq, V_2}} \rr
\ll 1- Q_{\, 1 2 3}\, x^{\, L^+_{\bsq, V_1}}\, y^{\, A^+_{\bsq, V_2}} \rr
}{
\ll 1 -              x^{\,L^+_{\bsq, \, V_2}}\, y^{\,   A_{\bsq, \, V_2}}\rr
\ll 1 -              x^{\,  L_{\bsq, \, V_2}}\, y^{\, A^+_{\bsq, \, V_2}}\rr
\ll 1 - Q_{\, 1 2}\, x^{\, L^+_{\bsq,\, V_1}}\, y^{\,    A_{\bsq,\, V_2}}\rr
\ll 1 - Q_{\, 2 3}\, x^{\,   L_{\bsq,\, V_1}}\, y^{\,  A^+_{\bsq,\, V_2}}\rr
}, 
\label{1.point.00.5D}
\end{multline}

\begin{multline}  
\ll \cS_{\bV\, \bV\, \bD} \rr_{\ll q, \, t\rr} = 
\prod_{n = 1}^\infty 
\prod_{i = 1}^2
\prod_{\wsq \in V_i} 
\frac{ 
\ll 1 - x^{\,L^+_{\wsq, \, V_i}}\, y^{\,   A_{\wsq, \, V_i}}\, q^{\, n - 1}\, t  \rr
\ll 1 - x^{\,  L_{\wsq, \, V_i}}\, y^{\, A^+_{\wsq, \, V_i}}\, t^{\, n - 1}\, q  \rr
}{ 
\ll 1 - x^{\,L^+_{\wsq, \, V_i}}\, y^{\,   A_{\wsq, \, V_i}}\, q^{\, n \phantom{-1}}\, \pt \rr
\ll 1 - x^{\,  L_{\wsq, \, V_i}}\, y^{\, A^+_{\wsq, \, V_i}}\, t^{\, n \phantom{-1}}\, \pq \rr 
}
\\
\prod_{\wsq \in V_1}
\frac{ 
\ll 1 - Q_{\, 1 2}\, x^{- L_{\wsq,\, V_2}}\, y^{- A^+_{\wsq,\, V_1}}\, q^{\, n-1}\, t \rr
\ll 1 - Q_{\, 2 3}\, x^{- L^+_{\wsq,\, V_2}}\, y^{- A_{\wsq,\, V_1}}\, t^{\, n-1}\, q \rr
}{ 
\ll 1 - Q_{\, 1 2}\, x^{- L_{\wsq,\, V_2}}\, y^{- A^+_{\wsq,\, V_1}}\, q^{\, n \phantom{-1}}\, \pt \rr
\ll 1 - Q_{\, 2 3}\, x^{- L^+_{\wsq,\, V_2}}\, y^{- A_{\wsq,\, V_1}}\, t^{\, n \phantom{-1}}\, \pq \rr	
}
\\
\prod_{\bsq \in V_2}
\frac{ 
\ll 1 - Q_{\, 1 2}\, x^{\, L^+_{\bsq,\, V_1}}\, y^{\, A_{\bsq,\, V_2}}\, q^{\, n-1}\, t \rr
\ll 1 - Q_{\, 2 3}\, x^{\, L_{\bsq,\, V_1}}\, y^{\, A^+_{\bsq,\, V_2}}\, t^{\, n-1}\, q \rr
}
{ 
\ll 1 - Q_{\, 1 2}\, x^{\, L^+_{\bsq,\, V_1}}\, y^{\, A_{\bsq,\, V_2}}\, q^{\, n\phantom{-1}}\, \pt \rr
\ll 1 - Q_{\, 2 3}\, x^{\, L_{\bsq,\, V_1}}\, y^{\, A^+_{\bsq,\, V_2}}\, t^{\, n\phantom{-1}}\, \pq \rr
}
\label{1.point.qt.5D}
\end{multline}

\subsection{The 4D limit of $W^{\, \cN = 2^\star}_{\bD}$}
Setting, 

\begin{equation}
x=e^{ R\, \eps_1},\, 
y=e^{-R\, \eps_2},\,
q=e^{-R\, \eps_3},\,
t=e^{-R\, \eps_4},\,
Q_i = e^{\, R\, \Delta_i},\, i = \ll 1,\, 2,\, 3 \rr,
\end{equation}
	
\noindent where $R$ is the circumference of the $M$-theory circle, 
the 4D limit of Equation \ref{1.point} is,

\begin{equation}
    \cS^{\, 4D}_{\bV\, \bV\, \bD}  = 
\ll \cS^{\, 4D}_{\bV\, \bV\, \bD} \rr_{\ll 0, \, 0\rr} 
\ll \cS^{\, 4D}_{\bV\, \bV\, \bD} \rr_{\ll q, \, t\rr}, 
\end{equation}
	
\begin{equation}
\ll \cS^{\, 4D}_{\bV\, \bV\, \bD} \rr_{\ll 0, \, 0\rr} = 
\prod_{\begin{subarray}{c} \wsq \in V_1 \\ \bsq \in V_2 \end{subarray}}
\frac{
\ll 
\begin{array}{c}
\ll \Delta_1     - \eps_1\, L_{\wsq,\,V_1}   + \eps_2\, A_{\wsq,\,V_1}   \rr
\ll \Delta_1     + \eps_1\, L_{\wsq,\,V_1}^+ - \eps_2\, A_{\wsq,\,V_1}^+ \rr
\\
\ll \Delta_2     + \eps_1\, L_{\bsq,\,V_1}   - \eps_2\, A_{\bsq,\,V_2}   \rr
\ll \Delta_2     - \eps_1\, L_{\wsq,\,V_2}^+ + \eps_2\, A_{\wsq,\,V_1}^+ \rr
\\
\ll \Delta_3     - \eps_1\, L_{\bsq,\,V_2}   + \eps_2\, A_{\bsq,\,V_2}   \rr
\ll \Delta_3     + \eps_1\, L_{\bsq,\,V_2}^+ - \eps_2\, A_{\bsq,\,V_2}^+ \rr
\\
\ll \Delta_{1 2 3} - \eps_1\, L_{\wsq,\,V_2}   + \eps_2\, A_{\wsq,\,V_1}   \rr
\ll \Delta_{1 2 3} + \eps_1\, L_{\bsq,\,V_1}^+ - \eps_2\, A_{\bsq,\,V_2}^+ \rr
\end{array}\rr
}{  
\ll\begin{array}{c}
\ll \eps_1\, L^+_{\wsq,\, V_1} - \eps_2\,   A_{\wsq,\, V_1} \rr
\ll \eps_1\,   L_{\wsq,\, V_1} - \eps_2\, A^+_{\wsq,\, V_1} \rr
\\
\ll \eps_1\, L^+_{\bsq,\, V_2} - \eps_2\,   A_{\bsq,\, V_2} \rr
\ll \eps_1\,   L_{\bsq,\, V_2} - \eps_2\, A^+_{\bsq,\, V_2} \rr
\\
\ll \Delta_{1 2} - \eps_1\,   L_{\wsq,\, V_2} + \eps_2 \, A^+_{\wsq,\, V_1} \rr
\ll \Delta_{2 3} - \eps_1\, L^+_{\wsq,\, V_2} + \eps_2 \,   A_{\wsq,\, V_1} \rr
\\
\ll \Delta_{1 2} + \eps_1\, L^+_{\bsq,\, V_1}  - \eps_2   A_{\bsq,\, V_2} \rr
\ll \Delta_{2 3} + \eps_1\, L_{\bsq,\, V_1}    - \eps_2 A^+_{\bsq,\, V_2} \rr
\end{array}
\rr
}, 
\label{1.point.00.4D}
\end{equation}  
	
\begin{multline}
\ll \cS^{\, 4D}_{\bV\, \bV\, \bD} \rr_{\ll q, \, t\rr} = 
\\
\prod_{n=1}^\infty
\prod_{\begin{subarray}{c} \wsq \in V_1  \\ \bsq \in V_2 \end{subarray}}
\frac{
\ll
\begin{array}{c}
\ll  n \eps_3 + \eps_4 -          \eps_1\, L_{\wsq, \,V_1}^+ + \eps_2\, A_{\wsq, \,V_1} \rr
\ll  n          \eps_4 + \eps_3 - \eps_1\, L_{\wsq, \,V_1} 
+ \eps_2\, A^+_{\wsq, \,V_1} \rr
\\
\ll  n          \eps_3 + \eps_4 - \eps_1\, L_{\bsq, \,V_2}^+ + \eps_2\, A_{\bsq, \,V_2} \rr
\ll  n          \eps_4 + \eps_3 - \eps_1\, L_{\bsq, \,V_2} 
+ \eps_2\, A^+_{\bsq, \,V_2} \rr
\\
\ll \Delta_{1 2} - n \eps_3 - \eps_4 - \eps_1\,   L_{\wsq, \, V_2} + \eps_2\, A^+_{\wsq, \, V_1} \rr
\ll \Delta_{2 3} - n \eps_4 - \eps_3 - \eps_1\, L^+_{\wsq, \, V_2} + \eps_2\,   A_{\wsq, \, V_1}  \rr
\\
\ll \Delta_{1 2} - n \eps_3 - \eps_4 + \eps_1\, L_{\bsq, \,V_1}^+ - \eps_2\, A_{\bsq, \,V_2} \rr
\ll \Delta_{2 3} - n \eps_4 - \eps_3 + \eps_1\, L_{\bsq, \,V_1}   - \eps_2\, A_{\bsq, \,V_2}^+ \rr
\end{array}
\rr
}{  
\ll
\begin{array}{c}
\ll  \ll n + 1 \rr  \eps_3 - \eps_1\, L_{\wsq, \,V_1}^+ + \eps_2\, A_{\wsq, \,V_1}  \rr
\ll  \ll n + 1 \rr  \eps_4 - \eps_1\, L_{\wsq, \,V_1} + \eps_2\, A^+_{\wsq, \,V_1}  \rr
\\
\ll  \ll n + 1 \rr  \eps_3 - \eps_1\, L_{\bsq, \,V_2}^+ + \eps_2\, A_{\bsq, \,V_2} \rr
\ll  \ll n + 1 \rr  \eps_4 - \eps_1\, L_{\bsq, \,V_2} + \eps_2\, A^+_{\bsq, \,V_2} \rr
\\
\ll \Delta_{1 2} - \ll n + 1 \rr  \eps_3 - \eps_1\,   L_{\wsq, \,V_2} + \eps_2\, A^+_{\wsq, \,V_1} \rr
\ll \Delta_{2 3} - \ll n + 1 \rr  \eps_4 - \eps_1\, L^+_{\wsq, \,V_2} + \eps_2\,   A_{\wsq, \,V_1} \rr
\\
\ll \Delta_{1 2} - \ll n + 1 \rr  \eps_3 + \eps_1\, L_{\bsq, \,V_1}^+ - \eps_2\, A_{\bsq, \,V_2}   \rr
\ll \Delta_{2 3} - \ll n + 1 \rr  \eps_4 + \eps_1\, L_{\bsq, \,V_1}   - \eps_2\, A_{\bsq, \,V_2}^+ \rr
\end{array}
\rr
} 
\label{1.point.qt.4D}
\end{multline}

\subsection{Comments on the structure of $W^{\, \cN = 2^\star}_{\bD}$}
As in the case of 
the 5D and 4D $\cN = 2$ $q t$-instanton partition functions that correspond to 
a 2D 4-point conformal blocks on the sphere, 
Equations \ref{1.point.00.5D}-\ref{1.point.qt.4D} show that 
the 5D and 4D $\cN = 2^\star$ $q t$-instanton partition functions that correspond 
to a 2D 1-point conformal block on the torus include $q t$-dependent terms that 
bring in new poles. In the following, we take the limit of 
$W^{\, \cN = 2^\star}_{\bD}$, which leads to an $\cN = 4$ instanton partition 
function. 
	
\subsection{The $\cN = 4$ limit of the $\cN = 2^\star $ instanton partition function}
In the conventions used in the present work, this limit is obtained by setting, 

\begin{equation}
\Delta_1 = \eps_2, \quad
\Delta_2 =  2 a - \eps_2, \quad
\Delta_3 =  \eps_2,
\label{massless.parameters}
\end{equation}

\noindent where $a$ is the Coulomb parameter of the $U \! \ll 2 \rr$ gauge group, 
or equivalently the charge of the Virasoro primary field that flows in the internal 
channel of the 1-point conformal block on the torus
\footnote{\,
The derivation of the parameters in Equation \ref{massless.parameters} is based 
on Equation 6.2 in \cite{foda.wu.01}
}. In 2D conformal filed theory terms, this limit corresponds to choosing the vertex 
operator insertion in the 1-point conformal block on the torus to be the identity operator.
In the absence of Macdonald parameters, the result is the character of the Virasoro highest 
weight representation that flows in the internal channel. We would like to check that, 
starting from $\cM^{\, q \, t}_{Y_1 Y_2 Y_3} \ll x, y\rr$, we reproduce this result in 
the limit $q \rightarrow t$, and check the $q t$-corrections. Using the parameters 
of Equation \ref{massless.parameters} in Equation \ref{1.point.00.4D}, 

\noindent 
\begin{equation}
\ll - 1 \rr^{|V_1|+|V_2|} \, 
\ll \cS^{\, 4D}_{\bV\, \bV\, \bD_{\, \pmb 0}} \rr_{\ll 0, \, 0\rr} = 1
\label{00.result}
\end{equation} 

\noindent If there were no $q t$-dependent terms, Equations \ref{1.point} 
and \ref{00.result} would combine to give,

\begin{equation}
W^{\, \cN = 2^\star}_{\bD} \bigr\vert_{\, q \rightarrow t}
=
\sum_{\bV}
\ll Q \rr^{ | V_1 | + | V_2 |}, 
\label{1.point.simple.00}
\end{equation}

\noindent which is the correct $\cR_{Y_1 Y_2 Y_3} \ll x, y \rr$ result. 
However, using $\cM^{\, q \, t}_{Y_1 Y_2 Y_3} \ll x, y \rr$, there is 
a $q t$-dependent term, 

\begin{multline}
\ll \cS^{\, 4D}_{\bV\, \bV\, \bD_{\, \pmb 0}} \rr_{\ll q, \, t\rr} = 
\\
\prod_{n=1}^\infty 
\prod_{\begin{subarray}{c} \wsq \in V_1  \\ \bsq \in V_2 \end{subarray}} 
\frac{
\ll
\begin{array}{c}
\ll  n \eps_3 + \eps_4 -          \eps_1\, L_{\wsq, \,V_1}^+ + \eps_2\,   A_{\wsq, \,V_1} \rr
\ll  n \eps_4 + \eps_3 - \eps_1\, L_{\wsq, \,V_1}   + \eps_2\, A^+_{\wsq, \,V_1} \rr
\\
\ll  n \eps_3 + \eps_4 - \eps_1\, L_{\bsq, \,V_2}^+ + \eps_2\,   A_{\bsq, \,V_2} \rr
\ll  n \eps_4 + \eps_3 - \eps_1\, L_{\bsq, \,V_2}   + \eps_2\, A^+_{\bsq, \,V_2} \rr
\\
\ll 2a + n \eps_3 + \eps_4 + \eps_1\, L_{\wsq, \, V_2} - \eps_2\, A^+_{\wsq, \, V_1} \rr
\ll 2a + n \eps_4 + \eps_3 + \eps_1\, L^+_{\wsq, \, V_2} - \eps_2\, A_{\wsq, \, V_1}  \rr
\\
\ll 2a + n \eps_3 + \eps_4 - \eps_1\, L_{\bsq, \,V_1}^+ + \eps_2\, A_{\bsq, \,V_2} \rr
\ll 2a + n \eps_4 + \eps_3 - \eps_1\, L_{\bsq, \,V_1} + \eps_2\, A_{\bsq, \,V_2}^+ \rr
\end{array}
\rr
}{  
\ll
\begin{array}{c}
\ll  \ll n + 1 \rr  \eps_3 - \eps_1\, L_{\wsq, \,V_1}^+ + \eps_2\, A_{\wsq, \,V_1}  \rr
\ll  \ll n + 1 \rr  \eps_4 - \eps_1\, L_{\wsq, \,V_1} + \eps_2\, A^+_{\wsq, \,V_1}  \rr
\\
\ll  \ll n + 1 \rr  \eps_3 - \eps_1\, L_{\bsq, \,V_2}^+ + \eps_2\, A_{\bsq, \,V_2} \rr
\ll  \ll n + 1 \rr  \eps_4 - \eps_1\, L_{\bsq, \,V_2} + \eps_2\, A^+_{\bsq, \,V_2} \rr
\\
\ll 2a + \ll n + 1 \rr  \eps_3 + \eps_1\, L_{\wsq, \,V_2} - \eps_2\, A^+_{\wsq, \,V_1} \rr
\ll 2a + \ll n + 1 \rr  \eps_4 + \eps_1\, L^+_{\wsq, \,V_2} - \eps_2\, A_{\wsq, \,V_1} \rr
\\
\ll 2a + \ll n + 1 \rr  \eps_3 - \eps_1\, L_{\bsq, \,V_1}^+ + \eps_2\, A_{\bsq, \,V_2} \rr
\ll 2a + \ll n + 1 \rr  \eps_4 - \eps_1\, L_{\bsq, \,V_1} + \eps_2\, A_{\bsq, \,V_2}^+ \rr
\end{array}
\rr
} 
\label{1.point.simple.qt}
\end{multline}

\subsection{Comments on the structure of $W^{\, \cN = 2^\star}_{\bD}$}
In the $q \rightarrow t$ limit, the right hand side of Equation \ref{1.point.simple.00} 
is an unevaluated version of the character of the Virasoro highest weight representation 
that flows in the internal channel of the 1-point function of the identity vertex operator 
on the torus, and $Q$ plays the role of the indeterminate $q$ that appears $q$-series 
expressions of Virasoro characters. 
In the presence of $q\ t$-terms, this is modified to,

\begin{equation}
W^{\, \cN = 2^\star}_{\bD} = \sum_{\bV} \ll Q \rr^{ | V_1 | + | V_2 |}
\, 
\ll \cS^{\, 4D}_{\bV\, \bV\, \bD_{\, \pmb 0}} \rr_{\ll q, \, t\rr}, 
\label{extended.virasoro.character}
\end{equation}

\noindent where $\ll \cS^{\, 4D}_{\bV\, \bV\, \bD_{\, \pmb 0}} \rr_{\ll q, \, t\rr}$ 
is defined in Equation \ref{1.point.simple.qt}, which has the interpretation of 
{\it \lq an extended Virasoro character\rq}, with infinitely-many additional states 
weighted by the factors on the right hand side of Equation \ref{1.point.simple.qt}

\section{Remarks}
\label{11}
{\it We conclude with a number of remarks on relevant literature, and open questions.}
\smallskip 

\subsection{Macdonald processes} The present work starts from the observation that 
there are two ways to refine MacMahon's partition function of plane partition, one
due to Iqbal {\it et al.} \cite{iqbal.kozcaz.vafa}, and one due to Vuleti\'c \cite{vuletic.02}, and 
these refinements are orthogonal in the sense that one can impose one or the other 
or both at the same time. 
The refinement of Iqbal {\it et al.} is natural in the sense that it is the right 
modification that leads to instanton partition functions that are in turn related 
to conformal blocks in 2D conformal field theories with $c \neq 1$. 
The refinement of Vuleti\'c is natural in the sense that it is the right 
modification that leads to Macdonald processes, a very rich topic in stochastic 
processes studied by A Borodin and collaborators in 
\cite{borodin.01, borodin.02, borodin.03, borodin.04}, and references therein.
We anticipate that the $q t$-vertex operators used in the present work, in addition 
to the $y$-refinement of Iqbal {\it et al}, can be useful in the study of Macdonald 
processes. 

\subsection{The physical meaning of the $q t$-deformation is an open problem}
The Macdonald vertex 
$\cM^{\, q\, t}_{Y_1 Y_2 Y_3} \ll x, y\rr$ depends explicitly on four parameters 
$\ll x, y, q, t\rr$, and implicitly on $R$, the radius of the M-theory circle. 
The original topological vertex 
$\cO_{Y_1 Y_2 Y_3} \ll x, y\rr$ is obtained in the limit 
$x \rightarrow y$, and 
$q \rightarrow t$. 
In the limit $R \rightarrow \infty$, gluing copies of 
$\cO_{Y_1 Y_2 Y_3} \ll x, y\rr$, one obtains 4D instanton partition functions, 
with Nekrasov parameters $\eps_1 + \eps_2 = 0$, which are equal to 
Virasoro conformal blocks with central charge $c=1$.

The refined topological vertex 
$\cR_{Y_1 Y_2 Y_3} \ll x, y\rr$ is obtained by keeping $x \neq y$, and taking 
the limit $q \rightarrow t$. 
In the limit $R \rightarrow \infty$, gluing copies of 
$\cR_{Y_1 Y_2 Y_3} \ll x, y\rr$, one obtains 4D instanton partition functions, 
with Nekrasov parameters $\eps_1 + \eps_2 \neq 0$, which are equal to 
Virasoro conformal blocks with central charge $c \neq 1$.
The point we wish to make is that, in 2D terms, in the limit $R \rightarrow 0$, 
the $y$-deformation of Iqbal {\it et al.} takes one from a 2D conformal field 
theory to another 2D conformal field theory, without affecting criticality.
At finite $R$, the above remarks remain the same, but the instanton partition 
functions are 5D, and the corresponding 2D objects are off-critical deformations 
of 2D conformal blocks. With reference \cite{shiraishi.02} and related works, 
$q t$-deformation definitely affects criticality. 

To discuss the physical meaning of the $q t$-deformations, it is simpler to work 
in the $R \rightarrow 0$ limit. In each of Equations \ref{A.norm.4D}--\ref{C.norm.4D}, 
there is a leading factor that remains intact as $\eps_3 \rightarrow \eps_4$, 
and infinitely-many factors, each of which $\rightarrow 1$, in the limit
$\eps_3 \rightarrow \eps_4$. We need to understand the $q t$-dependent terms.

\subsubsection{The 2D interpretation}
The $q t$-dependent factors, with their additional poles, are reminiscent of 
factors that appear in computations of $q t$-deformed expectation values in 
works such as \cite{shiraishi.02}. In the latter works, the $q t$-deformed 
expectation values are typically given in terms of contour integrals, and 
it is tempting to expect that the $q t$-conformal blocks in the present work 
are evaluations of the contour integrals, possibly in the context of statistical 
mechanical models based on Virasoro algebra in the presence of an additional 
$U \! \ll 1 \rr$ algebra, which typically greatly simplifies evaluations. 
More precise statements on this issue are beyond the scope of this work.

\subsubsection{The instanton partition function interpretation}
The infinitely-many additional poles are reminiscent of Kaluza-Klein particles 
that decouple as $q \rightarrow t$. If this is the case, then given that 
$\cR_{Y_1 Y_2 Y_3} \ll x, y\rr$ leads to 5D instanton partition functions, 
$\cM^{\, q\, t}_{Y_1 Y_2 Y_3} \ll x, y\rr$ leads to 6D instanton partition 
functions on Calabi-Yau threefolds given in terms of planar web diagrams. 

The problem with this 6D interpretation is that, in the $\cR_{Y_1 Y_2 Y_3} \ll x, y\rr$ 
literature \cite{hollowood.iqbal.vafa}, and in more recent works 
\cite{iqbal.kozcaz.yau, nieri, tan, mironov.morozov.zenkevich, kimura.pestun.01, kimura.pestun.02, kimura}, 
6D instanton partition functions are obtained by gluing copies of 
$\cR_{Y_1 Y_2 Y_3} \ll x, y\rr$ to form planar web diagrams, then gluing external legs 
to make closed loops and summing over all states that can propagate along these loops. 
This leads to expressions that are identified with 6D instanton partition functions. 
If the instanton partition functions obtained 
using $\cM^{\, q\, t}_{Y_1 Y_2 Y_3} \ll x, y\rr$ are 6D objects, then one needs 
to understand how they relate to the 6D expressions of 
\cite{iqbal.kozcaz.yau, nieri, tan, mironov.morozov.zenkevich, kimura.pestun.01, kimura.pestun.02, kimura}. 
This is another open problem that is beyond the scope of this work.

\subsection{Towards an elliptic extension of the Macdonald vertex}
In section {\bf \ref{ding.iohara.miki.algebra}}, we noted that the Macdonald vertex 
is related to Ding-Iohara-Miki algebra {\it via} the fact that the former is built 
from the same $q t$-vertex operators as the latter. 
In \cite{saito.01, saito.02, saito.03}, Saito presented an elliptic extension of 
the Ding-Iohara-Miki algebra, based on $p\, q t$-vertex operators that depend 
on an additional {\it \lq nome\rq\,} parameter $p$.

It is straightforward to extend the construction of the Macdonald vertex to 
an elliptic topological vertex that depends on $p$, such that we recover
the Macdonald vertex in the limit $p \rightarrow 0$, apart from the fact 
that it is not clear what the elliptic analogues of the Macdonald symmetric 
functions are. 

It is straightforward to propose an {\it ad hoc} solution to this problem by 
conjecturing that a class of symmetric functions that play the role of eigenstates 
of Saito's $p q t$-vertex operators exist, the same way that the Macdonald functions 
are eigenstates of the $q t$-vertex operators. As Saito's $p q t$-vertex operators 
depend on two free boson fields, the required elliptic Macdonald functions would be 
labelled by two Young diagrams \cite{foda.to.be.published}. 

\section*{Acknowledgements}
We thank M Bershtein, D Krefl, A Morozov, F Nieri, A Ram, J Shiraishi, Y Yamada, J Yang, 
and particularly A Polpitov, O Warnaar, and Y Zenkevich for comments and discussions on 
this work and on related topics, and H Segerman for help with the figures.

\end{document}